\documentclass[apj,twocolumn,twocolappendix,numberedappendix]{openjournal}
\usepackage{newtxtext,newtxmath,enumitem,multirow,ctable}
\usepackage[T1]{fontenc}
\usepackage[breaklinks,colorlinks,citecolor=blue,urlcolor=blue]{hyperref}

\newcommand{\Alf}{{Alfv\'en}}

\newcommand{\paperone}{Paper {\small I}}
\newcommand{\papertwo}{Paper {\small II}}
\newcommand{\paperthree}{Paper {\small III}}
\newcommand{\paperonetwo}{Papers {\small I} \&\ {\small II}}

\newcommand{\orcidauthor}[3]{\author{\href{http://orcid.org/#1}{#2$^{#3}$}}}

\shorttitle{FORGE'd in FIRE III}
\shortauthors{Hopkins et al.}

\begin{document}

\title{\vspace{-0.8cm}FORGE'd in FIRE III: The IMF in Quasar Accretion Disks from STARFORGE\vspace{-1.5cm}}

\orcidauthor{0000-0003-3729-1684}{Philip F. Hopkins}{1,*}
\orcidauthor{0000-0002-1655-5604}{Michael Y. Grudi{\'c}}{2,\dagger}
\orcidauthor{0000-0002-4086-3180}{Kyle Kremer}{1,\dagger}
\orcidauthor{0000-0003-1252-9916}{Stella S. R. Offner}{3}
\orcidauthor{0000-0001-5541-3150}{D\'avid Guszejnov}{3,\dagger}
\orcidauthor{0000-0003-4423-0660}{Anna L. Rosen}{4}
\affiliation{$^{1}$TAPIR, Mailcode 350-17, California Institute of Technology, Pasadena, CA 91125, USA}
\affiliation{$^{2}$Carnegie Observatories, 813 Santa Barbara St, Pasadena, CA 91101, USA}
\affiliation{$^{3}$Department of Astronomy, University of Texas at Austin, TX 78712, USA}
\affiliation{$^{4}$Center for Astronomy \&\ Space Sciences, University of California, San Diego, La Jolla, CA 92093, USA}

\thanks{$^*$E-mail: \href{mailto:phopkins@caltech.edu}{phopkins@caltech.edu}},
\thanks{$\dagger$NASA Hubble Fellow}

\begin{abstract}
Recently, we demonstrated self-consistent formation of strongly-magnetized quasar accretion disks (QADs) from cosmological radiation-magnetohydrodynamic-thermochemical galaxy-star formation simulations, including the full STARFORGE physics shown previously to produce a reasonable IMF under typical ISM conditions. Here we study star formation and the stellar IMF in QADs, on scales from $100\,$au to $10\,$pc from the SMBH. We show it is critical to include physics often previously neglected, including magnetic fields, radiation, and (proto)stellar feedback. Closer to the SMBH, star formation is suppressed, but the (rare) stars that do form exhibit top-heavy IMFs. Stars can form only in special locations (e.g.\ magnetic field switches) in the outer QAD. Protostars accrete their natal cores rapidly but then dynamically decouple from the gas and ``wander,'' ceasing accretion on timescales $\sim 100\,$yr. Their jets control initial core accretion, but the ejecta are ``swept up'' into the larger-scale QAD flow without much dynamical effect. The strong tidal environment strongly suppresses common-core multiplicity. The IMF shape depends sensitively on un-resolved dynamics of protostellar disks (PSDs), as the global dynamical times can become incredibly short ($\ll $\,yr) and tidal fields are incredibly strong, so whether PSDs can efficiently transport angular momentum or fragment catastrophically at $\lesssim 10\,$au scales requires novel PSD simulations to properly address. Most analytic IMF models and analogies with planet formation in PSDs fail qualitatively to explain the simulation IMFs, though we discuss a couple of viable models.
\end{abstract}


\maketitle

\section{Introduction}
\label{sec:intro}

How stars form, and what shapes their masses and other characteristics, are fundamental questions that inform almost every subfield of astrophysics. A key parameterization is the ``initial mass function'' (IMF) of stars - the distribution of stellar masses at the time of their formation. Observationally, the IMF exhibits remarkable uniformity across different observed environments \citep[for reviews, see e.g.][]{bastian:2010.imf.universality,kroupa:2011.imf.review,offner:2014.imf.review,hopkins:2018.imf.review}. In the local Solar neighborhood and similar ``typical'' galactic environments, a universal IMF is perhaps not surprising, as the properties of progenitor giant molecular clouds (GMCs) and the typical interstellar medium (ISM) fall along well-defined scaling relations within a relatively narrow parameter space \citep[e.g.][]{larson:gmc.scalings,solomon:gmc.scalings,rosolowsky:2003.gmc.rotation,bolatto:2008.gmc.properties,goodman:2009.gmc.column.dist,heyer:2009.gmc.trends.w.surface.density,pokhrel:2020.star.gas.correlation}. But this makes it all the more important to explore extreme environments which lie well outside of this space. And there are some hints of (modest) IMF variation in special environments: for example towards more bottom-heavy (with a lower-mass turnover) in some massive galaxy bulges \citep{van-dokkum:2011.imf.test.gcs,spiniello:2012.bottom.heavy.imf.massive.gal}, or towards more top-heavy (``standard'' but with a slightly shallower-than-Salpeter slope) in a couple of special environments very close to the supermassive black hole (SMBH) Sgr A$^{\ast}$ in our Galactic center \citep{nayakshin:2005.top.heavy.imf.mw.center,paumard:2006.mw.nucdisk.imf,maness:2007.mw.nucdisk.imf,espinoza:arches.massive.imf,bartko:2010.mw.nucdisk.imf,lockmann:2010.gal.center.imf,lu:2013.mw.center.imf,hosek:2019.arches.top.heavy.imf.1pt8.up.to.40.msun} and perhaps (much more tentatively) in similar environments around luminous quasars \citep[e.g.][]{toyouchi:2022.topheavy.imf.qso.disks.from.metals,fan.wu:2023.hints.of.top.heavy.imf.around.quasars}.

These Galactic center observations, in particular, have prompted a number of theoretical studies \citep[e.g.][]{nayakshin:sfr.in.clumps.vs.coolingrate,bonnell.rice:2008.gmc.infall.smbh.sf.sims,hobbs:2009.mw.nucdisk.sim,hocuk.spaans:2011.radiated.clouds.near.galactic.center.quasars.imf,hocuk:2012.b.fields.in.circumnuclear.smbh.imf.star.form,mapelli:2012.sf.around.sg.astar.sims,alig:2013.nuclear.disk.sims.idealized.hydro.only,davies:2020.massive.stars.via.accretion.in.agn.disks}, which have largely argued that the ``excess'' population of intermediate-mass stars in that neighborhood formed in some kind of accretion disk at $\lesssim 0.2\,$pc from the SMBH during an earlier accretion episode. Moreover, in recent years there has been a surge of theoretical interest in star formation in quasar\footnote{In this manuscript we will use the term ``quasar'' loosely to refer to any SMBH accreting at high luminosities, as opposed to e.g.\ the traditional optical luminosity and/or spectral criteria.} accretion disks (QADs) and the circum-quasar medium (CQM),\footnote{In this paper, for the sake of clarity we will refer to the medium on scales $\ll10\,$pc around the SMBH as the ``circum-quasar medium'' (CQM), and within the CQM we refer to the global, nuclear disk of gas and stars orbiting the SMBH in the galaxy center as the ``quasar accretion disk'' (QAD). The CQM we are careful to distinguish from the more ``typical'' interstellar medium (ISM or LISM) and giant molecular clouds (GMCs) observed in the Solar neighborhood. We are also careful to distinguish individual circum-stellar or proto-planetary or proto-stellar disks (themselves potentially embedded in the larger CQM or QAD) -- which we will refer to as ``protostellar disks'' (PSDs) throughout -- from the QAD itself. When we refer to a global ``position'' or ``radius'' $R$, we refer always to a coordinate frame in which the SMBH is at the origin (e.g.\ a radius $R$ from the SMBH), and we often refer to a cylindrical coordinate frame based on the orientation of the inner QAD (so $\phi$ is the prograde azimuthal angle rotating with the inner QAD and $z$ the vertical/angular momentum direction).} as a potential means to form hyper-massive and/or infinitely-long lived accreting stars or compact binaries which can act as a source for gravitational wave events (for LIGO/LISA) or tidal disruption events (TDEs) or other types of astrophysical transients \citep[e.g.][]{cuadra:agn.disks.as.bh.merger.host.sites,stone:2017.bh.binary.mergers.in.agn.disks,tagawa:2020.binaries.in.agn.disks.gw.sources,baruteau:2011.gal.center.binaries.in.disk.events,bortolas:2022.tdes.in.agn.acc.disks}

But star formation -- even in ``normal'' Solar-neighborhood environments -- is a highly non-linear, chaotic, multi-physics problem which necessitates detailed numerical simulations \citep[see discussion in][and references therein]{mckee:2007.sf.theory.review,offner:2014.imf.review}. It has only recently become possible for ``first principles'' simulations of star formation under Solar neighborhood/local ISM (LISM) conditions to reproduce the full behavior of the IMF after ``completion'' of star formation events in a GMC, and those simulations have argued that this depends on physics including self-gravity; hydrodynamics and magnetic fields; explicit radiation-hydrodynamics accounting for multiple wavelengths and self-consistently able to handle the transition between optically thin and thick cooling; detailed (potentially non-equilibrium) thermo-chemistry of dusty, partially-ionized atomic and molecular gas; self-consistent driving of turbulence, GMC surface densities, and other large-scale outer boundary or initial conditions; and detailed feedback from proto-stars and (pre)-main-sequence stars including (proto)stellar jets, stellar radiation, and stellar mass-loss \citep[e.g.][]{bate:2012.rmhd.sims,lee:2019.imf.magnetic.field.fx,hennebelle:2020.stellar.radiative.feedback,guszejnov:2020.starforge.jets,grudic:starforge.methods,lee:2021.non.ideal.mhd.fx.planet.disks,grudic:2022.sf.fullstarforge.imf,guszejnov:2022.starforge.cluster.assembly,guszejnov:environment.feedback.starforge.imf,guszejnov:starforge.environment.multiplicity}. Meanwhile, simulating the outer regions of QADs (where star formation may be possible) clearly requires all of these physics as well, with the structure of the QAD sensitive to magnetic field structure (both strengths and geometry/topology), and QADs clearly necessitating treatments that can handle radiation-pressure-dominated limits; dust sublimation and its effects on thermo-chemistry; and the ability to follow global modes such as spiral arms and gravito-turbulence (see \citealt{davis.tchekhovskoy:2020.accretion.disk.sims.review} for a recent review).

This highlights two major areas for improvement on previous-generation simulations of star formation in QAD/CQM environments. (1) All of the previous QAD IMF studies cited above used ad-hoc initial/boundary conditions, as they were modeling some portion of an arbitrarily analytically constructed QAD, as opposed to predicting these conditions self-consistently from simulations of larger scales actually producing inflows and forming the QAD. (2) None of the QAD IMF studies cited above included a combination of physics and numerical methods which has been actually shown to reproduce the Solar neighborhood IMF. Indeed, the vast majority of them considered only hydrodynamics (i.e.\ ignored magnetic fields), with idealized equilibrium thermochemistry (or even further simplified ideal gas equations-of-state) neglecting explicit multi-band radiation hydrodynamics, and did not include any kind of explicit stellar evolution or feedback. It is well-established that this combination of physics alone will, even in Solar-neighborhood environments, produce a systematically incorrect IMF, incorrect star formation rates/efficiencies, and far too much cloud-to-cloud or cluster-to-cluster IMF variation compared to observations \citep[e.g.][]{guszejnov.2015:feedback.imf.invariance,guszejnov:protostellar.feedback.stellar.clustering.multiplicity,guszejnov:imf.var.mw,guszejnov:2018.isothermal.nocutoff,guszejnov:2019.imf.variation.vs.galaxy.props.not.variable,guszejnov:2020.mhd.turb.isothermal.imf.cannot.solve}. 

In this paper, we therefore present a first attempt to address both of these issues. Regarding issue (1), in \citet{hopkins:superzoom.overview} (hereafter \paperone), we presented the first fully-cosmological galaxy and star formation simulations which employed an adaptive hyper-refinement technique to self-consistently resolve the formation of a QAD (the properties of which were studied in detail in \citealt{hopkins:superzoom.disk}, hereafter \papertwo) around a SMBH of $\sim 10^{7}\,M_{\odot}$ during a bright quasar phase (resolving the interior structure of the QAD down to scales of $<80$\,au, or $\sim 300$ Schwarzchild radii, from the SMBH). Regarding issue (2), these simulations utilize the STARFORGE modules and physics, including all of the specific physics discussed above (e.g.\ detailed non-equilibrium ion/atomic/molecular/dust thermo-chemistry coupled to multi-band explicit radiation transport; magnetic fields; proto-stellar and main-sequence stellar evolution with feedback from jets, radiation, winds, and supernovae), which has been explicitly shown to reproduce reasonable IMFs with minimal variation and higher-order properties (e.g.\ clustering, metallicity, star formation efficiencies) plausibly consistent with observational constraints for ``typical'' Solar-neighborhood-like environments \citep{guszejnov:2022.starforge.cluster.assembly,guszejnov:environment.feedback.starforge.imf,guszejnov:starforge.environment.multiplicity}. In \paperone\ and \papertwo, we showed that we resolve the transition to a weakly star-forming (i.e.\ not catastrophically-fragmenting) QAD, as required to sustain high accretion rates for quasar activity, and that the QAD is strongly magnetized with magnetic fields playing a fundamental role in the dynamics in a manner qualitatively distinct from the assumptions in the historical numerical and analytic studies cited above. 
Similar ``hyper-magnetized'' and magnetically ``flux-frozen'' disks have since been seen by subsequent attempts to self-consistently form QADs from large-scale gas flows under quite different conditions (and with different numerical methods/codes), including low-accretion-rate massive ellipticals \citep{guo:2024.fluxfrozen.disks.lowmdot.ellipticals} and accretion onto intermediate-mass BHs in star clusters \citep{shi:2024.imbh.growth.feedback.survey}. Moreover, such disks can naturally explain a wide variety of features in quasar spectra which cannot be produced in more traditional thermal-pressure-dominated accretion disks \citep{hopkins:multiphase.mag.dom.disks}. This suggests such conditions may be ubiquitous in nature, and provides yet more reason to explore the IMF in these conditions. Here, we present a first systematic study of the stellar IMF in both the QAD and CQM environments in these simulations, and explore how it depends qualitatively on physics not included in most previous studies.

In \S~\ref{sec:methods}, we review the numerical methods used in these simulations (\S~\ref{sec:methods:ov}), with special attention to how un-resolved PSDs/accretion flows and their accretion/feedback processes are treated around individual sink particles (\S~\ref{sec:methods:accretion}). In \S~\ref{sec:ism.diff}, we present and review from \paperonetwo\ some of the qualitative properties of the QAD and CQM and how they differ from the LISM, including: their surface density, density, and opacity scales (\S~\ref{sec:ism.diff:rho}); magnetic field strengths (\S~\ref{sec:ism.diff:B}); dust, thermal gas, and radiation temperature structures (\S~\ref{sec:ism.diff:T}); turbulent properties (\S~\ref{sec:ism.diff:V}); and tidal fields (\S~\ref{sec:ism.diff:tidal}). In \S~\ref{sec:results} we present and discuss the key results of the simulations. We first focus on the sites of star formation (\S~\ref{sec:where}), in infalling CQM filaments (\S~\ref{sec:where:filament}) and the outer QAD (\S~\ref{sec:where:disk}); we then discuss how these special regions are able to overcome magnetic support and collapse to form stars (\S~\ref{sec:how}). We explore the timescales over which stars form and accrete in \S~\ref{sec:speed}, and show how this relates to the formation of a large population of ``wandering stars'' in \S~\ref{sec:wandering} and has crucial effects on stellar accretion and feedback in \S~\ref{sec:accretion}. We discuss the global influence of (proto)stellar feedback on the environment in \S~\ref{sec:feedback}. We then consider the IMF directly (\S~\ref{sec:imf}), showing its general properties (\S~\ref{sec:imf:shape}), dependence on distance from the SMBH (\S~\ref{sec:imf:distance}) and evolution in time (\S~\ref{sec:imf:time}), exploring how it depends on physics such as the un-resolved treatment of PSDs on $\sim$\,au scales (\S~\ref{sec:imf:accmodel}, with further discussion of their fragmentation and the system-versus-single star IMF in \S~\ref{sec:imf:frag}) and magnetic fields (\S~\ref{sec:imf:bfields}). We briefly discuss multiplicity in \S~\ref{sec:imf:multiplicity}. We then discuss the physics driving the IMF shape (\S~\ref{sec:imf:physics}), specifically reviewing a large number of simple analytic scalings that have been proposed to explain the dependence of the IMF on environment which are clearly {\em not} describing our simulations (\S~\ref{sec:imf:physics:bad}) as well as physical explanations which appear to reproduce the simulations in both the weak feedback/slow PSD-accretion limit (\S~\ref{sec:imf:physics:good:turbfrag.for.slow}) and strong feedback/fast PSD-accretion limit (\S~\ref{sec:imf:physics:good:feedback.for.fast}). With this in mind we then review the extent to which star formation in the QAD is (or is {\em not}) analogous to planet formation in a ``typical'' PSD (\S~\ref{sec:sf.qad.planets.psd}). We discuss some caveats of resolution in \S~\ref{sec:resolution}, and then conclude in \S~\ref{sec:conclusions}.

\begin{figure*}
	\centering\includegraphics[width=0.92\textwidth]{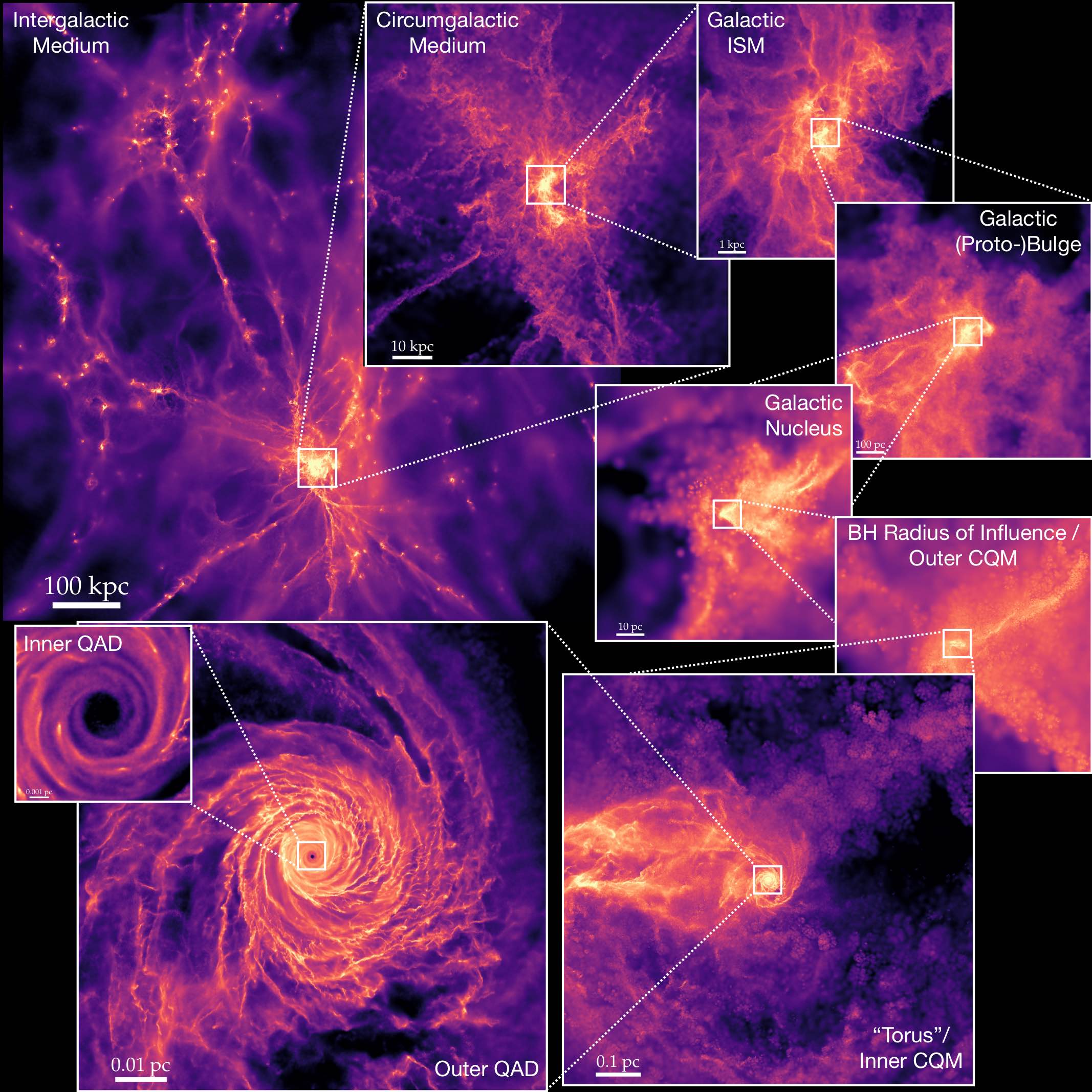} \\
	\centering\includegraphics[width=0.92\linewidth]{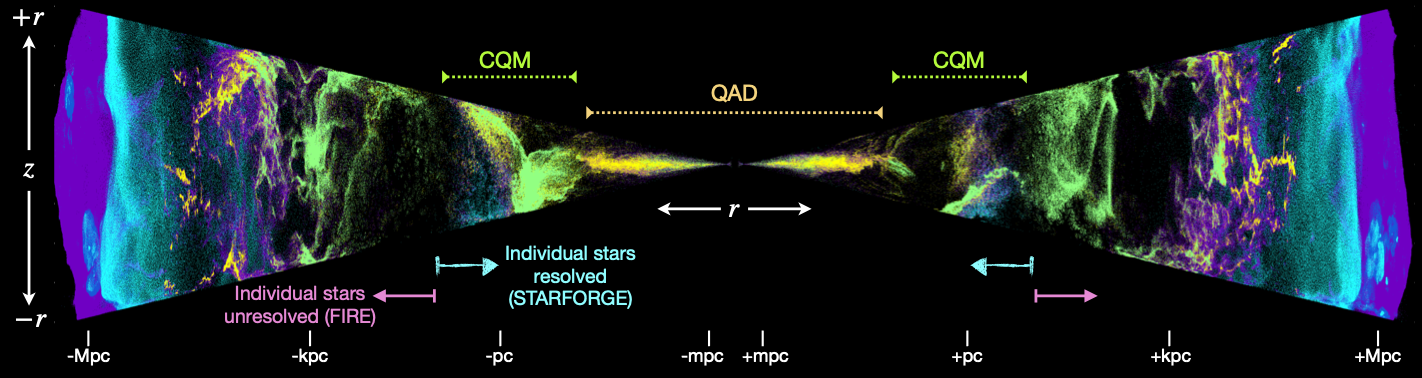}
	\caption{Projected simulation gas density (\S~\ref{sec:methods}) at one moment in time (redshift $z\approx 4.4$) when we ``zoom in'' around the QSO accretion disk (QAD) and circum-quasar medium (CQM).
	{\em Top:} Face-on, color encodes surface density increasing dark-to-light on a log scale (median pixels in the largest-scale panel have $N_{H} \sim 10^{19}\,{\rm cm^{-2}}$ or density $n_{H} \sim 10^{-5}\,{\rm cm^{-3}}$, while in the smallest-scale panel $N_{H} \sim 10^{27}\,{\rm cm^{-2}}$, $n_{H} \sim 10^{12}\,{\rm cm^{-3}}$). Adapted from Fig.~1 in \paperone\ (re-labeled with the names of characteristic scales used here),  to illustrate the ``parent'' galaxy (a merging, clumpy, gas-rich starburst with galactic SFR $>200\,{\rm M_{\odot}\,yr^{-1}}$) which provides the initial/boundary conditions for our study of SF here (where a star-forming complex of mass $\gtrsim 10^{8}\,M_{\odot}$ passes by a $\sim 10^{7}\,M_{\odot}$ SMBH and is partially tidally disrupted around the BHROI). Gas falls in as a tidally compressed stream in the CQM, which circularizes at $\sim 0.1\,$pc to form the QAD, which we follow down to $r=80\,$au from the SMBH. 
	{\em Bottom:} Edge-on, showing distance from SMBH ($r$) along the QAD midplane in a wedge of azimuthal angle ($|\sin{\phi}|<0.3$) versus vertical distance ($z$), both stretched logarithmically. Colors encode gas temperature: $T<10^{3}\,$K ({\em green}), $10^{3}<T<10^{4}$\,K ({\em yellow}), $10^{4}<T<10^{5}\,$K ({\em magenta}), $10^{5}<T<10^{6}\,$K ({\em purple}), $T>10^{6}\,$K ({\em cyan}). Adapted from Fig.~9 in \paperone, but labeling radii corresponding to the QAD and CQM, and denoting where the resolution is sufficiently high for our STARFORGE resolved-star-formation model. At larger radii, individual stars are un-resolved and the simulation uses the FIRE model for un-resolved fragmentation, so this part of the simulation informs our boundary conditions but cannot be used to predict the IMF. Subsequently, we focus exclusively on star formation interior to the BHROI, in the STARFORGE regime. 
	\label{fig:images.faceon.stylized}}
\end{figure*}

\section{Methods}
\label{sec:methods}

\subsection{Review and Key Physics}
\label{sec:methods:ov}

The simulations studied here are presented and extensively described in \paperone. Briefly, we begin from a $\sim (100\,{\rm cMpc})^{3}$ cosmological periodic box at redshift $z\sim 100$ with a primordial trace magnetic field, and follow it as a cosmological galaxy formation simulation following the {\em combined} Feedback In Realistic Environments (FIRE) project methods from \citet{hopkins:fire2.methods,hopkins:fire3.methods} and STARFORGE physics treatment from \citet{grudic:starforge.methods,guszejnov:2020.starforge.jets}. At a redshift $z\sim 4.5$ when a period of violent activity induces large inflows into the central $\sim\,$kpc of the galaxy, we hyper-refine \citep{daa:20.hyperrefinement.bh.growth} to go to higher and higher resolution, reaching sufficiently high resolution to resolve {\em individual} (proto)star formation, accretion and evolution and PSD structure in the central $\lesssim 10-100\,$pc of the galaxy. We continue to integrate self-consistently for a mix of ``individual star'' (STARFORGE) and ``stellar population'' (FIRE) particles given the methodology in \paperone, and continue to refine to a target resolution of $\Delta m <0.01\,{\rm M_{\odot}}$ in the central $\lesssim 1-10\,$pc, to follow gas inflows and QAD formation down to $<300$ Schwarzschild radii around the super-massive black hole of mass $\sim 1.3\times10^{7}\,{\rm M_{\odot}}$. Fig.~\ref{fig:images.faceon.stylized} shows an image of the gas properties on some of the wide range of scales simultaneously resolved in the simulation, illustrating the circum-BH QADs which form on sub-pc scales and form the focus of our study here.

The simulations include a wide range of physics including magnetic fields (using the high-order constrained-gradient method from \citealt{hopkins:mhd.gizmo,hopkins:cg.mhd.gizmo}), with anisotropic \citet{spitzer:conductivity}-\citet{braginskii:viscosity} viscosity and conduction using the mesh-free method in \citep{su:2016.weak.mhd.cond.visc.turbdiff.fx,hopkins:gizmo.diffusion}; a variable local cosmic-ray background \citep{hopkins:m1.cr.closure,hopkins:cr.spectra.accurate.integration,hopkins:cr.multibin.mw.comparison,hopkins:2021.sc.et.models.incompatible.obs} here modeled using the simple sub-grid method from \citet{hopkins:2022.cr.subgrid.model}; self-gravity with adaptive self-consistent softenings scaling with the resolution \citep[following][]{price:2007.lagrangian.adaptive.softening}, and Hermite integrators \citep{makino.aarseth.1992:hermite.integrator} capable of integrating $\gtrsim 10^{5}$ orbits in hard binaries \citep{grudic:2020.tidal.timestep.criterion,grudic:starforge.methods,grudic:2021.accelerating.hydro.with.adaptive.force.updates,hopkins:tidal.softening}; metal enrichment and dust and dust destruction/sublimation \citep{ma:2016.disk.structure,gandhi:2022.metallicity.dependent.Ia.rates.statistics.fire,choban:2022.fire.dust.growth.destruction.chemistry}; super-massive black hole seed formation and growth via gravitational capture of gas (\citealt{hopkins:qso.stellar.fb.together,shi:2022.hyper.eddington.no.bhfb,wellons:2022.smbh.growth}; following the physics in \citealt{hopkins:inflow.analytics,shi:2022.hyper.eddington.no.bhfb}); (proto)star formation and accretion and explicit feedback from stars in the form of protostellar jets, main-sequence stellar mass-loss, radiation, and supernovae \citep{grudic:2022.sf.fullstarforge.imf,guszejnov:2022.starforge.cluster.assembly,guszejnov:environment.feedback.starforge.imf,guszejnov:starforge.environment.multiplicity}. The simulations evolve explicit multi-band radiation-hydrodynamics with adaptive-wavelength bands \citep{hopkins:radiation.methods,hopkins:2019.grudic.photon.momentum.rad.pressure.coupling,grudic:starforge.methods} using the M1 method \citep{levermore:1984.FLD.M1} coupled explicitly to all the thermo-chemical processes, together with radiative cooling and thermo-chemistry incorporating cosmic backgrounds, radiation from local stars, re-radiated cooling radiation, dust/molecular/atomic/metal-line/ionized species, cosmic rays, and other processes. This allows us to self-consistently model the thermochemistry and opacities in gas with densities from densities $n \sim 10^{-8} - 10^{16}\,{\rm cm^{-3}}$ and temperatures $\sim 1-10^{10}\,$K in a range of radiation and cosmic ray environments (see \paperone). As we show below, the most important physics on these scales are gravity, magnetohydrodynamics (MHD), and radiation-thermodynamics, coupled to accretion onto and feedback from (proto)-stars.

Protostellar sink particle formation and accretion follow the methods in \citet{grudic:starforge.methods} for STARFORGE. Briefly, sink particles form with a mass of their parent gas cell, if and only if (1) they are above some minimum density; (2) they represent the density maximum among interacting gas neighbors; (3) there is no other sink within the interaction radius; (4) the cell density is increasing and velocity divergence is negative; (5) the cell is fully bound/self-gravitating at the resolution scale, including its turbulent, magnetic, and thermal support; (6) the free-fall time for collapse is faster than the tidal timescale or accretion/orbital timescale onto any other sink; and (7) the tidal tensor at the cell center-of-mass is fully compressive (possesses three negative eigenvalues), which ensures it is the local gravitational center/focus and can collapse even in extreme tidal environments like those here (for discussion of these criteria, see e.g.\ \citealt{federrath:2010.sink.particles,grudic:starforge.methods}). Once formed, sinks capture and remove from the gas flow neighboring gas cells if and only if: (1) those cells fall within the sink capture radius ($= 80\,$au); (2) the gas cell is bound to the sink (including its kinetic, thermal, and magnetic energy relative to the sink); (3) the cell possesses less angular momentum than a circular orbit around the sink at its separation; (4) its apocentric orbital two-body radius also falls inside $R_{\rm sink}$; (5) the volume of the cell is smaller than the volume enclosed by $R_{\rm sink}$; and (6) the gas cell is not eligible for capture by any other sink with a shorter free-fall time onto that sink.

On all scales we study in detail in this paper, the refinement has reached the target resolution of $< 0.01\,{\rm M_{\odot}}$, and so the simulation is fully in the STARFORGE limit (with no FIRE ``stellar population'' particles). These and other resolution properties are shown and discussed explicitly in Appendix~\ref{sec:resolution}. This is necessary for any meaningful predictions about the IMF, since the FIRE model used at much lower resolution in the Galactic radii $\gg 10\,$pc assumes a well-sampled Milky Way IMF, from which to model stellar populations.  In the most-dense regions around collapsing proto-stars, this mass resolution corresponds to a local spatial resolution $\sim 10\,{\rm au}$ and time resolution as small as $\sim$\,days. We discuss some potential effects of resolution on our conclusions in \S~\ref{sec:resolution}.

\subsection{Sub-Grid PSD/Reservoir Accretion Rates}
\label{sec:methods:accretion}

In the simulations here, when gas is captured by a (proto)stellar sink particle, it is first added to a ``reservoir'' of mass $M_{\rm PSD}$ which represents the sub-grid (by definition un-resolved in the simulation) portion of the PSD/accretion flow. This then accretes onto the (proto)star of mass $M_{\ast}$ at some (necessarily) prescribed rate $\dot{M}_{\ast} = f_{\rm acc}\,M_{\rm PSD}/t_{\rm dep}$, where $t_{\rm dep}$ is the assigned depletion time, and $f_{\rm acc}$ is the fraction which actually is retained in the (proto)star as opposed to ejected in outflows/protostellar jets. In previous studies of GMCs with typical Solar-neighborhood-like conditions, \citet{guszejnov:2018.isothermal.nocutoff,guszejnov:2020.starforge.jets,guszejnov:2022.starforge.cluster.assembly,grudic:starforge.methods} showed that predictions for stellar properties (the IMF, multiplicity) as well as cloud-scale properties (global star formation efficiencies, cloud dynamics) were relatively insensitive to the details of the model for $t_{\rm dep}$, so long as $t_{\rm dep}$ is relatively ``short'' compared to the global timescales of interest (e.g.\ cloud lifetimes or timescales for massive stars to accrete all of their mass from large scales around their initial core, of order $\sim$\,Myr). The default STARFORGE model is therefore a relatively simple prescription based on an isothermal \citet{shu:isothermal.sphere.collapse} sphere in LISM conditions (though for alternatives which depend on the accreted angular momentum explicitly, see discussion in \citealt{grudic:starforge.methods}), which -- inserting typical values for our resolution and the assumed default simulation parameters -- gives a more or less constant $t_{\rm dep}$ between $2000-5000\,{\rm yr}$. This easily satisfies the ``sufficiently fast' condition in LISM conditions. 

However, as we note below, on the scales we follow, the global dynamical/orbital times around the SMBH can be as short as months or even days. This also necessarily means any PSD/PPD/accretion reservoir at this radius must have an even shorter internal dynamical time or else it would be tidally disrupted. So assuming the same ``default'' $t_{\rm dep}$ leads to accretion timescales from the PSD/reservoir to the star which are very large compared to the global dynamical time, and thus mass can rapidly accumulate and the ``reservoir'' mass $M_{\rm PSD}$ can greatly exceed $M_{\ast}$, which is probably unphysical. We therefore denote this as the ``slow'' accretion regime. Given that this is, by definition, unresolved behavior in our simulations, a detailed study of the ``correct'' accretion rates or $t_{\rm dep}$ -- and an ultimate determination of whether these PSDs fragment efficiently (indeed whether a PSD even forms at all, or whether the accretion flow on small scales remains in some sort of quasi-spherical isothermal collapse) -- necessarily requires much higher-resolution idealized simulations which resolve the PSD scales and angular momentum transfer in detail given the conditions {\em predicted} by our simulations here. 

Therefore we bracket the range of interesting behaviors by re-running our fiducial simulation (restarting just before the first resolved STARFORGE sink actually appears) with identical physics except assuming $t_{\rm dep} \equiv {\rm MIN}[\,50\,{\rm yr}\ ,\ 100\,t_{\rm dyn,s}^{\rm max} ]$ where $t_{\rm dyn,s}^{\rm max} \equiv t_{\rm tidal} \equiv (\| {\bf T} \|^{2} / 6 )^{-1/4}$. Here $\| {\bf T} \|$ is the Frobenius norm of the gravitational tidal tensor ${\bf T} \equiv \nabla {\bf g}$ and the normalization is defined such that in a Keplerian potential (as we approximately have here with the SMBH dominant), $ (\| {\bf T} \|^{2} / 6 )^{-1/4} \approx 1/\Omega \approx (G\,M_{\rm BH}/R^{3})^{-1/2}$. As discussed below $t_{\rm tidal}$ is roughly the maximum possible dynamical time of the un-resolved PSD/accretion flow. We denote this as the ``fast'' accretion regime.

We have also experimented (re-running for a short period of time) with models using even shorter/faster $t_{\rm dep}$ (or even longer/slower than our ``slow'' model), as well as a couple of intermediate cases. We find that {\em on the scales resolved in our simulations} the key behaviors split into just two regimes, essentially depending on whether we assume the un-resolved accretion occurs rapidly (``fast'') compared to the global QAD/CQM dynamical time, or slowly (``slow''), but are insensitive to the details of $t_{\rm dep}$ within each regime. We therefore treat these two simulations as representative of these two broad limiting cases.

\begin{figure*}
	\centering\includegraphics[width=0.75\textwidth]{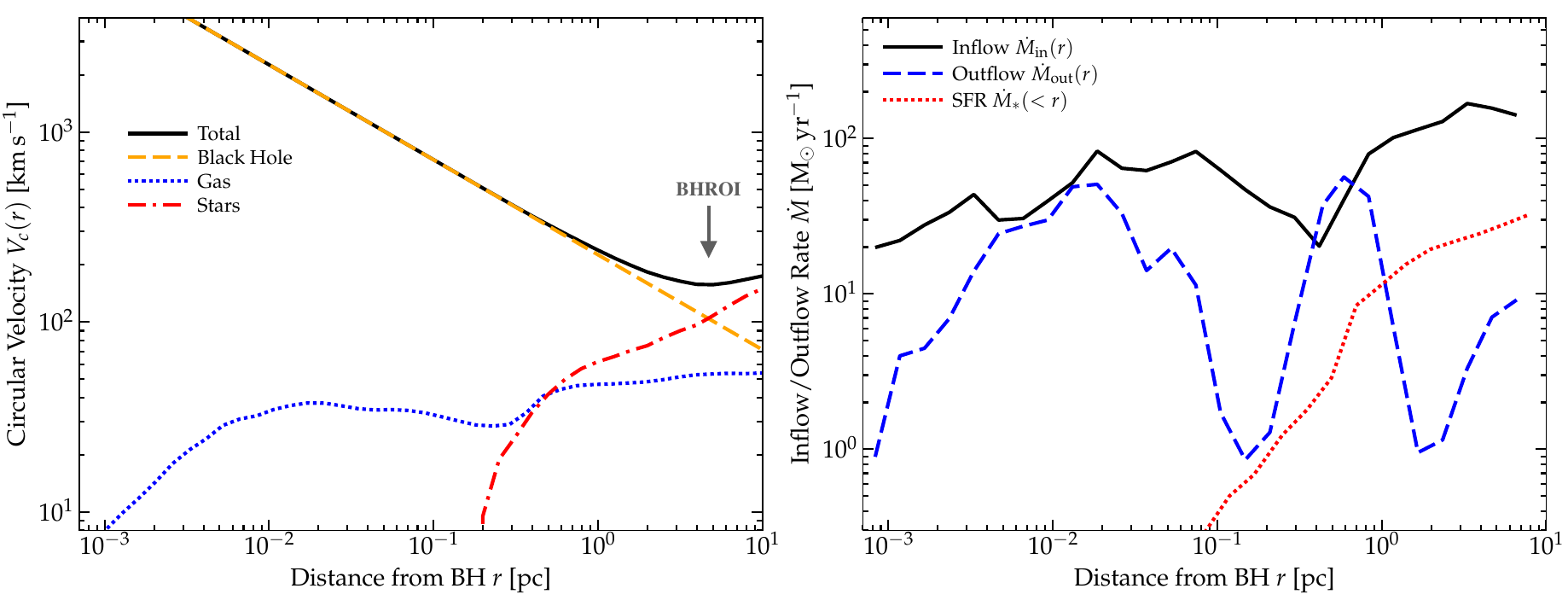}
	\centering\includegraphics[width=0.75\textwidth]{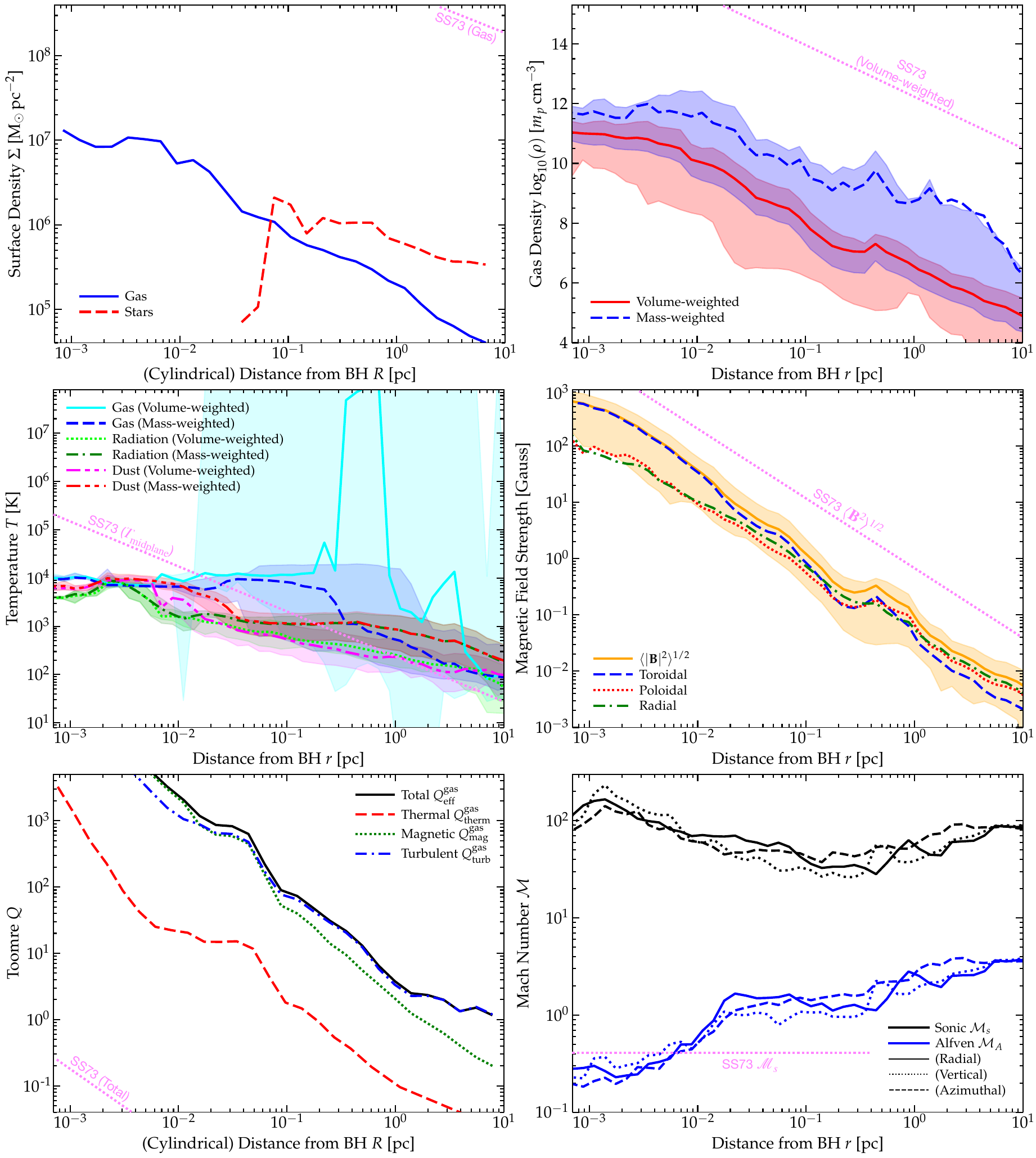}
	\caption{Properties of the QAD/CQM versus radius $r$ from SMBH (\S~\ref{sec:ism.diff}). 
	We focus exclusively on radii where STARFORGE physics applies inside our maximum-refinement region (Fig.~\ref{fig:images.faceon.stylized}).
	In salient panels, we compare the simulations to the analytic prediction for a standard \citet[][SS73]{shakurasunyaev73} $\alpha$-disk QAD ($\alpha=0.1$) extrapolated to the same radii.
	{\bf (1)} Circular velocity $V_{\rm c} \equiv \sqrt{G M_{\rm enc}(<r)/r}$, with contributions from different mass components (dark matter is present but negligible). The SMBH excision radius truncates the gas at $\lesssim 0.001\,$pc. We label the BHROI.
	{\bf (2)} Inflow $\dot{M}_{\rm in}$ and outflow $\dot{M}_{\rm out}$ rates through each annulus, and cumulative SFR (averaged over the previous $t_{\rm dyn}$, summed in radii $<r$). Inflow and outflow co-exist, but a stable $\dot{M}_{\rm in} \sim 10-100\,{\rm M_{\odot}\,yr^{-1}}$ persists. SF is suppressed relatively to inflow on small scales.
	{\bf (3)} Gas and stellar surface density $\Sigma$ in cylindrical shells. Surface densities are uniformly $\gtrsim 10^{4}\,{\rm M_{\odot}\,pc^{-2}}$, much higher than LISM GMCs.
	{\bf (4)} Gas density $\rho$, showing volume and mass-weighted means ({\em lines}) and $90\%$ inclusion intervals ({\em shaded}; mean can exceed this if the distribution has large ``tails'').
	{\bf (5)} Temperatures: we plot means and $90\%$ inclusion radii for gas, radiation, and dust temperatures (all explicitly evolved). 
	Temperatures in the QAD approach blackbody equilibrium (different opacities and $\Sigma$ explain differences with SS73). Dust sublimates in-code at $T_{\rm dust} \gtrsim 1500\,$K. 
	{\bf (6)} Magnetic field strengths: energy-weighted field strength and radial/toroidal/poloidal components. In and outside the CQM fields are crudely isotropic with a mild radial bias, but the mean toroidal component comes to dominate in the QAD.
	{\bf (7)} Toomre $Q$ accounting for thermal ($Q_{\rm therm}$), magnetic ($Q_{\rm mag}$), turbulent ($Q_{\rm turb}$), or combined support.  The outer CQM has turbulent $Q\sim 1$ with thermal $Q<1$, but even the thermal-only $Q\gg1$ in the QAD. 
	{\bf (8)} Sonic ($\mathcal{M}_{s} \equiv \delta v_{\rm turb}/c_{s}$) and \Alf{ic} ($\mathcal{M}_{A} \equiv \delta v_{\rm turb}/v_{A}$) Mach numbers (mass-weighted), for each component of the random motions in the midplane. Dispersions are highly supersonic, trending from super-\Alf{ic} to sub-\Alf{ic} at decreasing $r$. 
	While $\Sigma$, $\rho$, and $|{\bf B}|$ are vastly larger than in the LISM, the QAD is vastly lower-density compared to the SS73 model, owing to strong magnetic support (\paperthree; note even though $|{\bf B}|$ is lower than in SS73, $v_{A}$ and $v_{A}/c_{s}$ are orders-of-magnitude larger, owing to the different densities). This means $Q$ is larger here by a factor $\gtrsim 10^{5}$ and $\mathcal{M}_{s}$ larger by a factor $\gtrsim 100$.
	\label{fig:profile.dynamics}}
\end{figure*}

\section{Summary of Key Differences Between the CQM/QAD and the ``Normal'' ISM and $\alpha$-Disk QAD Models}
\label{sec:ism.diff}

Before going forward, it is useful to review and differentiate the extreme conditions in the simulations here from even the ISM of galactic nuclei, let alone the Solar neighborhood LISM (the focus of the vast majority of IMF literature). We refer to the gas interior to the BHROI at $\lesssim 5\,$pc scales as the CQM, and the material circularized into an accretion disk around the SMBH on $\lesssim 0.1$\,pc scales as the QAD. The properties discussed below are studied in more detail in \paperone\ and \papertwo, but our intent here is to provide necessary context for our results. Some zeroth-order quantitative properties of the medium are presented in Fig.~\ref{fig:profile.dynamics}.\footnote{Adapted from portions of Figs.~7,\,10,\,12 in \paperone, and Figs.~5 \&\ 22 in \papertwo, but consolidating the relevant information, plotting it at the same times analyzed here, adding additional data/comparisons, and zooming in to the range of radii we actually study.}

Some useful ``global'' numbers including the enclosed gas mass and dynamical times $t_{\rm dyn} \equiv r/V_{c}(r)$ as a function of radius $r$ can be read off from Fig.~\ref{fig:profile.dynamics} (and Appendix~\ref{sec:resolution}), given $\Omega \equiv V_{c}/r$ (and noting the global tidal field scales as $\sim \Omega^{2}$), alongside inflow, outflow, and star formation rates. We focus here on scales interior to the BH radius of influence (BHROI; defined as $R_{\rm BHROI} \sim G M_{\rm BH}/\sigma_{\rm gal}^{2}$ in a galactic potential with velocity dispersion $\sigma_{\rm gal}$) at several pc, interior to which (by definition) the BH dominates the global gravitational forces. As shown in \papertwo, properties such as the inflow rates in Fig.~\ref{fig:profile.dynamics} are time-steady over the duration of our simulation ($\sim 10^{5}\,t_{\rm dyn}^{\rm inner}$). 

In Fig.~\ref{fig:profile.dynamics} we also compare the simulation properties to the scalings predicted by the traditional \citet{shakurasunyaev73} (henceforth SS73) $\alpha$-disk QAD model, assuming the typical $\alpha=0.1$, with the same $\dot{M}$ as the simulations from Fig.~\ref{fig:profile.dynamics}. This comparison, and the fundamental differences in accretion disk properties between the magnetically-dominated disks seen in the simulations and SS73-like models (which fundamentally assume relatively weak magnetic support, with \Alf\ speeds much smaller than thermal sound speeds), are discussed at length in \papertwo\ and \citet{hopkins:superzoom.analytic} (from which the scalings in Fig.~\ref{fig:profile.dynamics} are taken, henceforth \paperthree). We show this because a number of recent analytic studies assuming properties of stars forming in QADs (see references in \S~\ref{sec:intro}) have {\em assumed} the QAD to have properties given by an SS73-like model, which Fig.~\ref{fig:profile.dynamics} shows can be orders-of-magnitude different from the conditions here, in ways that have important consequences below.

These QAD properties are also radically distinct from those assumed for ``self-gravitating QADs'' in most analytic studies \citep[e.g.][]{paczynski:1978.selfgrav.disk,kolykhalov.syunyaev:1980.qso.acc.disk.outer.frag.sf,collins.zahn:1999.qso.selfgrav.disk.sf,collins.zahn:1999.qso.selfgrav.disk.sf.evol,levin:2003.selfgrav.disk.sf,goodman:qso.disk.selfgrav,levin:2007.sefgrav.disk.sf,goodman.tan:2004.qso.disk.supermassive.stars}. Specifically, those self-gravitating QADs are much more similar to SS73 (often taking SS73 as an initial/boundary condition), in the properties shown in Fig.~\ref{fig:profile.dynamics}.

\subsection{High (Surface and 3D) Densities and Optical Depths}
\label{sec:ism.diff:rho}

Per Fig.~\ref{fig:profile.dynamics}, on sub-pc scales, the surface and 3D midplane gas density in the QAD/CQM scales approximately as $\langle \Sigma_{\rm gas} \rangle \sim 2\times10^{5}\,({\rm pc}/R)\,{\rm M_{\odot}\,{\rm pc}^{-2}}$ and $\langle \rho\rangle \sim 3\times 10^{6}\,m_{p} {\rm cm^{-3}}\,({\rm pc}/R)^{2}$ (in a volume-weighted sense, or about $\sim 30-100$ times larger in a mass-weighted sense). 
Thus the least-dense CQM structure that could be considered an ``overdensity'' in our simulations is already $\sim 5000$ times higher column density than a typical Solar-circle GMC, and $\gtrsim 1000$ times more dense in 3D than a typical protostellar core. This, in turn, means that even in the least-dense outer regions, the ``diffuse'' CQM is opaque to optical/UV/EUV radiation and already marginally optically thick to IR cooling radiation. 

That said, the densities are also much {\em lower} than those of a canonical SS73 disk with the same $\dot{M}$. As discussed at length in \papertwo\ and \paperthree, this owes to the combination of strong magnetic fields and turbulence ``puffing up'' the disk but also providing strong  stresses, so the same $\dot{M}$ corresponds to much lower surface densities. This leads to very different thermal structure, as discussed in \citet{hopkins:multiphase.mag.dom.disks}.

\subsection{Extremely Strong Magnetic Field Scales}
\label{sec:ism.diff:B}

Typical magnetic field strengths in the CQM scale as $|{\bf B}| \sim 0.1\,{\rm G}\,({\rm pc}/R)^{4/3}$ (Fig.~\ref{fig:profile.dynamics}). Thus the {\em ambient} gas before collapse into a PSD can already have $|{\bf B}|\gg {\rm G}$. These are strong in an absolute sense, but also relative to other sources of pressure. As shown in \papertwo, the plasma $\beta \equiv c_{s}^{2}/v_{A}^{2} \sim 10^{-4}$ in the diffuse CQM -- much smaller than typical GMCs. This is also orders-of-magnitude smaller than assumed in the SS73 model (which by definition assumes $\beta \gg 1$).\footnote{As discussed in \paperthree, the {\em absolute} value of $|{\bf B}|$ is actually larger in an SS73 disk (estimated here as in \paperthree\ by assuming the Maxwell $R\phi$ stress driving $\alpha$ is comparable to or larger than the Reynolds stress, so similar to the value ``$H$'' therein, but SS73 disks could have even stronger mean vertical fields), because the densities are so enormous by comparison to the simulations here that producing even a modest Maxwell stress to explain $\alpha \sim 0.1$ requires extremely large $|{\bf B}|$. But the \Alf\ speeds $v_{A}$ and ``magnetic support'' of the disk (e.g.\ $v_{A}/V_{c}$) are vastly smaller in SS73 compared to the simulations here.}

Fig.~\ref{fig:profile.dynamics} shows the magnetic Toomre $Q_{\rm mag}$ parameter is $\gtrsim 100$ everywhere in the QAD and $\gtrsim 1$ throughout the entire CQM. Related, we show below that the magnetic critical mass is larger than the entire QAD mass on sub-pc scales. This means the field is strong enough to ensure the QAD is stable against gravito-turbulence in the classical sense even where the thermal-only Toomre $Q_{\rm thermal}$ is modest. In contrast, for a ``typical'' thermal-pressure supported SS73 disk, the predicted $Q_{\rm SS73}$ is $\sim 10^5-10^{7}$ times smaller. Indeed, absent magnetic fields, the QAD fragments catastrophically at $\sim$\,pc scales (\paperone).

As shown in \papertwo\ and visually below, the fields within the QAD are primarily toroidal/azimuthal, albeit with non-negligible radial and poloidal/vertical components. However there can be some regions of locally lower $|{\bf B}|$, and some regions where the field switches direction (owing to its being the flux-frozen relics of originally more isotropically turbulent ISM fields amplified as they fall around the BH). In the CQM, where the inflow has the form of an infalling filamentary, tidally-disrupted molecular cloud complex, the magnetic fields are more isotropically turbulent/disordered (see \papertwo\ for details), with a mild radial bias, more akin to LISM molecular clouds in geometry.

\subsection{Warm Thermal Dust, Gas, \&\ Radiation Temperatures}
\label{sec:ism.diff:T}

Most of the QAD mass is in a warm neutral medium (WNM)-like phase of atomic gas with temperatures of a few thousand Kelvin and free electron fractions $\sim 1-10\%$, and metallicities close to ($\sim 1/2$) Solar (Fig.~\ref{fig:profile.dynamics}). As shown in \papertwo, cooling times are short compared to dynamical times ($t_{\rm cool} \ll t_{\rm dyn}$) throughout the CQM/QAD, like in the LISM. However, given the high densities and optical depths noted above, the dust/radiation/gas temperature of the midplane phases are coupled, and scale as approximately $T_{\rm rad} \sim T_{\rm dust} \sim 300\,{\rm K}\,({\rm pc}/R)^{1/2}$ (Fig.~\ref{fig:profile.dynamics}). There is some gas cooler than this but not much, so much of the molecular gas in the diffuse QAD is actually quite warm at $\gtrsim 1000\,$K. 

The warm temperatures owe to a combination of optical depth effects, heating by stellar irradiation (this is the center of a massive starburst galaxy with star formation rate $\gtrsim 100\,{\rm M_{\odot}\,yr^{-1}}$), and heating by the accretion luminosity itself ($\sigma_{\rm B}\,T_{\rm eff}^{4} \sim \dot{M}\,\Omega^{2}$). Indeed, the ``interstellar radiation field'' (ISRF) in the CQM here has an energy density $\sim 10^{7}\,({\rm pc}/R)^{2}\,{\rm eV\,cm^{-3}}$. Moreover, because of its coupling to these large temperatures, dust begins to sublimate in the QAD interior to $R \lesssim 0.1\,$pc.

\subsection{Large Turbulent Velocities}
\label{sec:ism.diff:V}

The turbulence in the QAD/CQM is highly super-sonic (sonic Mach numbers $\mathcal{M}_{s} \sim 30-300$, related physically to the rapid cooling noted above) and trans-\Alf{ic} (\Alf\ Mach numbers $\sim 0.3-3$), shown in Fig.~\ref{fig:profile.dynamics}. In these dimensionless terms this is not so different from some cold ISM phases. However recalling the large temperature and magnetic field scales, this translates to turbulent velocities at the driving scale (of order the QAD scale height $H \sim (0.1-0.3)\,R$) of $\gtrsim 200\,{\rm km\,s^{-1}}$ at all radii, rising to as large as $\sim 1000\,{\rm km\,s^{-1}}$ in the innermost regions at $\lesssim 0.001\,{\rm pc}$.  So there can be extremely large post-shock temperatures and compression ratios. This differs dramatically from an SS73 disk, where (by assumption) the turbulence is always subsonic.

\subsection{Extremely Strong Tidal Forces}
\label{sec:ism.diff:tidal}

Of special importance in a circum-BH environment, the tidal field is extremely strong, with $|{\bf T}| \equiv |\nabla {\bf g}| \equiv |\nabla \otimes \nabla \Phi | \sim G\,M_{\rm BH}/R^{3} \sim \Omega^{2}$. The dynamical time $t_{\rm dyn} \sim 1/\Omega \sim 4000\,{\rm yr}\,(R/{\rm pc})^{3/2}$,  $\sim 1\,$month at $R\sim 0.001\,{\rm pc}$ (compare this to $1/\Omega \sim 40\,{\rm Myr}$ in the Solar neighborhood). This in turn means the tidal field $\sim \Omega^{2}$ is stronger than that in the Solar neighborhood by a factor of $\sim 10^{8}\,({\rm pc}/R)^{-3}$.

This tidal field has immediate consequences for the structure of PSDs/PPDs in the CQM. 
As already noted, these short timescales force us to consider different prescriptions for the depletion timescale $t_{\rm dep}$ from the un-resolved PSDs onto (proto)stars. But they also have important implications for PSD structure and stability and stellar evolution, which we will discuss in more detail below.

\section{Results \&\ Discussion}
\label{sec:results}

Having described the simulation methods (\S~\ref{sec:methods}) and emergent properties of the QAD/CQM (\S~\ref{sec:ism.diff}) in which we will study SF, we now present the results. 

\begin{figure*}
	\centering\includegraphics[width=0.95\textwidth]{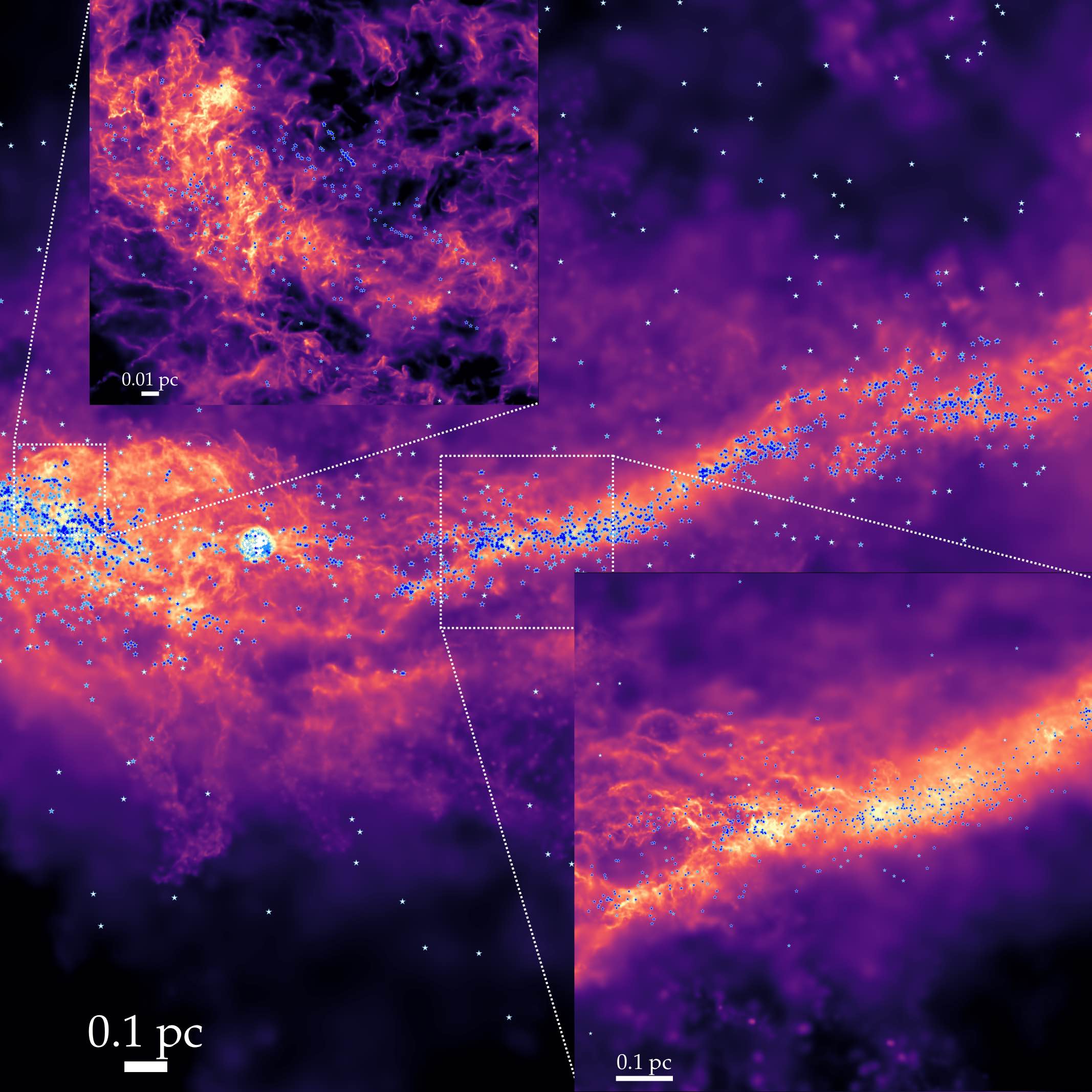} 
	\caption{Image illustrating the locations of star formation in the CQM/QAD (\S~\ref{sec:where}). We plot the projected density of gas (face-on to the inner QAD) akin to Fig.~\ref{fig:images.faceon.stylized}, and superimpose locations of individual (proto)stars ({\em points}). Stars are colored by ``dynamical age'': time since formation $t_{\ast} < t_{\rm dyn}$ at their current radius ({\em dark blue}), or $<10\,t_{\rm dyn}$ ({\em light blue}) or older ({\em white}). Inset images highly two sub-regions. SF primarily occurs in two regions here: (1) the free-falling, tidally compressed filament fueling the QAD (tidally stripped from larger-scale massive cloud-complexes) in the CQM, and (2) in the outer QAD. The filamentary regions are analogous to LISM-like dense filamentary star-forming regions, though they exhibit structure on much smaller absolute physical owing to the extreme densities and tidal fields (individual ``cores'' and shock sub-structures have scales $<0.001$\,pc, as compared to $\sim 0.1$\,pc). 
	\label{fig:wheresf.global}}
\end{figure*}

\begin{figure*}
	\centering\includegraphics[width=0.95\textwidth]{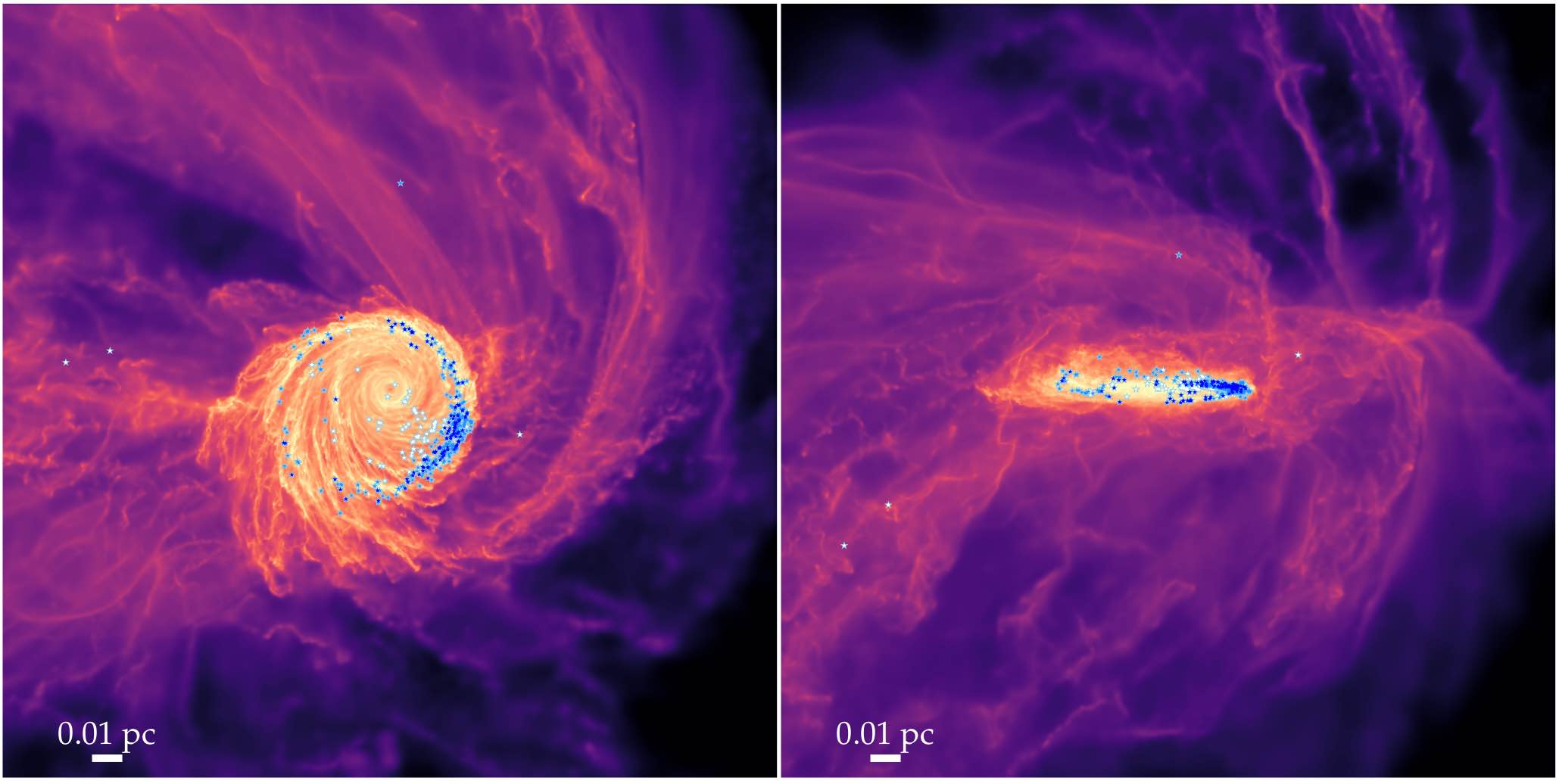} 
	\caption{As Fig.~\ref{fig:wheresf.global}, but focusing on the QAD (projected face-on and edge-on). SF occurs around where gas circularizes and is stabilized by magnetic fields but is still appreciably gravito-turbulent with a modest {\em thermal-only} Toomre $Q_{\rm thermal} \sim 1-10$ (Fig.~\ref{fig:profile.dynamics}). In the inner QAD at $\ll 0.01\,$pc, where even the thermal-only $Q_{\rm thermal}$ rises to $\gg 10^{2}-10^{3}$, star formation becomes nearly impossible: all stars at these radii are dynamically ``old,'' and formed at larger radii on highly-eccentric orbits.	\label{fig:wheresf.qad}}
\end{figure*}

\begin{figure*}
	\centering\includegraphics[width=0.95\textwidth]{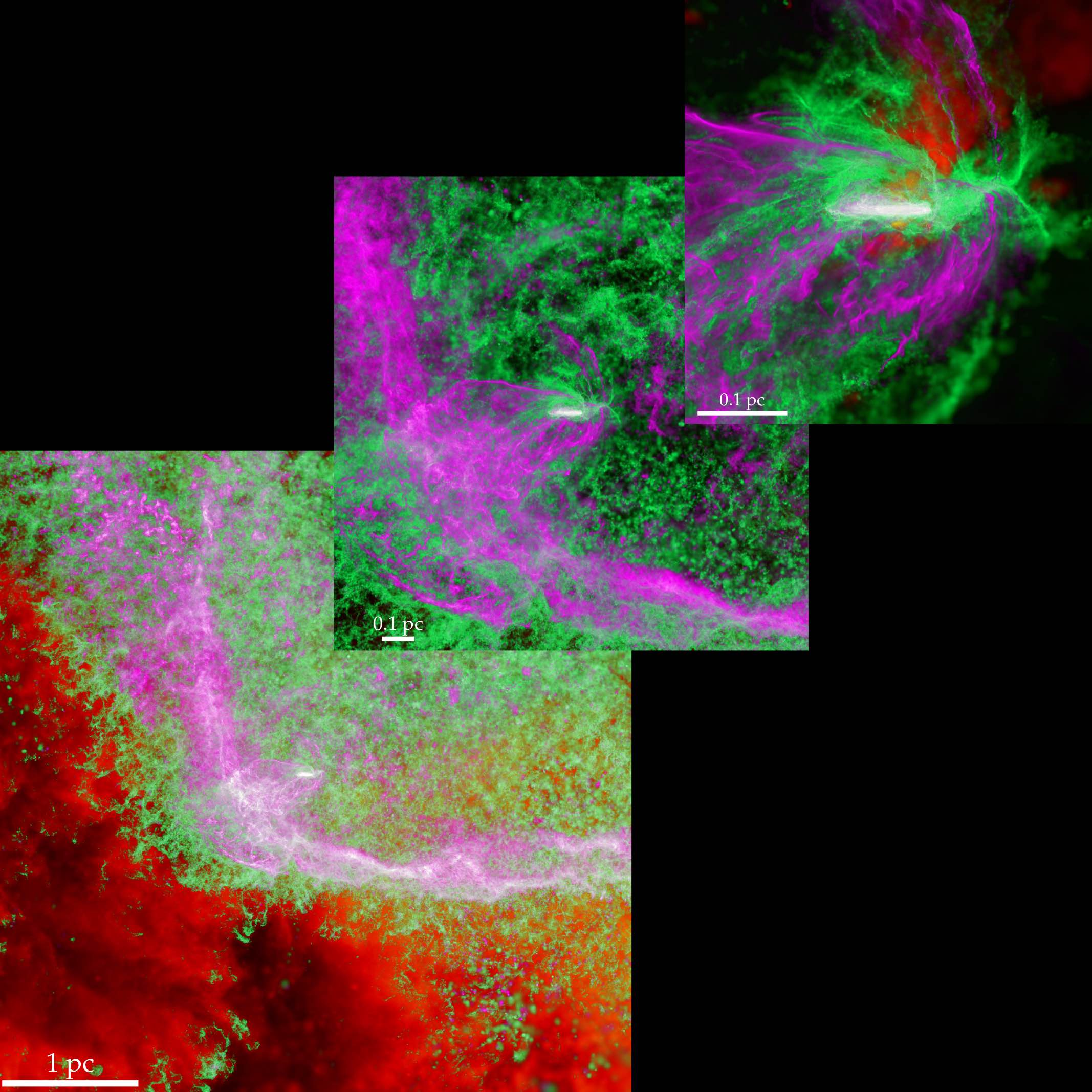} 
	\caption{Images highlighting the multi-phase structure of the gas over the same range of scales as Figs.~\ref{fig:wheresf.global}-\ref{fig:wheresf.qad}. We plot projected density edge-on, colored by gas temperature: $\ll 10^{4}\,$K ({\em magenta}), $\sim 10^{4}\,$K ({\em green}), and $\gg 10^{4}$\,K ({\em red}). Warm/hot, low density gas is present in the CQM from shocks and supernovae explosions, and vents from the inner QAD in a jet-like structure. The outer CQM also exhibits colder ``clouds'' embedded in warm gas, but the tidally-disrupted GMC-like complex is much more visible as the extended cold ``stream,'' with some material free-falling onto the QAD, until it is heated via accretion and inefficient radiative cooling to temperatures $\gtrsim 1000\,$K. 
	\label{fig:wheresf.phase}}
\end{figure*}

\subsection{Where do stars form in the QAD \&\ CQM?}
\label{sec:where}

Fig.~\ref{fig:wheresf.global} ``zooms in'' to show the gas and stars in the CQM, and Fig.~\ref{fig:wheresf.qad} does the same in the QAD, at a time near the end of the evolved simulation duration (well after it reaches a quasi-steady-state in the QAD). We highlight some key features (some of which are also clear in the gas-only morphology shown in Fig.~\ref{fig:images.faceon.stylized}). First, the gas exhibits significant inhomogeneity. Fig.~\ref{fig:wheresf.phase} shows that this is reflected as well in the gas phase structure. Second, the positions of the (proto)stars do not appear particularly correlated with the gas structure. We discuss this in more detail below in \S~\ref{sec:wandering}, but it fundamentally relates to the fact that the dynamical times at these radii are vastly shorter than even proto-stellar evolution times (\S~\ref{sec:ism.diff}), so stars can move far from their ``birth'' locations in timescales as short as years.

If we wish to understand ``where star formation occurs,'' it is more useful to look at stars which have just formed. Highlighting the youngest stars/sinks in Figs.~\ref{fig:wheresf.global}-\ref{fig:wheresf.qad}, we see that there are two qualitatively distinct ``sites'' of star formation.

\subsubsection{Star Formation in Infalling CQM Filaments}
\label{sec:where:filament}

At radii between $\sim 0.1-10$\,pc, star formation is largely occurring within relatively large-scale filamentary structures composed of gas captured by the BH, falling onto it in tidal streams, highlighted in Figs.~\ref{fig:wheresf.global} \&\ \ref{fig:wheresf.phase}. On these scales there is no QAD to speak of: as discussed in \paperone, the host galaxy is a chaotic, highly turbulent, clumpy, massive, gas-rich system undergoing a major merger at $z\sim 4.4$, and the quasar event here is related to a close passage (induced by the merger) of one extremely massive molecular cloud complex (gas mass $> 10^{8}\,M_{\odot}$) by the galactic nucleus containing a SMBH with mass $\sim 10^{7}\,M_{\odot}$. Some material is tidally stripped off of the cloud and captured in this event, so it initially free-falls onto the BH from the BH radius of influence (where the BH begins to dominate the potential) at a few pc. This forms the ``parent'' filament we see. 

This infalling filament is {\em already} part of a massive star-forming complex when it is captured from the ISM, so it is not surprising that star formation continues. As discussed in \papertwo, as it falls in, it is tidally stretched in the radial/infall direction while being compressed in both perpendicular directions, so it becomes more visibly elongated/filamentary but the local 3D density of the dense clumps within it is not suppressed and is actually tidally enhanced/compressed. More formally, it already has a sufficiently low virial/Toomre parameter ($Q_{\rm eff} \sim 1$ even including turbulent+magnetic support at $r \gtrsim 1\,$pc, with thermal-only $Q_{\rm thermal} \ll 1$; see Fig.~\ref{fig:profile.dynamics}), and sufficiently rapid cooling ($t_{\rm cool} \ll t_{\rm dyn}$), and relatively low optical depth to its own cooling radiation, such that we should expect fragmentation. Likewise while the magnetic fields are strong, the turbulence is still trans-to-super \Alf{ic} (Fig.~\ref{fig:profile.dynamics}), and the magnetic field geometry is not strongly toroidal on these scales, but more quasi-isotropic (meaning the fields will be less efficient at preventing collapse).

In this regime, therefore, the star formation is at least ``expected,'' and reasonably analogous to star formation in dense filamentary structures in the LISM \citep[see e.g.][for a review]{pineda:2023.ppvii.star.formation.structures.ism}. We stress that {\em absolute values} of most parameters here -- densities, magnetic field strengths, turbulent velocities, etc.\ -- are orders-of-magnitude different from the LISM (\S~\ref{sec:ism.diff}), so the problem is rescaled and not exactly analogous. For example, (1) the high surface density/acceleration scales mean we should be in the regime where stellar feedback has a weak effect on the dynamics \citep{grudic:sfe.cluster.form.surface.density} and we confirm this below; (2) the high optical depths and strong radiation background lead to $T_{\rm dust} \sim T_{\rm rad} \sim T_{\rm gas} \sim 100-1000$\,K; (3) the absolute size scales of structures are much smaller, with for example the sonic scale of the turbulence (scale below which rms turbulent velocity fluctuations become sub-sonic, and density fluctuations become small, approximately $\sim H/\mathcal{M}_{s}^{2}$ in a supersonically turbulent disk)  extending to $< 0.001\,$pc.

\subsubsection{Star Formation in the Rotating QAD}
\label{sec:where:disk}

At radii $\ll 0.1\,$pc, the focus of Fig.~\ref{fig:wheresf.qad}, the infalling gas circularizes and forms a QAD that persists in to the smallest radii we can resolve ($\ll 0.001\,$pc, with no obvious reason it could not continue down to scales of order the horizon around the SMBH; see \paperthree). As discussed at length in \paperone\ and \papertwo\ and briefly reviewed above, star formation becomes strongly suppressed at small radii below these scales, for several reasons. Strong magnetic fields produce rapid accretion through an annulus while providing a ``magnetic'' $Q \gg 1$ or magnetic critical mass much larger than the entire QAD mass, and even the thermal-only $Q$ rises to $\sim 10-30$ on scales down to $R \sim 0.005\,$pc (before rising very rapidly to even much larger values at still smaller $R$), the turbulence becomes mildly sub-\Alf{ic}, the temperatures of the dust and radiation and ``cold/warm'' phases continue to rise to $\gtrsim 10^{3}-10^{4}\,$K and most of the dust sublimates by $\sim 0.01\,$pc. As a result, the global SFR integrated within the QAD+CQM drops precipitously and at all radial annuli $\ll$\,pc it is more than an order of magnitude smaller than inflow/accretion rates onto the SMBH, and the CQM becomes a gas-dominated QAD in the usual sense (Fig.~\ref{fig:profile.dynamics}). However, this does not mean there is {\em zero} star formation, just that in e.g.\ \papertwo\ where we study the properties of the QAD, we can neglect star formation and stellar feedback and stellar dynamics as significant perturbations to the gas dynamics/inflow/accretion physics. 

Indeed, we see here that there are still some dense star-forming structures in this QAD zone. They tend to occur at larger radii within the QAD: unsurprisingly, there is essentially no detectable star formation at the smallest radii $\lesssim 0.01\,$pc, where accretion maintains the QAD at temperatures $>10^{4}$\,K with a thermal-only Toomre $Q$ parameter of $\gtrsim 1000$ (Fig.~\ref{fig:profile.dynamics}). Stars might still dynamically make their way into these radii, however, as we discuss below. Star formation occurs primarily in the disk midplane as expected, and in radius it occurs vaguely around spiral-arm like structures which are clearly ``shearing out'' in the outer disk, but we show below that in detail the picture is more complicated. It is clear that stars form even within these structures in some preferred locations of the QAD (the youngest stars are not uniformly distributed with radius or even azimuthal angle). We discuss this further below.

\begin{figure*}
	\centering\includegraphics[width=0.77\textwidth]{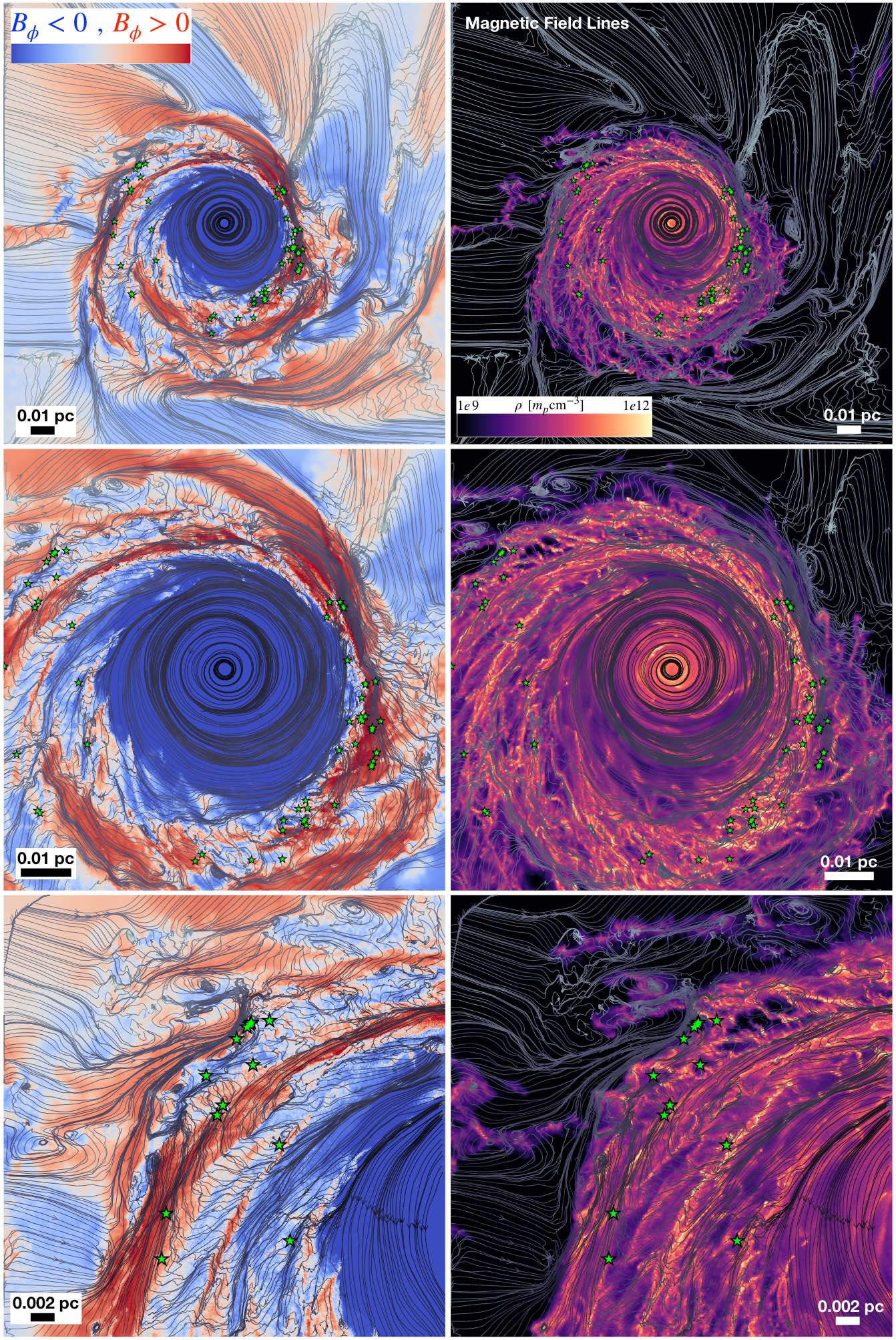} 
	\caption{Comparison of where stars form in the QAD to the magnetic field structure.
	{\em Left:} Magnetic field lines (in-plane projection, so $B_{\phi}-B_{R}$), overlaid on a colormap showing the sign of the toroidal magnetic field ($B_{\phi}>0$ points in the direction of rotation), on different scales (top-to-bottom). To emphasize where stars form we show only dynamically young stars ($t_{\ast} < t_{\rm dyn}(r)$). 
	{\em Right:} Same, but overlaid on the gas density projection from Fig.~\ref{fig:wheresf.qad}. While there is some density sub-structure throughout the QAD, we see a striking correspondence between where stars form and ``switches'' in the mean toroidal field sign. Per \papertwo, these ``switches'' arise from flux-freezing of turbulent ISM fields which are stretched into the dominant toroidal component and advect inwards with the gas. The strong coherent toroidal field strongly suppresses gravitational collapse in the vertical/radial directions, while azimuthal modes are sheared out by the extreme differential rotation, so together this prevents star formation throughout most of the QAD. However, at the ``switches,'' the field is both locally weaker and oriented primarily in the radial direction, so radial (Toomre-like) collapse becomes possible. 
	\label{fig:wheresf.bfield}}
\end{figure*}

\begin{figure*}
	\centering\includegraphics[width=0.77\textwidth]{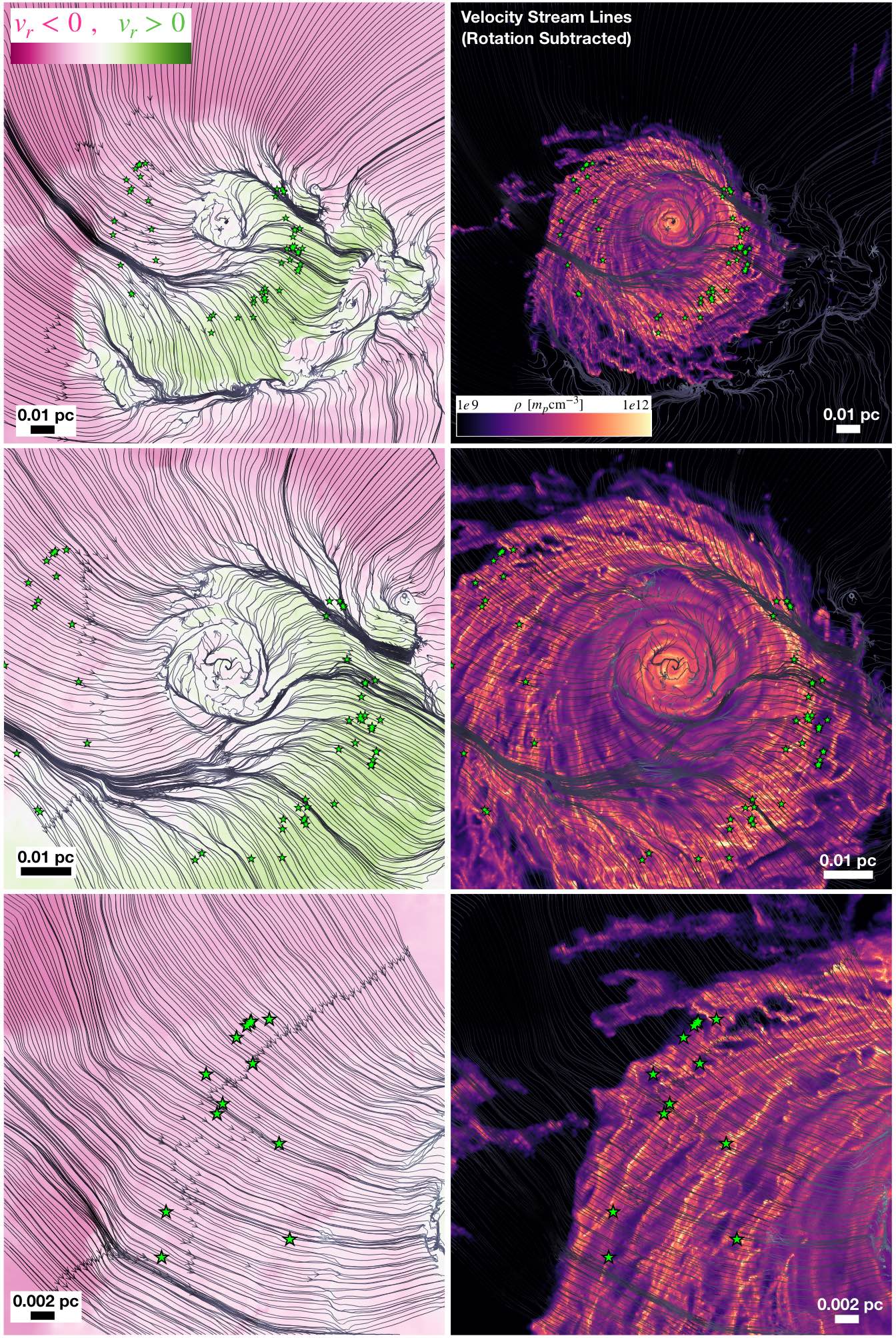} 
	\caption{Comparison of where stars form in the QAD to the velocity field structure; we show young stars and density/field line maps as Fig.~\ref{fig:wheresf.bfield} but color code by radial velocity $v_{r}$. If we plot the velocity stream lines directly, no small-scale structure is visible, as the velocity is totally dominated by rotation at these scales, so instead we plot the streamlines of the velocity after subtracting the expected mean circular velocity: $\Delta {\bf v} \equiv {\bf v} - V_{c}(r)\,\hat{\phi}$. To leading order the large-scale residual and $v_{r}<0$, $>0$ sign pattern is dominated by the global eccentricity of the QAD (visible in the density map), though in the inner QAD the velocity perturbations from individual spiral arms becomes visible. In either case, there is no correlation between these large-scale velocity modes and star formation (indeed SF appears in especially smooth regions of the velocity flow), so SF is {\em not} primarily driven by spiral modes in the QAD, but rather by the magnetic structure in Fig.~\ref{fig:wheresf.bfield}.
	\label{fig:wheresf.vfield}}
\end{figure*}

\begin{figure*}
	\centering\includegraphics[width=0.99\textwidth]{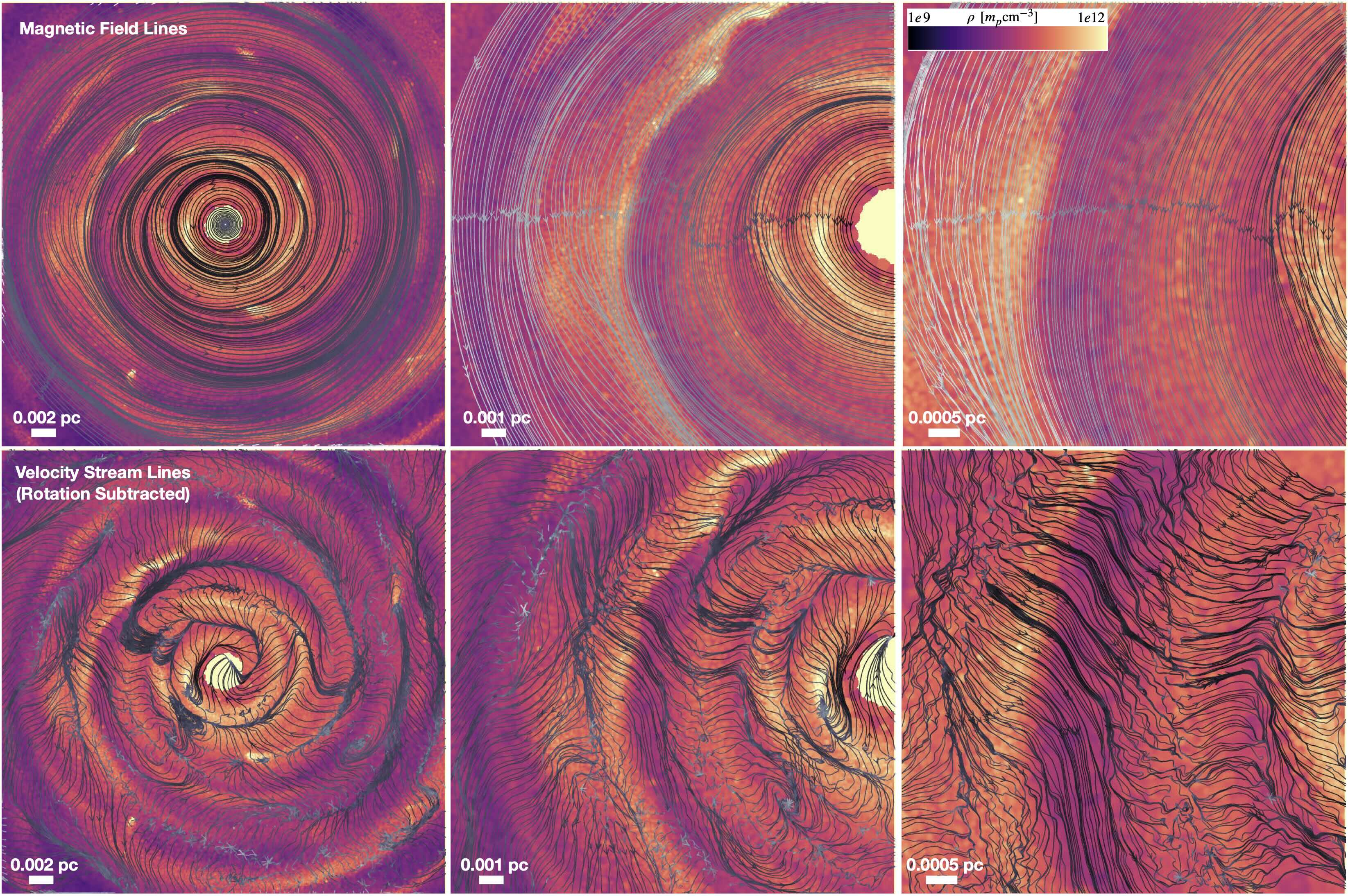} 
	\caption{Magnetic ({\em top}) and (rotation-subtracted $\Delta {\bf v}$) velocity ({\em bottom}) field line structure as Figs.~\ref{fig:wheresf.bfield}-\ref{fig:wheresf.vfield} overlaid on gas density in the inner QAD where there is no SF in the last $\sim t_{\rm dyn}$ here. The coherence of $B_{\phi}$ is visible, and prevents the spiral overdensities (which clearly correlate with the $\Delta {\bf v}$ field) from non-linearly collapsing.
	\label{fig:wheresf.bv.innerdisk}}
\end{figure*}

\begin{figure*}
	\centering\includegraphics[width=0.95\textwidth]{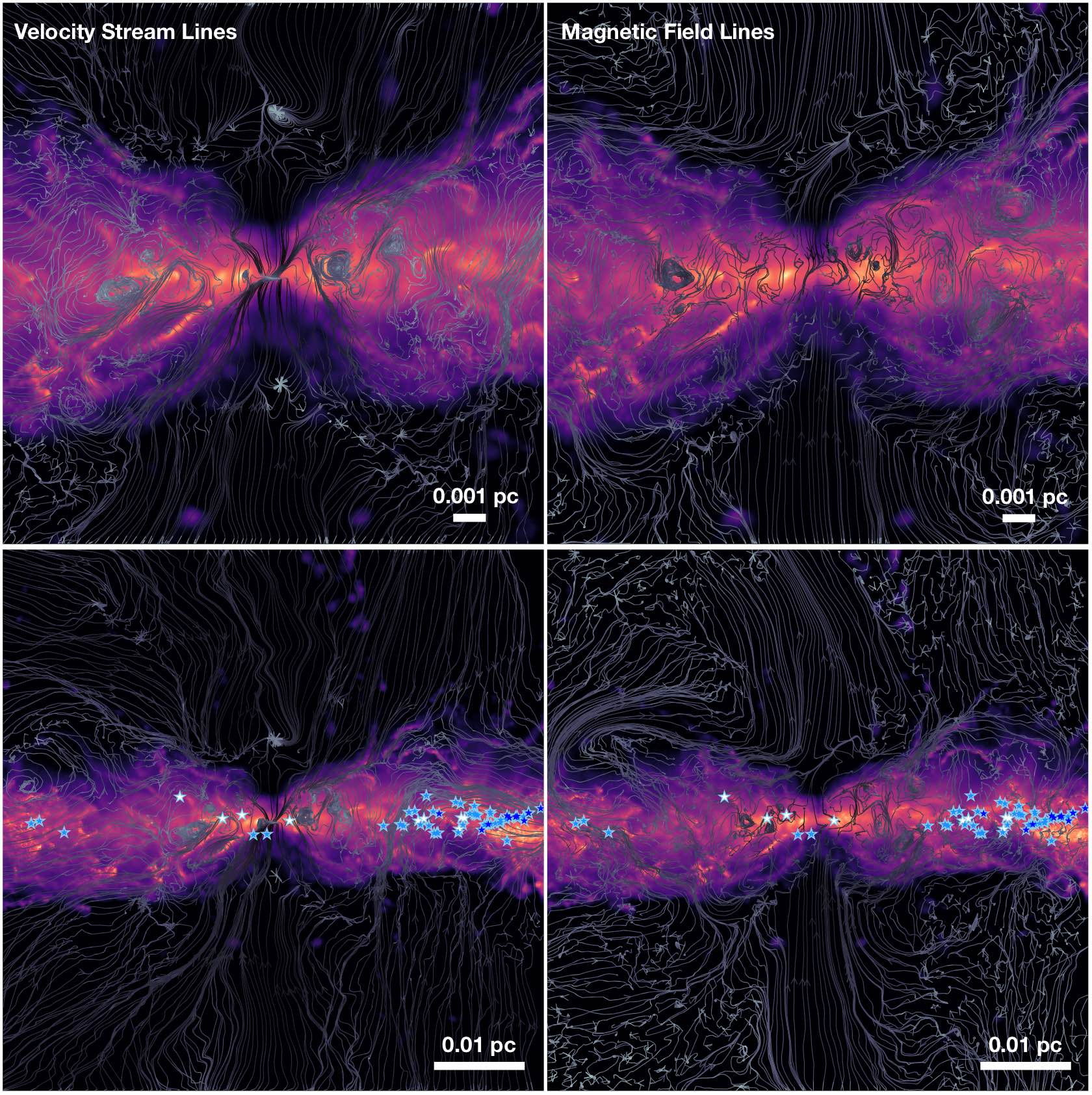} 
	\caption{Magnetic and velocity fields face-on in the QAD as Fig.~\ref{fig:wheresf.bfield}-\ref{fig:wheresf.vfield}, but in an edge-on projection in cylindrical coordinates ($\hat{R}$-$\hat{z}$ plane, selecting a wedge with $|\sin{\phi}|<0.3$). The field lines are again in the projected plane ($B_{R}-B_{z}$, $v_{R}-v_{z}$). The dynamically-young stars clearly form in the denser midplane. QAD is thick and turbulent, with largest eddies of size $\sim H \sim 0.1\,R$, which are trans-\Alf{ic} but highly super-sonic so generate strong density enhancements in the shocks seen, but these are rapidly sheared out into the diagonal ``fans'' seen here as they form.
	\label{fig:edgeon.fieldlines}}
\end{figure*}

\begin{figure*}
	\centering\includegraphics[width=0.95\textwidth]{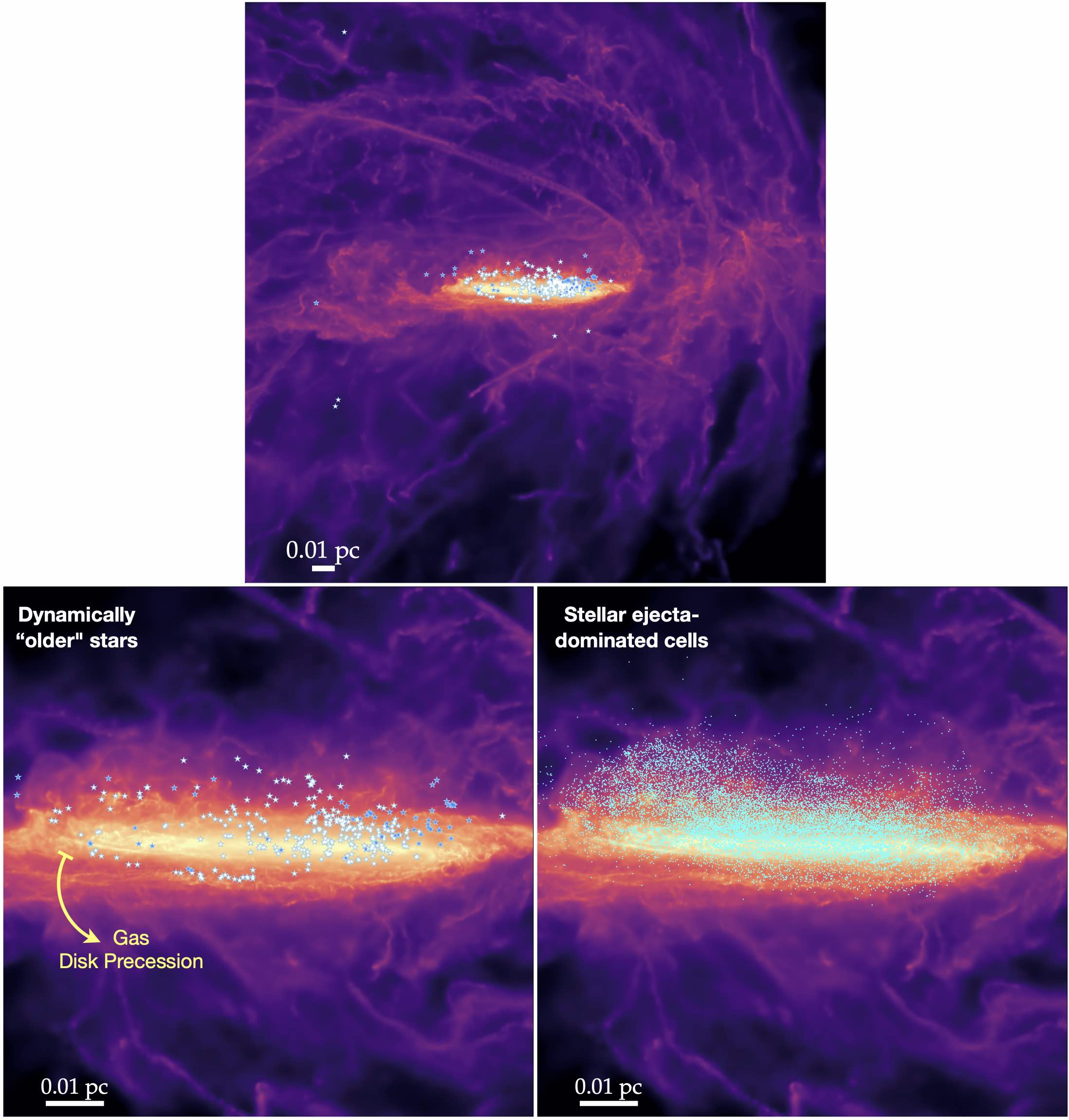} 
	\caption{Edge-on projection of the QAD on larger scales compared to Fig.~\ref{fig:edgeon.fieldlines}, now in a Cartesian projection (not taking a wedge in cylindrical coordinates), and showing the dynamically ``older'' stars ({\em white stars} at {\em left/top}), as well as all gas cells which are primarily composed of (proto)stellar ejects (winds and jets from material accreted by sink particles; {\em cyan points} at {\em right}). 
	This allows us to better see that while the youngest stars form in the midplane, the older stars and their ejecta are out-of-midplane with an asymmetric skew ``above'' the QAD. This is real and owes to the continuous precession of both the QAD eccentricity axis {\em and} the QAD plane as the angular momentum vector of the ``accretion stream'' (visible at {\em top} and Figs.~\ref{fig:wheresf.global}) wraps around the QAD at a continuously varying angle (as expected, since the disrupted cloud/stream is still moving), on a timescale of order the dynamical time at the outer QAD radius ($\sim 100\,$yr). The dynamically older stars and their ejecta reflect the {\em previous} QAD orbits.
	\label{fig:edgeon.oldstars.and.ejecta}}
\end{figure*}

\subsection{Overcoming Support to Form Stars: The Critical Role of Toroidal Magnetic Fields and Tidal Forces}
\label{sec:how}

As discussed above (\S~\ref{sec:where:filament}), in the ``infall/filament'' CQM region,  just like in a typical GMC, there is not really a dramatic ``barrier'' to fragmentation that needs to be overcome. As such, the formation of bound/collapsing sub-clumps can occur via much the same processes: trans/super-\Alf{ic}, highly super-sonic turbulence creates a spectrum of dense sub-structures that can be internally Jeans/Toomre unstable and collapse/fragment \citep[see e.g.][and references therein for a theoretical review]{hopkins:frag.theory}. 

In the disky, rotating QAD region, per \S~\ref{sec:where:disk}, there is such a barrier, which prevents most of the gas from fragmenting. While there are many contributing factors, as reviewed above and shown in detail in \paperone, the most important factor is the combination of strong toroidal magnetic fields in a strong tidal environment. This is illustrated in Figs.~\ref{fig:wheresf.bfield} \&\ \ref{fig:wheresf.vfield} which plot the face-on magnetic and velocity field lines in the QAD alongside the gas density field and locations of the most-recently-formed stars, with Figs.~\ref{fig:wheresf.bv.innerdisk} \&\ \ref{fig:edgeon.fieldlines} showing the same for the inner QAD and in edge-on projection, respectively.

{\em Without} magnetic fields, much of the QAD outside of $\gg 0.001\,$pc would easily meet the conditions for efficient gravito-turbulent fragmentation \citep{paardekooper:2012.stochastic.disk.frag,meru:2012.nonconvergence.disk.frag,hopkins:2013.turb.planet.direct.collapse,deng:gravito.turb.frag.convergence.gizmo.methods}: it features a cooling time short compared to the dynamical time ($t_{\rm cool} \sim \beta\,t_{\rm dyn} \sim (10^{-5}-10^{-2})\,t_{\rm dyn}$; see \papertwo), supersonic turbulence, and the thermal-only Toomre $Q \sim 10$ is modest. This is important for understanding the behaviors relative to historical analytic models for SF in a QAD, almost all of which assumed negligible magnetic fields \citep[references above and e.g.][]{kolykhalov.syunyaev:1980.qso.acc.disk.outer.frag.sf,collins.zahn:1999.qso.selfgrav.disk.sf,collins.zahn:1999.qso.selfgrav.disk.sf.evol,levin:2003.selfgrav.disk.sf,goodman:qso.disk.selfgrav,levin:2007.sefgrav.disk.sf,goodman.tan:2004.qso.disk.supermassive.stars}.
We confirmed this makes a dramatic difference in \papertwo, where we compared simulations with magnetic fields artificially removed, which experienced runaway fragmentation and star formation on these scales. A strong mean toroidal field directly prevents collapse in the vertical and radial directions (i.e.\ suppresses axisymmetric/Toomre-like modes). Collapse would in principle be possible in the azimuthal direction but any modes/structures with finite radial extent or radial velocities are rapidly sheared apart in this direction owing to the strong tidal field. For the typical field strengths here, this would be sufficient to totally suppress gravito-turbulent collapse if e.g.\ the field were idealized as a coherent, cylindrical pure-azimuthal mean field everywhere \citep{lizano:2010.magnetized.ppd.linear.considerations,lin:2014.linear.stability.magnetized.selfgrav.ppds,riols:2016.mhd.ppd.gravitoturb,forgan:2017.mhd.gravitoturb.sims}. The strong field also enables strong torques which accelerate accretion and prevent mass from ``piling up'' in the QAD (\papertwo, \paperthree). And it provides vertical support, greatly increasing $Q$ and decreasing the midplane gas density (Fig.~\ref{fig:profile.dynamics}).

However, although there is a strong toroidal preference, the magnetic fields here are not perfectly cylindrical, uniform, azimuthal structures. In \papertwo\ and Fig.~\ref{fig:edgeon.fieldlines} we show for example that there is also a clear turbulent field within the QAD midplane with fluctuating components not {\em too} much smaller than the mean toroidal field (see Fig.~\ref{fig:profile.dynamics}), but this generally does not change its qualitative ability to resist collapse (the mean field is still strong and primarily toroidal). More relevant, at certain radial annuli in the QAD (which move inwards with the accreting gas as it advects), the mean toroidal field experiences coherent sign flips (Fig.~\ref{fig:wheresf.bfield}, see \papertwo\ for details). This can occur semi-stochastically if the field is driven by instabilities like the MRI \citep{kudoh:2020.strong.b.field.agn.acc.disk.sims.compare}, but we show in \papertwo\ that here it owes to an even simpler explanation: the mean field ``dynamo'' is powered primarily by advection of magnetic flux as material accretes from larger radii. As such, initially tangled fields from ISM scales (where the turbulence is super-\Alf{ic}) are tidally stretched and shorn into a radial (at large BH-centric radii) then toroidal (inside the QAD) configurations with the sign flips and ``zones'' of weaker field reflecting the initial small-scale ISM turbulent magnetic field conditions. 

We can see directly in Fig.~\ref{fig:wheresf.bfield} that star formation in the QAD is clearly strongly associated with these ``flips'' in the toroidal field. The young sinks are not homogeneously distributed throughout the QAD, and they do not correspond to obvious global structures in the velocity field in Fig.~\ref{fig:wheresf.vfield} or Fig.~\ref{fig:wheresf.bv.innerdisk}. There we plot the residual velocity field lines, after subtracting the mean rotational motion, to highlight deviations from circular orbits, which are strongly dominated by the global coherent eccentric structure of the disk and the spiral arms, but these structures do not actually appear to play the key role in local SF in the QAD. Likewise, in Fig.~\ref{fig:edgeon.fieldlines}, while we see the stars form vaguely around the midplane/width of the disk (with $|z| \lesssim H \sim v_{A}/\Omega$), there is no sharp association with some structure in either the $v_{R}-v_{z}$ or $B_{R}-B_{z}$ fields, as there is for the $B_{\phi}$ field.

In Fig.~\ref{fig:wheresf.bfield}, we see that where the mean field is coherent, fragmentation is almost totally suppressed, as we expect from the arguments above. But near the sign flips, there must be regions where the toroidal field vanishes entirely (the fields become primarily radial, in this case), and we see gas collapse radially along the mean field lines to form structures, just like we would expect for simple axisymmetric Toomre instabilities in a gravitoturbulent QAD. In these regions, the field strength is also somewhat weaker, but this is a smaller effect (the magnetic critical mass, using just $|{\bf B}|$, is often still large compared to the condensing gas mass), so the more important effect is the local field {\em geometry}, which when non-toroidal means that gas can collapse along field lines in the strongly-magnetized QAD (and thus not feel the strong magnetic pressure which would otherwise resist collapse). In contrast, in the CQM at $r \gtrsim 1\,$pc, where the field is more isotropically turbulent and the gas is not in an ordered disk, there are essentially ``switches'' located throughout, so there are always directions for collapse to proceed even if the local magnetic field value is strong/magnetic critical mass is large.

Note that while we see essentially all SF in the QAD associated with switches in the polarity of the toroidal field, the converse is not true: not every polarity switch produces star formation. In the inner disk, even the thermal-only $Q$ is much too large. The SF is most vigorous in the outer disk ``flips,'' where $Q_{\rm thermal}$ is relatively low, and the most intense episodes (evident in Fig.~\ref{fig:wheresf.bfield}) tend to coincide with the overlap of a polarity switch and spiral overdensity in the outermost QAD. 

\begin{figure}
	\includegraphics[width=0.98\columnwidth]{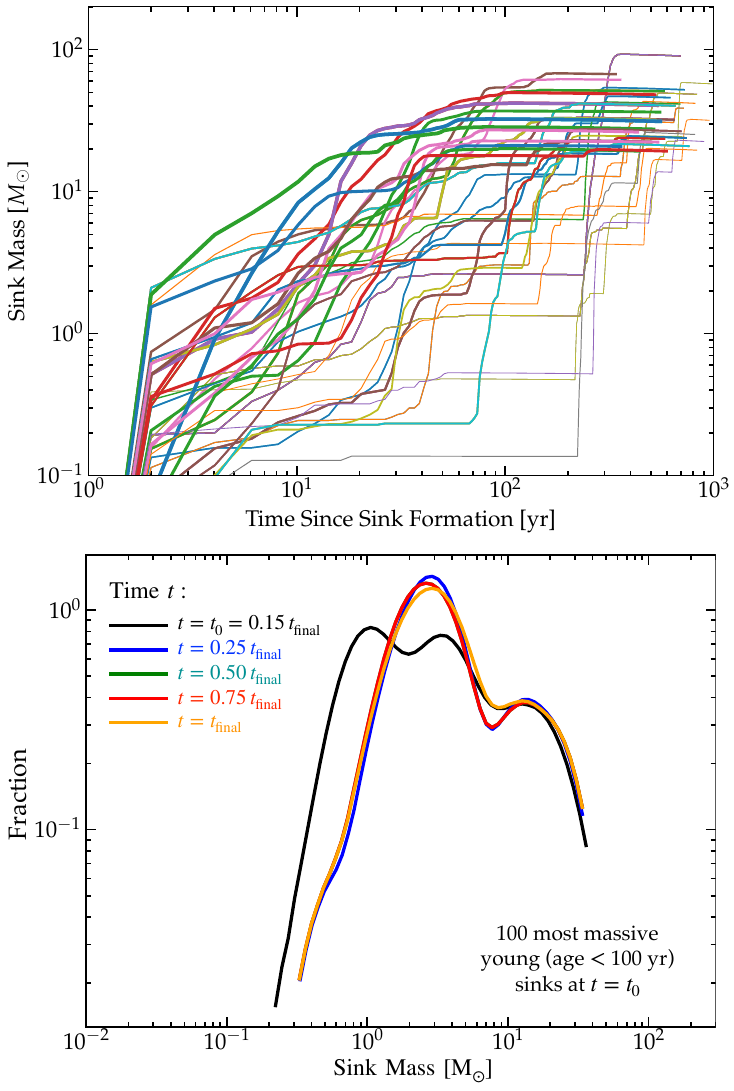} 
	\caption{Growth of massive stars. 
	{\em Top:} We select the $\sim 50$ most massive stars which form at radii $0.02<r<0.05\,$pc in the ``slow'' PSD accretion simulation, and plot total mass $M_{\rm s}$ versus time since the sink particle was created in the simulation ($\Delta t \equiv t - t_{\rm sink}$). Linewidths are ordered by the time when the star reaches $1/2$ of its final mass. Since this simulation has weak (proto)stellar accretion/feedback and more massive stars compared to the ``fast'' PSD accretion run, this represents an ``upper limit'' to the duration of star formation. We see that almost all of the growth occurs in an extremely short period of time (\S~\ref{sec:speed}), with the rapid rise in $M_{\rm s}$ occurring on of order the dynamical time at this radius $t_{\rm dyn}=\Omega^{-1}(r) \sim 50\,$yr (median time to $1/2$ or $1/6$ of the final mass is $\sim 3$ or $\sim 1\,t_{\rm dyn}$). Even those stars with ``late'' growth often grow via a delayed secondary ``burst'' with duration $\lesssim t_{\rm dyn}$. In this strong tidal environment, growth is limited by the dynamical separation of the initial cores/sinks and surrounding gas and the extremely short dynamical times relative to normal pre-main sequence stellar evolution timescales.
	{\em Bottom:} We select the 100 most massive recently-formed ($\Delta t < 100\,$yr) sinks at an early time $t=t_{0}$, and plot the mass of the same sinks at subsequent times until the final simulation time $t_{\rm final}$. We see almost all growth occurs in a short time after sinks form (either $<100\,$yr or $<0.1\,t_{\rm final}$), with negligible late-time sink growth especially for the most massive sinks.
	\vspace{-0.2cm}
	\label{fig:dt.accretion}}
\end{figure}

Importantly, most of the assumptions of historical analytic models for star formation in QADs like \citet{kolykhalov.syunyaev:1980.qso.acc.disk.outer.frag.sf,levin:2003.selfgrav.disk.sf,goodman:qso.disk.selfgrav,goodman.tan:2004.qso.disk.supermassive.stars} do not apply here. In addition to neglecting the role of magnetic fields directly supporting the QAD, providing strong global torques, and suppressing both star formation and stellar accretion (orders-of-magnitude effects), those papers implicitly assumed SS73-like boundary conditions with weak magnetic pressure ($c_{s} \gg v_{A}$), sub or trans-sonic turbulence ($v_{\rm turb} \lesssim c_{s}$), slow cooling ($t_{\rm cool} \gg t_{\rm dyn}$), thin scale-heights ($H/R\sim c_{s}/V_{\rm c} \ll 1$), and orders-of-magnitude larger midplane densities, compared to the models here. As a result they generally predicted fragmentation into Toomre-mass ``stars'' which would contain a large fraction of the QAD mass and would accrete with an extremely large Bondi-Hoyle rate ($\gtrsim M_{\rm clump} \Omega$) until they would clear gaps (have tidal radii $r_{t} > H$ and masses $M_{\ast} \gtrsim 0.001\,M_{\rm BH}$; \citealt{goodman.tan:2004.qso.disk.supermassive.stars}) and ``capture'' most of the QAD inflows. But none of these conditions and subsequent scalings apply here. The Keplerian QAD does not fragment in this fashion (it has $Q \gg 1$ and strong toroidal fields; Fig.~\ref{fig:profile.dynamics}), the effects of non-thermal (turbulent, magnetic, star-gas drift) terms suppress the Bondi-Hoyle accretion rate by factors of $\sim 10^{10}-10^{15}$ (\S~\ref{sec:accretion} and Yashvardan et al., in prep.),\footnote{Calculating the Bondi-Hoyle rate $\dot{M}_{\ast}^{\rm Bondi} \sim 4\pi G^{2}M_{\ast}^{2}\rho\,\delta v^{-3}$ accounting for the combined star-gas drift, magnetic, and turbulent terms in $\delta v$ (Figs.~\ref{fig:profile.dynamics} \&\ \ref{fig:profile.fragmass}) gives $\dot{M}_{\ast}^{\rm Bondi} \sim 10^{-7}\,{\rm M_{\odot}\,yr^{-1}}\,(M_{\ast}/100\,M_{\odot})^{2}$ at the radii of interest. So even the most massive stars we see form in the simulations would grow by only $\sim 1\%$ from continuous Bondi-Hoyle accretion over the typical $\sim 10^{7}\,$yr lifetime of the QAD \citep{hopkins:lifetimes.letter,hopkins:lifetimes.interp,hopkins:lifetimes.methods}. Indeed Fig.~\ref{fig:dt.accretion} shows that the little accretion present is not Bondi-Hoyle like (occurring through encounters with discrete clumps).} 
and the criteria for gap opening/isolation/runaway accretion are impossible to satisfy.\footnote{From \citet{lin.papaloizou:1986.migration}, even if we ignore magnetic fields, two necessary criteria for gap opening are: (1) the tidal radius exceeds the QAD scale-height $r_{t} \sim (M_{\ast}/4 M_{\rm BH})^{1/3} R > H$ or $M_{\ast}/M_{\rm BH} \gtrsim 4\,(H/R)^{3}$, and (2) that the gap cannot be ``filled,'' which amounts to $M_{\ast}/M_{\rm BH} > (40\nu_{\rm eff}/R^{2}\Omega)^{1/2} (H/R)^{3/2} \gtrsim 6.3\, (H/R)^{5/2}$, where the latter expression uses the appropriate $\nu_{\rm eff}$ for a super-sonically turbulent QAD (\paperthree) as opposed to the SS73 assumption usually adopted. Given the large $H/R \sim 0.3$ here, this amounts to $M_{\ast} \gtrsim 0.1\,M_{\rm BH} \gtrsim 10^{6}\,M_{\rm odot}$ -- i.e.\ it would require a major merger with another SMBH to cause appreciable gap-opening in such a thick, turbulent, magnetized QAD.}

\subsection{Timescales of Star Formation and Growth}
\label{sec:speed}

In Fig.~\ref{fig:dt.accretion}, we examine how rapidly collapse and star formation actually occurs in these environments. Specifically, we can examine how the total mass of a typical sink particle grows in time (for sinks forming at some position in the QAD). 

Recall from \S~\ref{sec:methods:ov}, sinks form with the mass of their parent gas cell ($\sim (1-5)\times10^{-3}\,{\rm M_{\odot}}$ here), then capture only gas which is bound and has essentially un-resolved orbits around the sink. This is added to a reservoir representing any un-resolved accretion flow or a PSD (or, if the PSD fragments, un-resolved multiples and/or planets) which can also accrete onto the proto-star or be ejected (representing protostellar jets, magnetic outflows from the PSD, main-sequence mass-loss, etc.). Rather than focus on the growth of just the ``star'' itself, which depends on these un-resolved assumptions about transfer between ``reservoir'' and ``star'' labels within the sink (see \S~\ref{sec:methods:accretion}), in Fig.~\ref{fig:dt.accretion} we just examine the total sink masses which represent the total accreted (minus expelled) mass.

We see the sink masses initially rise extremely rapidly, approaching a maximum on a timescale as short as a few years, even at $\sim1\,$pc from the BH where the dynamical time $t_{\rm dyn} \sim 1/\Omega \sim 4000\,{\rm yr}\,(R/{\rm pc})^{3/2}$. Then they flatten or asymptote to a final mass, or even begin to decline slightly (as e.g.\ some mass is lost from the systems owing to jets).\footnote{Note that the very early initial rise time of some of the sink masses in Fig.~\ref{fig:dt.accretion} being much faster than the local dynamical time should not be taken too literally and does not necessarily mean cores form much faster than $\sim 1/\Omega$ (though they can in principle, e.g.\ in strong shocks). Rather, what we often see is that the dense gas condensations (which will form sinks) form as described in \S~\ref{sec:how}, on timescales broadly $\sim 1/\Omega$, and (by construction) only after a resolved collapsing core has formed will the code (numerically) convert the gas cell at the density maximum (roughly the center of the core) into a sink particle. The sink can then very rapidly numerically accrete the neighbor cells which were already part of its bound, resolved collapsing core mass at the time of the sink particle creation.}
Almost all of the growth comes at early times, and over the longer duration of the simulation, we see negligible growth in mass of the most massive sinks formed early. We have verified that this qualitative behavior is independent of the initial radius of sink formation in the QAD, and independent of the ``fast'' or ``slow'' PSD accretion prescription.

The important thing is that after this initial core is rapidly accreted and converted into a sink, there is very little on-going growth/accretion from the medium, especially for the most massive stars formed here. This is very much {\em unlike} the situation for massive stars in STARFORGE simulations (with identical physics and numerics) of ``typical'' Solar neighborhood GMCs, where \citet{grudic:2022.sf.fullstarforge.imf} showed the most massive stars in particular have the most extended accretion histories often extending over an appreciable fraction of the GMC lifetime, i.e.\ on the order of the large-scale dynamical time or multiple Myr (also, notably longer than the protostellar lifetimes of such massive stars in isolation). The behavior in our simulations (Fig.~\ref{fig:dt.accretion}) is also essentially opposite the predictions of some analytic models for star formation in QADs (e.g.\ \citealt{levin:2003.selfgrav.disk.sf,goodman:qso.disk.selfgrav,goodman.tan:2004.qso.disk.supermassive.stars}) which predict massive stars/sinks should (a) accrete continuously throughout the lifetime of the QAD, and (b) that their accretion rate should rise super-linearly with sink mass (e.g.\ with a Bondi-Hoyle or isolation-mass-like $\dot{M} \propto M_{\rm sink}^{2}$). 

Physically, the lack of continuous, late-time accretion onto sinks relates directly to the dynamics of the sink orbits, and their accretion/mass ejection efficiencies after formation, which we discuss below. Also as discussed above (\S~\ref{sec:how}), the differences between the simulations and the specific historical analytic models predicting strong late-time accretion owes in part to those models assuming very different QAD properties (e.g.\ neglecting magnetic fields and assuming the QAD is razor-thin). 
Altogether, this also means that the IMF in our QAD/CQM environment is more directly ``set at birth'' by said QAD/CQM conditions, as compared to being more strongly modified or self-regulated via subsequent accretion in more typical GMC environments.

\begin{figure*}
	\centering\includegraphics[width=0.98\textwidth]{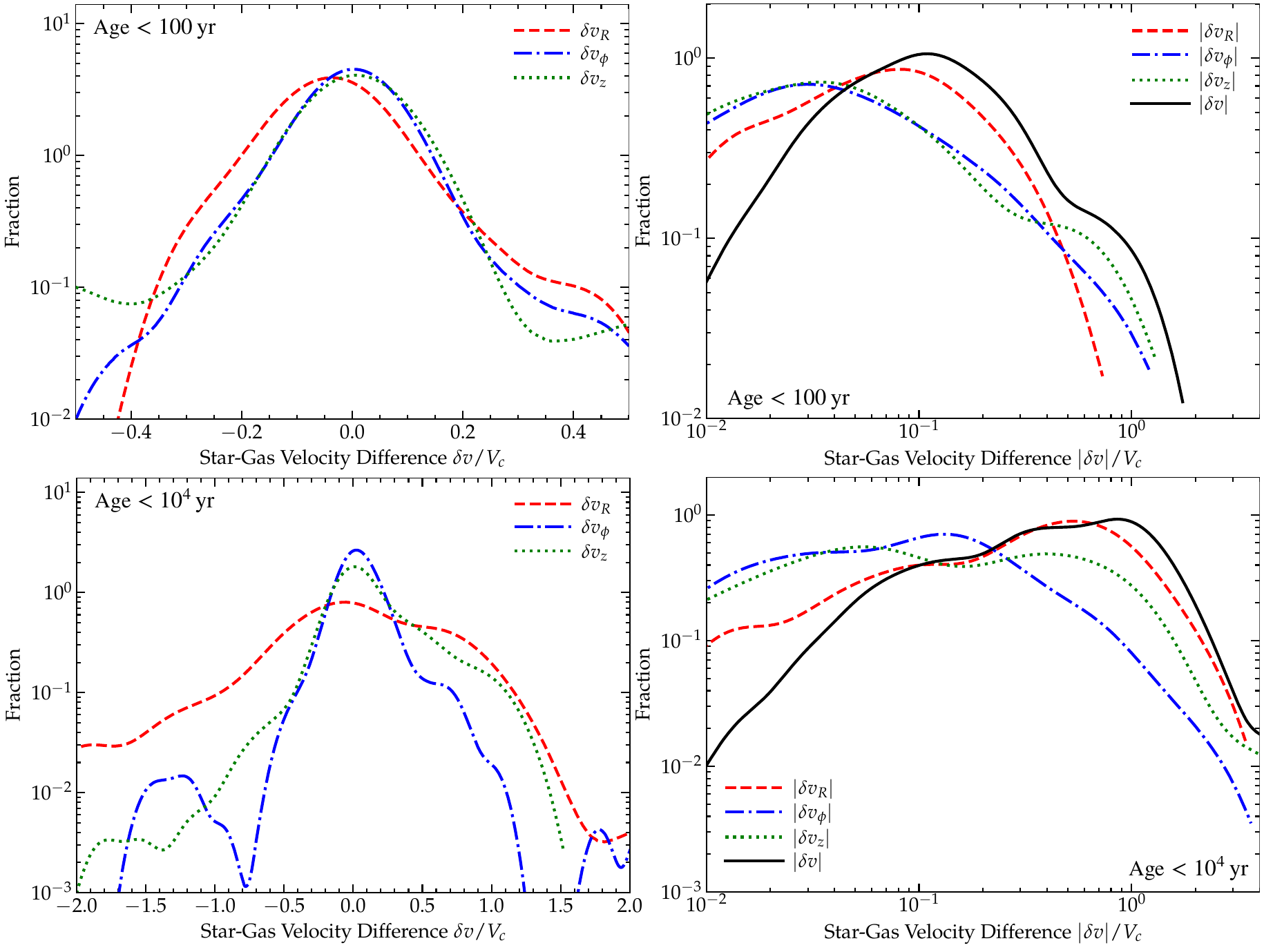} 
	\caption{Distribution of velocity differences $\delta {\bf v} \equiv {\bf v}_{\ast} - \langle v_{\rm gas}({\bf r}_{\ast})\rangle$ between individual stars (${\bf v}_{\ast}$) and their surrounding gas ($\langle v_{\rm gas}({\bf r}_{\ast})\rangle$, defined as the mass-weighted mean velocity of gas within a resolution element $\Delta x$ surrounding each star). We plot both the signed $\delta {\bf v}$ ({\em left}; divided into radial/azimuthal/vertical components) and absolute value $|\delta {\bf v}|$, both for very young stars ({\em top}; time since sink formation $<100\,$yr) and all stars ({\em bottom}). Even within $<100\,$yr after sink formation, stars have acquired a typical velocity difference of order $\sim 10\%$ of the circular velocity $V_{c}$ at their formation radius from the gas around them. Per \S~\ref{sec:wandering} this must happen owing to the extremely short dynamical times at these radii: stars and gas cannot be co-orbital owing to pressure/magnetic/radiation forces on the gas, which generate a velocity difference $|\delta {\bf v}| \sim (H/R)\,V_{c}$ on a timescale $\lesssim \mathcal{O}(\Omega^{-1})$. Physically, this amounts to several thousands (hundreds) of ${\rm km\,s^{-1}}$ in the QAD (CQM). This also explains the very rapid suppression of accretion in Fig.~\ref{fig:dt.accretion}. The older stars with velocity offsets $\sim V_{c}$ are formed on highly eccentric orbits and now ``plunging'' near pericenter through the inner QAD, whose precession (Fig.~\ref{fig:edgeon.oldstars.and.ejecta}) produces velocity offsets $\mathcal{O}(V_{c})$. 
	\label{fig:stellar.velocities}}
\end{figure*}

\subsubsection{The Formation of ``Wandering Stars''}
\label{sec:wandering}

The suppression of stellar accretion just after cores form (\S~\ref{sec:speed}) is closely related to the phenomenon of ``wandering stars'' which we see are ubiquitous in these simulations. Specifically, we see, following the positions of progressively ``older'' stars (in e.g.\ Figs.~\ref{fig:wheresf.global}, \ref{fig:wheresf.qad}, \ref{fig:edgeon.fieldlines}, and more directly in Fig.~\ref{fig:edgeon.oldstars.and.ejecta}), that very quickly after formation (on timescales as short as $\lesssim$\,years in the QAD, or hundreds of years at $\sim1\,$pc) stars ``separate'' from the dense gas structures in which they form. On somewhat longer timescales $\gtrsim \Omega^{-1}$, they end up in completely distinct orbits from the dense gas. This is illustrated in Fig.~\ref{fig:edgeon.oldstars.and.ejecta} where we plot all the stars (including more of the old stars), in the QAD (where the dynamical times are shorter so this is even more evident), where we see that the ongoing precession of the QAD has left behind a population of dynamically older stars whose orbits take them further out of the QAD plane the older the star is (in units of $\Omega^{-1}$), even though the youngest/newest sinks are clearly forming in the midplane (Fig.~\ref{fig:edgeon.fieldlines}). We stress that these are not stars scattered by dynamical processes (e.g.\ interacting triples, sub-cluster merging) as often occurs in LISM GMCs \citep{grudic:cluster.properties,guszejnov:2022.starforge.cluster.assembly,farias:2023.starforge.sfe.vs.bound.clusters.cluster.evol.nbody6}: if it were so, we would see a much more isotropic and symmetric stellar distribution, and it would occur over many dynamical times, not $\mathcal{O}(1)$ dynamical time. 
We can also immediately rule out processes like dynamical friction or Type I migration, as these would drag stars into the denser gas disk, not leave them behind, and would require timescales \citep{lin.papaloizou:1986.migration} $t_{\rm DF,\,1} \sim 2\,(M_{\rm BH}/M_{\ast})\,(M_{\rm BH}/\Sigma_{\rm gas}R^{2})\,(H/R)^{2}\,\Omega^{-1} \sim 10^{10}\,\Omega^{-1}\,(M_{\odot}/M_{\ast})\,(R/0.1\,{\rm pc})^{1/2}$ or $\sim 1500\,{\rm Gyr}\,(R/0.1\,{\rm pc})\,(M_{\odot}/M_{\ast})$ much longer than the Hubble time (let alone the QAD lifetime or simulation duration). 
Instead we see clear ``one sided'' examples where old stars are all in the orbits that the QAD/CQM gas {\em used to occupy} but the gas and stars have drifted systematically relative to one another, beginning almost immediately after star formation.

We can also directly test this by measuring the relative motions between stars and the gas surrounding them in Fig.~\ref{fig:stellar.velocities}, and validate it by checking in the formal orbital parameters of the stars. There, we clearly see strong systematic velocity offsets between stars and neighboring gas. For dynamically-old stars, this is less surprising, but we see even for the youngest stars (we obtain similar results in Fig.~\ref{fig:stellar.velocities} whether we define this by $t_{\ast}<\Omega^{-1}(r)$ or $t_{\ast}<100\,{\rm yr}$) a clear offset, in their orbits. And we can further see this reflected in the ``cometary trails'' made by stellar ejecta (jets and winds) as the stars traverse through the gaseous medium as shown in Fig.~\ref{fig:edgeon.oldstars.and.ejecta} and discussed further below.

This is further enhanced by the gravitational structures giving rise to overdense modes in the QAD. Broadly speaking, we see modes on two scales: first, a very obvious global lopsided coherent eccentric ($m=1$) mode at all radii (see details in \paperone). Such modes are generically expected around quasi-Keplerian potentials like accreting SMBHs \citep{sellwood:1994.counterrot.bend.instab,jacobs:longlived.lopsided.disk.modes,sambhus:m31.nuclear.disk.model,hopkins:m31.disk,hopkins:cusp.slopes,hopkins:inflow.analytics}, and the same physics necessarily produces a phase offset between the patterns for the collisional and collisionless components, which contributes (coherently) to the velocity drift between stars and gas as soon as stars form \citep[see e.g.][]{noguchi:merger.induced.bars.gas.forcing,wada:gas.orbits.in.weak.bars,barneshernquist96,hopkins:inflow.analytics,berentzen:gas.bar.interaction}. This global eccentricity also naturally explains why the radial velocity offsets in Fig.~\ref{fig:stellar.velocities} are especially notable. On small scales, we see local collapsing dense gravito-turbulent structures in the star-forming zones (\S~\ref{sec:how}). But these are at least partially density waves and shocks (they are formally more related to ``slow modes'' as in \citealt{tremaine:slow.keplerian.modes,hopkins:zoom.sims,hopkins:slow.modes}), so material compressed in them ``moves through them'' at a large fraction of the Keplerian velocity. As such, the timescale for separation is of order the size of the structure $\sim \ell$, divided by some appreciably fraction of the Keplerian speed $\sim \alpha\,V_{\rm c}$, or $\sim (\ell/\alpha\,r)\,t_{\rm dyn}$. This is quite familiar from the well-studied problem (both observationally and theoretically) of star formation in GMCs and especially spiral arms in the galaxy \citep[e.g.][]{tasker:2009.gmc.form.evol.gravalone,foyle:2010.spiral.arms.conc.gas.not.sf,leroy:2013.molecular.sf.law.residual.correlations,Colombo_2014_PAWS_survey,2019MNRAS.483.4707G,meidt:ism.turb.jeans.scale.key.for.structure.sizes,lee:2022.phangs.jwst.treasury.paper}. As has been studied for decades \citep{sellwood:spiral.from.acc,youngscoville:1991.mol.gas.review}, density modes lead to compression and shocks, which produces rapid star formation, then stars drift through and separate from the overdense gaseous arm region. 

The physics causing this is actually quite straightforward. Recall, the gas feels strong anisotropic stresses from pressure forces and, in particular, magnetic fields on these scales (\papertwo). As cores collapse into protostars, they effectively de-couple from the background gas MHD forces, and become ``collisionless'' (or at least ``pressure-free'') in their gravitational dynamics. This means that as soon as this force diminishes, they must begin to separate from the gas positions, with typical relative velocities of order the ratio of pressure forces to gravitational forces times the circular velocity -- in practice, up to tens of percent of the circular velocity at a given radius (tens to hundreds of ${\rm km\,s^{-1}}$; see Fig.~\ref{fig:profile.dynamics}). This is consistent with the behavior of the young stars in Fig.~\ref{fig:stellar.velocities}: we see a characteristic velocity offset of order $\sim 0.1\,V_{c}$ imprinted at formation. With time, a combination of eccentric orbits (e.g.\ the stars on ``plunging'' orbits through the inner QAD), precession of the QAD, systemic drift, and turbulence/new shocks creating new or shifting the locations of filamentary structures in the outer CQM, lead to non-linear offsets in velocities/orbits. This in turn means the local star-gas velocity difference for stars more than a few dynamical times old is more like $\sim V_{c}$ (order-unity, in a relative sense).\footnote{Interestingly, unlike the simple case of e.g.\ a planet in a nearly-circular PSD, in many cases the star-gas velocity offset is such that the gas is actually locally moving {\em faster} in the mutual direction of motion than the star, so the ejecta can be swept ``forward.'' This can occur for many reasons, depending on e.g.\ where the stars are in their non-linearly eccentric orbits and the precession of the QAD.}

The key difference between the situation in the Solar neighborhood and the simulations here is simply that the timescales for these offsets to occur, and the magnitude of the gravitational velocities, are wildly different. Stars drift out of e.g.\ spiral arms in the LISM on timescales of tens to hundreds of Myr ($\sim 1/\Omega$ at Solar circle Galactocentric radii) -- very long compared to the protostellar collapse/accretion timescales of massive stars. But here, separation on a fraction of $\sim 1/\Omega$ means timescales of years, much shorter than typical protostellar evolution timescales. And in the QAD, star-gas velocity separation at even $\sim 10\%$ of $V_{c}$ does not mean a relative velocity of $\sim 10\,{\rm km\,s^{-1}}$ as in the LISM, but more like several hundred ${\rm km\,s^{-1}}$.

\begin{figure*}
	\centering\includegraphics[width=0.95\textwidth]{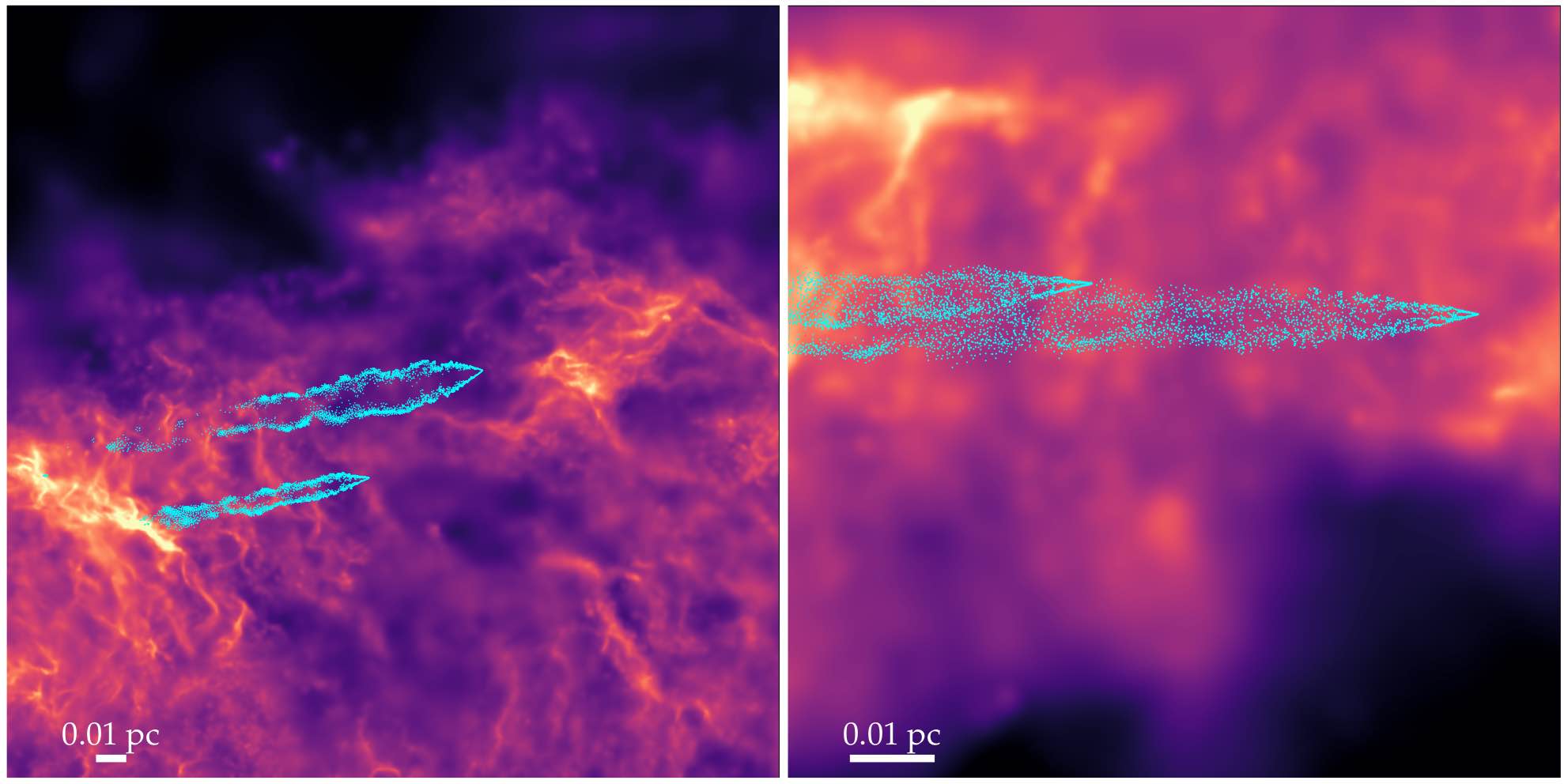} 
	\caption{Extreme examples of ``cometary trails'' of (proto)stellar outflows/jets. We show images of the total gas density in a couple of the filamentary regions from Fig.~\ref{fig:wheresf.global}, and highlight gas cells ({\em cyan}, as Fig.~\ref{fig:edgeon.oldstars.and.ejecta}) whose mass is dominated by the direct ejecta from a couple individual protostars in each sub-region. Owing to the rapid motion of the stars relative to the gas even for quite young stars (Fig.~\ref{fig:stellar.velocities}), the ejected jet material immediately encounters an extremely high ram and magnetic pressure ($\sim \rho\,V_{c}^{2}$) which produces a bow shock/stream/tail morphology.
	\label{fig:trails}}
\end{figure*}

\subsubsection{Consequences for Stellar Accretion and Feedback}
\label{sec:accretion}

The separation between stellar positions and their formation sites in gas (\S~\ref{sec:wandering}) has immediate important consequences for their ability to continue accreting gas, which is reflected in the time histories in \S~\ref{sec:speed}. Consider the Bondi-Hoyle-Lyttleton capture rate onto a core (moving through a statistically homogeneous, isotropic ambient medium with no self-gravity and no other forces), which scales roughly as $\dot{M} \sim 4\pi\,G^{2}\,M_{\rm core}^{2}\,\rho_{\rm gas}/v_{\rm eff}^{3}$ where $\rho_{\rm gas}$ is the ambient gas density and $v_{\rm eff}^{2} \sim  c_{s}^{2} + \langle (1-\cos^{2}{\theta})\, \rangle\,v_{A}^{2} +  |{\bf v}_{\rm core}-\langle {\bf v}_{\rm gas} \rangle|^{2} + |\delta {\bf v}_{\rm turb}|^{2}$ collects contributions from the thermal sound speed ($c_{s}$), magnetic support (the \Alf\ speed $v_{A}$, where $ \langle (1-\cos^{2}{\theta})\, \rangle$ depends on the geometry and is $\sim 2/3$ for tangled fields, see \citealt{lee:2014.mhd.bondihoyle.stellar.accretion}), the bulk relative velocity between core and gas $|{\bf v}_{\rm core}-\langle {\bf v}_{\rm gas} \rangle|$, and small-scale turbulence $|\delta {\bf v}_{\rm turb}|$. 

If we assumed a strictly laminar, non-magnetized, non-turbulent, thermal-pressure supported QAD with zero global/eccentric structure and zero drift velocity or asymmetric drift/pressure support of the gas and the stars perfectly co-orbital with the gas on aligned co-rotating circular orbits, then $v_{\rm eff} \sim c_{s}$. But in our simulations, the thermal sound speed is small compared to $v_{A}$ by a typical factor of $\sim 100$ (i.e.\ $\beta \sim 10^{-6}-10^{-2}$ is very small). The turbulence is trans-\Alf{ic}, so $|\delta {\bf v}_{\rm turb}| \sim v_{A}$. And the effects described in \S~\ref{sec:wandering} each naturally lead to systematic/drift velocity offsets of $|{\bf v}_{\rm core}-\langle {\bf v}_{\rm gas} \rangle| \sim (H/R)\,V_{c} \sim v_{A} \sim |\delta{\bf v}_{\rm turb}|$ (see Fig.~\ref{fig:stellar.velocities}) and add coherently. So we expect $v_{\rm eff} \sim {\rm a\ few} \times v_{A}$. And since the accretion rate is suppressed by $v_{\rm eff}^{-3}$, this means that the accretion rates will typically be suppressed relative to the laminar, co-moving expectation by a factor of $\sim (0.01-0.1)\times\beta^{3/2} \sim 10^{-8}-10^{-5}$. Thus almost as soon as the cores begin to ``de-couple'' after their density becomes sufficiently high as described in \S~\ref{sec:wandering}, the accretion rates will drop by orders of magnitude (and we have even neglected the additional linear suppression factor that will arise from $\rho_{\rm gas}$ decreasing as the stars move away from the local density maxima where they initially formed).

Worse yet, stars from outer radii on eccentric orbits intersect  the ``inner'' QAD (at their pericentric passage) with $\mathcal{O}(1)$ angle between their eccentric velocity vector and the local mean gas orbital velocity vector, owing to the fact that the QAD+CQM structure, eccentricity, and orbital plane are not perfectly independent of radius $r$. So in these intersection cases although the density rises, the velocity also rises to $\sim \mathcal{O}(1)\times V_{c}(r)$, which net suppresses the accretion rate by an additional factor of $\sim 10^{-4}-10^{-3}$ or so. 

The same relative velocity argument also immediately explains why we see the bow shocks and almost-cometary ``trails'' left behind by ejecta from (proto)stars, which we show some typical examples of in Fig.~\ref{fig:trails}. As soon as material is ejected from a star, it re-couples to these ambient RMHD forces, introducing a drift velocity relative to the star which ranges from $\sim 10\%$ to a couple times the circular velocity. Since the star always feels some coherent gas velocity from its point of view (analogous to e.g.\ a planetesimal or rocky body in a circum-stellar PSD), the ejecta are ``swept up'' in the stellar frame. Of course some examples of this (albeit much less common and with much lower absolute velocity scales) are known in protostars and YSOs (as well as AGB stars), such as L1551 IRS 5 or PV Cep \citep{goodman:2004.pvcep.star.high.vel.bowshock}, but the phenomenon here is more pronounced and ubiquitous owing to the extreme dynamical conditions.

\begin{figure*}
	\centering\includegraphics[width=0.95\textwidth]{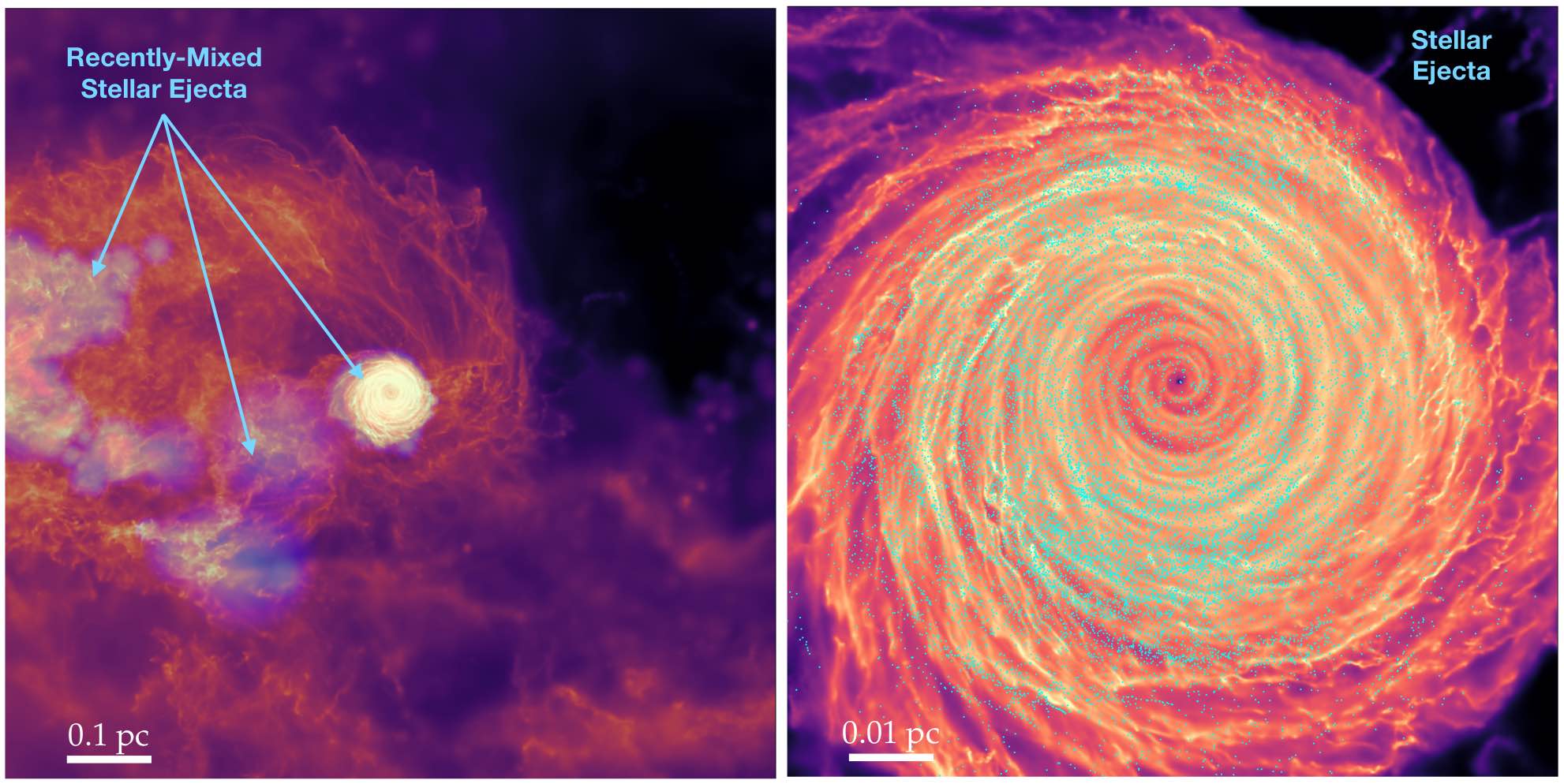} 
	\caption{Distribution of stellar ejecta (jets/outflows/winds) in the broader QAD/CQM, as Fig.~\ref{fig:trails} but focusing on larger scales and showing more of the ejecta mass. 
	{\em Left:} Gas density map as Fig.~\ref{fig:wheresf.global}, but with blue colorscale overlay highlighting gas weighted by the mass fraction of ejecta cells (direct ejecta cells ``spawned'' by protostellar particles to represent winds and jets) recently merged into the ``ambient'' (standard) gas cells. These correspond to regions of the QAD/CQM with particularly recent very massive star formation.
	{\em Right:} Zooming in on the QAD from {\em left}, and overplotting all un-merged ejecta cells as Fig.~\ref{fig:trails}. 
	In both cases the ambient gas interacting with or recently affected by ejecta show no deviation in their kinematics or density structure from the larger-scale ``background'' flow, clearly indicating that stellar feedback has very little dynamical effect on the CQM/QAD. This is expected (\S~\ref{sec:feedback}), owing to the extremely deep potential combined with suppressed/very low overall SFR.
	\label{fig:ejecta}}
\end{figure*}

\subsection{The (Weak) Effects of Stellar Feedback On the Environment}
\label{sec:feedback}

The simulations exhibit very weak {\em global} effects of stellar feedback on environment. In Fig.~\ref{fig:ejecta}, we see that stellar ejecta in both the CQM filamentary zone and disky QAD zone are simply mixed/incorporated into the surrounding gas with orbits typical of the ambient gas that has not been directly inflenced by feedback. We have evolved the system for many local dynamical times (up to $\gtrsim 10^{5}\,\Omega^{-1}$ at the innermost radii simulated) so this remains true in a ``local steady state'' sense. And we show in \paperone\ and \papertwo\ that the contribution of feedback on sub-pc scales to the turbulent or dynamical or binding energy budget of the QAD itself and/or infalling gas into the QAD is negligible. 

This is anticipated by many previous studies of stellar feedback processes in circum-quasar environments (see \citealt{thompson:rad.pressure,wada:torus.mol.gas.hydro.sims,hopkins:qso.stellar.fb.together,grudic:max.surface.density,kawakatu:2020.obscuration.torus.from.stellar.fb.in.torus,daa:20.hyperrefinement.bh.growth}, or for a review, \citealt{hopkins:2021.bhs.bulges.from.sigma.sfr}). So we only briefly review the reasons here. To begin, note that inside of $r\sim (0.01,\,0.1,\,1)\,{\rm pc}$, the star formation+stellar accretion rate is $\dot{M}_{\ast}(<r)\sim (10^{-4},\,0.1,\,10)\,{\rm M_{\odot}\,yr^{-1}}$, while the gas inflow rate is $\dot{M}_{\rm in} \sim (40,\,70,\,100)\,{\rm M_{\odot}\,yr^{-1}}$ (see \S~\ref{sec:ism.diff} or \paperone), so clearly the SFR and stellar mass-loss/jets (even if mass-loss returned the entire accreted stellar mass budget) are small perturbations to the inflow mass budget. 

In terms of momentum/force balance, recall that for our ``slow'' PSD accretion models, accretion from the PSD to the star (which is what actually determines the feedback rates) is very slow compared to the local dynamical time, so feedback is strongly suppressed. We therefore instead consider the opposite ``fast'' PSD accretion regime, where one can approximate accretion from PSD to star as instantaneous so that for calculating IMF-integrated fluxes from e.g.\ jets, we simply have $\dot{M}_{\rm jets} \sim f_{\rm jet}\,\dot{M}_{\ast}$. With that in mind, we can directly compute the momentum flux in jets $\dot{P}_{\rm jets} \sim \langle f_{\rm jet}\,\dot{M}_{\ast}\,v_{\rm jet} \rangle$ (with $f_{\rm jet} \sim 0.3$ the mass fraction going into jets and $\langle v_{\rm jet} \rangle \sim 0.3\,\sqrt{G\,M_{\odot}/R_{\odot}}$, roughly), or from main sequence winds plus radiation pressure $\dot{P}_{\ast} \sim 10^{3}\,L_{\odot}\,M_{\ast}/(M_{\odot}\,c)$, and compare to the gravitational force on the gas $\dot{P}_{\rm grav} \sim G\,M_{\rm BH}\,M_{\rm gas}(<r)/r^{2}$, and obtain $\dot{P}_{\rm jets}/\dot{P}_{\rm grav} \sim (10^{-7},\,10^{-4},\,0.01)$ and $\dot{P}_{\ast}/\dot{P}_{\rm grav} \sim (4\times10^{-8},\,10^{-5},\,5\times10^{-4})$, at $r\sim (0.01,\,0.1,\,1)\,{\rm pc}$. This is consistent with numerous studies (see references above and \citealt{fall:2010.sf.eff.vs.surfacedensity,colin:2013.star.cluster.rhd.cloud.destruction,2017MNRAS.472.4155G,grudic:sfe.cluster.form.surface.density,2018ApJ...859...68K}) which have shown that simply balancing the momentum injection rate for young stellar populations against the gravitational force per unit area (here $\sim |{\bf a}_{\rm grav}|\,\Sigma_{\rm gas} \sim G\,M_{\rm bh}\,\Sigma_{\rm gas}/r^{2}$) means that feedback becomes sub-dominant when the acceleration scale exceeds $ |{\bf a}_{\rm grav}| \gg 10^{-8}\,{\rm cm\,s^{-2}}$ (equivalent to an ``effective surface density'' $(M_{\rm bh} + M_{\rm gas} + M_{\rm other})/r^{2} \gg 1000\,{\rm M_{\odot}\,pc^{-2}}$). But here $ |{\bf a}_{\rm grav}| \sim G\,M_{\rm bh}/r^{2} \sim 2\times10^{-4}\,{\rm cm\,s^{-2}}\,({\rm pc}/R)^{2}$ exceeds this critical threshold by enormous ratios of $\gtrsim 10^{4}$. In other words, in the vicinity of the SMBH, the gravitational forces competing against feedback are equivalent to those of a cloud with effective surface density $\sim 10^{7}\,{\rm M_{\odot}\,pc^{-2}} \times ({\rm pc}/R)^{2}$.

\begin{figure*}
	\centering\includegraphics[width=0.95\textwidth]{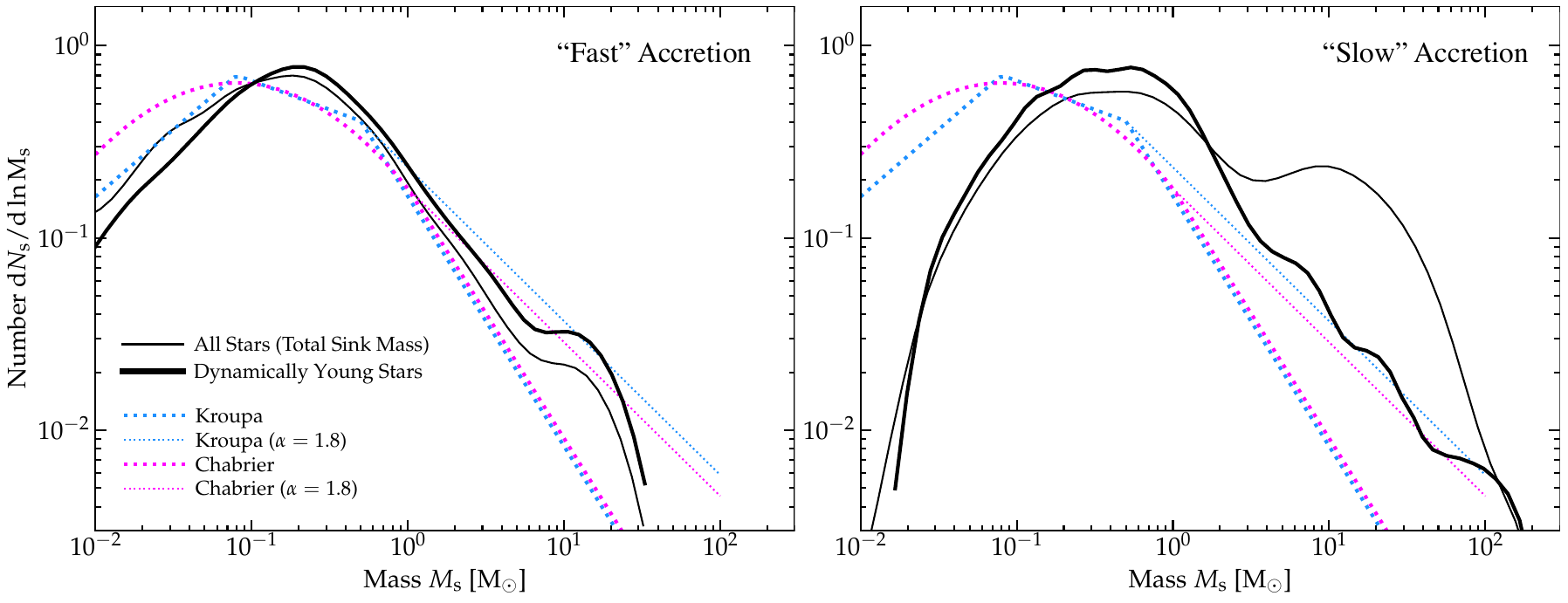} 
	\caption{Stellar IMFs (\S~\ref{sec:imf}) in the simulation CQM+QAD. Here we plot the (normalized) mass function $N_{s}^{-1} dN_{s}/d\ln{M_{s}}$ (including the total mass of the sink, both that assigned to ``star'' and accreted ``reservoir''), including all stars within the CQM and QAD (radii within the BHROI, $r \lesssim 5\,$pc), at the final time to which the simulation is run ($\sim 10^{5}$ dynamical times at the innermost radii). The normalization of the IMFs is arbitrary. We include all stars formed during the simulation {\em after} it reaches the highest refinement level (so we exclude contamination by low-resolution from earlier pre-refinement stages, leaving $\sim 10^{4}$ QAD stars). We compare the IMF of just the dynamically-young stars with $t_{\ast} < \Omega^{-1}$. We compare simulations which adopt the ``fast'' ({\em left}) or ``slow'' ({\em right}) PSD-to-protostar accretion model (\S~\ref{sec:methods:accretion}): as discussed in \S~\ref{sec:imf:accmodel}, this has an obvious effect on how ``top-heavy'' the IMF is, where the ``slow'' model, by suppressing accretion from PSD to star also suppresses (proto)stellar feedback and leads to accretion of more initial bound clump mass (hence a more top-heavy IMF). We compare to the canonical observed ``universal'' single-star IMFs from \citet{kroupa:2001.imf.var} and \citet{chabrier:2005.review2}, in their standard form (with a \citealt{salpeter:imf}-like high-mass slope $\alpha=2.3$) or with a modified top-heavy high-mass slope $\alpha=1.8$ (\S~\ref{sec:imf:shape}). The lack of the ``bump'' at $\sim 10-20\,M_{\odot}$ for young stars in the ``slow'' accretion run owes to the lack of recent star formation in the outer QAD zone (where the IMF is most top-heavy) at this particular moment.
	\label{fig:imf.toplevel}}
\end{figure*}

\begin{figure*}
	\centering\includegraphics[width=0.95\textwidth]{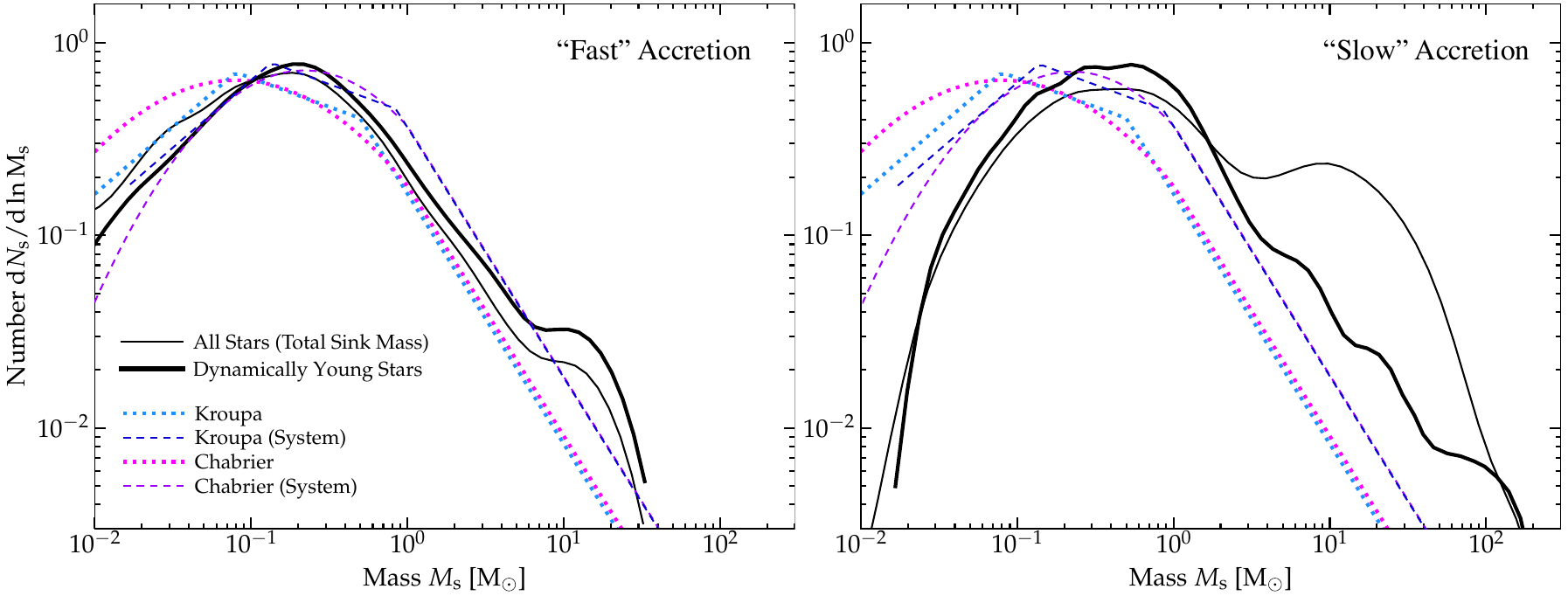} 
	\caption{IMFs as Fig.~\ref{fig:imf.toplevel}, but comparing with the \citet{kroupa:2001.imf.var} and \citet{chabrier:2005.review2} LISM system IMFs as well as single-star IMFs. As we show below, fragmentation and multiplicity are strongly suppressed by the extreme tidal fields in the simulations. This may explain the modest shift in the peak of the IMF for the ``fast'' accretion model, but both the high-mass slope and IMF peak/turnover mass are shifted higher than even the LISM system IMF. 
	\label{fig:imf.toplevel.vs.system}}
\end{figure*}

In terms of energetics and heating, noting that the circular velocities at these radii are $V_{c} \sim (2500,\,750,\,250)\,{\rm km\,s^{-1}}$, and that the gravitational luminosity (energy released by accretion, $\dot{E}_{\rm grav} \sim \dot{M}_{\rm in}\,V_{c}^{2}$), turbulent dissipation rate ($\dot{E}_{\rm cool} \sim M_{\rm gas}\,v_{\rm turb}^{2}\,\Omega$), and radiative cooling losses ($\dot{E}_{\rm cool}$) are all similar (as expected for an accretion-regulated system), we have $\dot{E}_{\rm grav} \sim  (10^{44},\,3\times 10^{43},\,5\times10^{42})\,{\rm erg\,s^{-1}}$ at $r \sim (0.01,\,0.1,\,1)\,{\rm pc}$. We can directly compare this to the kinetic luminosity of jets (or O/B winds, which are comparable) $\dot{E}_{\rm jets} \sim \langle f_{\rm jet}\,\dot{M}_{\ast} v_{\rm jet}^{2} \rangle  \sim (5\times10^{35},\,5\times10^{38},\,5\times10^{40})\,{\rm erg\,s^{-1}}$, and see that the latter is negligible in the global energy budget. Essentially, the escape velocities from the SMBH potential are larger than typical jet velocities, so the SFR would have to be much larger than QAD inflow rates (the opposite of what we see) for the jets to impact the kinematics globally. For comparison, a typical GMC would have a gravitational luminosity of $\sim 10^{37}\,{\rm erg\,s^{-1}}\,(M_{\rm cloud}/10^{6}\,M_{\odot})^{5/4}\,(\Sigma_{\rm cloud}/100\,{\rm M_{\odot}\,pc^{-2}})^{5/4}$, so again the extreme potential of the SMBH makes this more akin to the gravitational dynamics of a ``cloud'' with effective surface densities $\gtrsim 10^{7}\,{\rm M_{\odot}\,pc^{-2}}$. 

But what about radiative heating? Summing the accretion/contraction/main sequence luminosities of the stars, we obtain a total radiative luminosity of stars $L_{\ast} \sim (10^{38},\,10^{41},\,10^{43})\,{\rm erg\,s^{-1}}$ at these radii. Recall that at $\sim 1$\,pc the QAD/CQM has an IR optical depth $\gtrsim 1$, so this will mostly be re-processed locally (hence considering the local $L_{\ast}$ and $\dot{E}_{\rm grav}$ at each $R$, instead of external illumination by either the $\sim$\,kpc-scale galactic starburst or inner QAD, whose luminosity here can approach $\sim 10^{47}\,{\rm erg\,s^{-1}}$). Thus at $R \ll$\,pc, the heating by stars is negligible in the total QAD/CQM thermal budget, but at $R\gtrsim\,$pc, stellar heating becomes comparable to or larger than accretion luminosities, contributing importantly to the thermal balance of the medium. Indeed, it was noted in \paperone\ that the typical CQM temperatures at these larger radii ($\gtrsim$\,a few pc) were clearly elevated compared to those predicted from models with only accretion luminosity (e.g.\ traditional QAD models, extrapolated to $\gtrsim$\,pc scales, using the surface densities/optical depths actually appearing in the simulations\footnote{In Fig.~\ref{fig:profile.dynamics}, this is evident in the somewhat hotter mass-weighted (closer to midplane) temperatures compared to the SS73 model at $\gtrsim$\,pc scales, but it is worth noting that SS73 predicts a higher surface density and optical depth as noted therein, so the extra heating from e.g.\ ambient stars on gas is even more significant than this might naively imply at $r \gg 1\,$pc.}). 

Finally, outside of a few pc (the BHROI), stars begin to dominate the potential (\paperone) and the SFRs continue to grow, with the medium resembling a typical starburst environment, rather than QAD/CQM. There, stellar feedback may play a much more significant role.

\subsection{The IMF}
\label{sec:imf}

\subsubsection{Overview \&\ Shape}
\label{sec:imf:shape}

Fig.~\ref{fig:imf.toplevel} shows the resulting initial mass function of stars in the simulations, at the latest times we evolve. 
Below, we break down the IMF by region, time, and physics model in more detail. It makes very little difference if we plot the ``present-day mass function'' (PDMF) at this time, or the ``peak'' IMF where we define the ``mass'' of each sink by the maximum mass it achieves. This is because the dynamical timescales of interest and duration of the simulations are quite short compared to stellar evolution timescales. Here we plot the ``total sink mass'' defined as the total bound mass associated with the sink particle -- both what the code would actually call the ``(proto)star'' and the ``circum-stellar PSD mass'' or ``accretion reservoir.'' This is more robust, given our uncertainty over the rate at which material accretes from reservoir to ``star.'' Moreover, if there is unresolved fragmentation into e.g.\ close binaries which cannot be captured explicitly in our simulations, this mass will be robust to that effect.

\begin{figure*}
	\centering\includegraphics[width=0.95\textwidth]{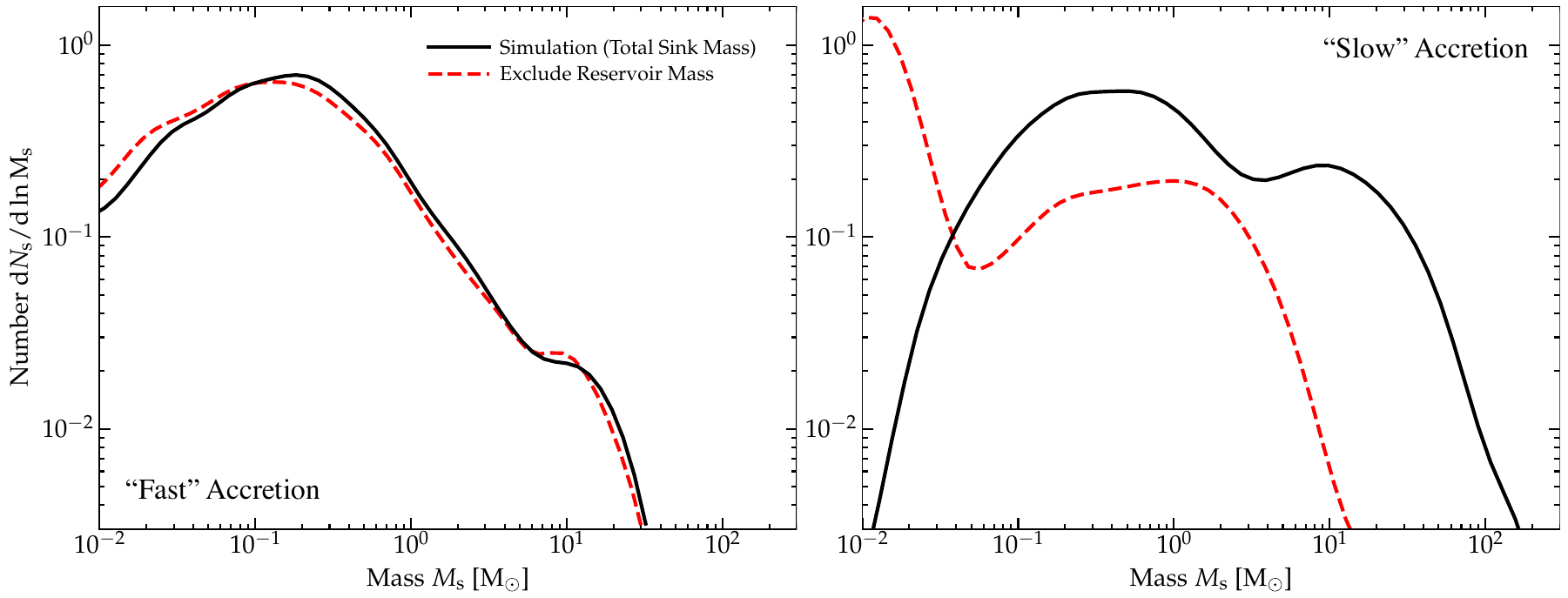} 
	\caption{IMF for the ``fast'' ({\em left}) and ``slow'' ({\em right}) PSD accretion models as Fig.~\ref{fig:imf.toplevel}, comparing the ``total sink mass'' $M_{s}$ (explicitly-evolved ``stellar'' mass $M_{\ast}$ plus the ``reservoir'' accreted by the sign but not yet assigned to the star $M_{\rm PSD}$) to the mass assigned just to the ``star'' $M_{\ast}$ (excluding the reservoir). We show this for all stars. As expected, with the ``fast'' PSD accretion model (\S~\ref{sec:imf:accmodel}), the PSD mass rapidly accretes on to the ``star'' so $M_{s} \approx M_{\ast}$, though at the smallest radii even the very fast assumed accretion can lead to some differences. In contrast, with the ``slow'' PSD accretion model nearly all ($>90\%$) of the accreted sink mass stays in the PSD ``reservoir'' and does not reach the ``star'' over the entire duration of the simulation ($\sim 10^{4}\,{\rm yr}$, or $\sim 10^{5}\,\Omega^{-1}$ at the innermost resolved QAD radii) meaning that there is much weaker local stellar feedback/ejection/suppression of accretion of the original core (\S~\ref{sec:imf:accmodel}), explaining the more top-heavy IMF.
	\label{fig:imf.system.vs.sink}}
\end{figure*}

Integrated over the nuclear region, we see star formation with (proto)stellar system masses ranging from $<0.01\,{\rm M_{\odot}}$ to $\gtrsim 200\,{\rm M_{\odot}}$ is possible. It is also a robust result that the IMF is ``top-heavy'' (in the sense of having a shallower high-mass slope) when compared to either the ``mean'' IMF or observed variation in IMFs  in nearby clouds in various compilations \citep{kroupa:2001.imf.var,chabrier:imf,chabrier:2005.review2,bastian:2010.imf.universality,offner:2014.imf.review,hopkins:2018.imf.review}.
It is less robust whether the lower-mass end of the IMF is also ``top heavy'' (i.e.\ shifted towards higher masses) -- this appears to be more sensitive to our assumptions and where/when we measure it (as well as whether we compare to the individual-star or system IMFs in e.g.\ \citealt{chabrier:imf,chabrier:2005.review2}), as we discuss in more detail below.  

It is interesting to note that this is at least qualitatively similar to the inferred IMF in the sub-pc region around Sgr A$^{\ast}$ in the Galactic center \citep{paumard:2006.mw.nucdisk.imf,bartko:2010.mw.nucdisk.imf,lu:2013.mw.center.imf,hosek:2019.arches.top.heavy.imf.1pt8.up.to.40.msun}. In fact, fitting a slope to the ``fast sub-grid accretion'' simulation gives a high-mass slope $dN/dM \propto M^{-\alpha}$ with $\alpha \sim 1.6$ (where Salpeter is $\sim 2.3$), quite similar to that inferred for the central $<0.5\,$pc around Sgr A$^{\ast}$ in \citet{lu:2013.mw.center.imf} (and also broadly similar to the slightly-less-top-heavy claim for Arches in \citealt{hosek:2019.arches.top.heavy.imf.1pt8.up.to.40.msun}). The ``slow sub-grid accretion'' model predicts an even more top-heavy IMF, though still not quite as top-heavy as the most top-heavy estimates from e.g.\ \citet{bartko:2010.mw.nucdisk.imf} ($\alpha\sim 0.5$) for the Galactic center \citep[though see][]{lockmann:2010.gal.center.imf}. This qualitative agreement -- that both are top-heavy -- is encouraging, though we caution against taking any quantitative comparison too literally. Not only is the detailed IMF shape in the Galactic center still controversial, and clearly physics-dependent here, but we very much are {\em not} attempting to model anything like the conditions in the Milky Way center. While the SMBH mass here is only a factor of $\sim 3$ larger than Sgr A$^{\ast}$, our default simulation represents an extremely bright quasar in a massive starburst galaxy at redshift $z\sim 4.4$, quite distinct from the event which formed the observed stars around Sgr A$^{\ast}$. And even our default LISM-cloud STARFORGE simulations often produce a slightly flat ($\alpha\sim2$) IMF \citep{guszejnov:environment.feedback.starforge.imf}. Still, at least some of the physics arguments below should apply to both regimes.

\subsubsection{System versus Single-Star IMFs and Dependence on the Sub-Grid Model for Accretion from Reservoir to (Proto)Star}
\label{sec:imf:accmodel}

In Fig.~\ref{fig:imf.toplevel}, we clearly see systematic differences between the simulations where we adopt a ``fast'' versus ``slow'' sub-grid model for accretion from the captured PSD ``reservoir'' onto the (proto)star within the sink particles. Fig.~\ref{fig:imf.toplevel.vs.system} compares the same IMFs from the simulations to both the individual-star and system IMFs from the Milky Way/LISM in \citet{kroupa:2001.imf.var} and \citet{chabrier:2005.review2}. For the ``slow'' accretion case, the IMF is top-heavy both in terms of the high-mass slope and in terms of the location of the peak/turnover mass (here at $\sim 0.25-1\,{\rm M_{\odot}}$), even compared to the system IMF. For the ``fast'' accretion case the results are a bit more ambiguous. The high-mass slope is shallower than Salpeter, but as noted above not nearly as shallow as in the ``slow'' accretion case. The IMF peak in the ``fast'' accretion case at $\sim 0.2\,{\rm M_{\odot}}$ is clearly somewhat higher than the LISM individual-star IMF, but notably similar to the system IMF peak. This may suggest that the major difference here is the suppression of fragmentation/binarity (an effect we examine explicitly below). However, resolution may also be playing a role for these more modest offsets at sink masses which are only $\sim 40-60$ times our median mass resolution in the QAD (see \S~\ref{sec:resolution}).

Turning to why this may occur, in Fig.~\ref{fig:imf.system.vs.sink} we can see that this choice of accretion model has a huge effect on the in-simulation sub-grid model breakdown between ``star'' versus total sink mass (including the ``reservoir''), as speculated in \S~\ref{sec:methods}. For the ``fast'' accretion model, where the timescale for accretion from the reservoir to star is, by construction, always faster than the local dynamical/tidal timescale, the ``star'' masses closely track the ``system'' or ``total'' masses. For the ``slow'' accretion model, which uses a constant accretion time $\sim 2000-5000\,$yr from PSD onto the star, often longer than the local dynamical time in the CQM, we see a much larger offset, where the ``star'' mass often is $\sim 30$ or so times smaller than its ``reservoir'' mass. This trend is expected: the key difference is when $t_{\rm acc} \ll t_{\rm dyn} \sim 1/\Omega$, the mass is rapidly transferred to the sink (so sink and total masses are similar) while when $t_{\rm acc} \gg t_{\rm dyn}$ the mass simply ``piles up'' in the reservoir/PSD. This in itself is neither surprising nor predictive: we are simply getting out what we put in, and any value of $t_{\rm acc}$ is explicitly sub-grid, so the ratio of ``sink'' to ``reservoir'' mass simply reflects the analytic prescription chosen.

But what is at least somewhat interesting and predictive is that, even if we consider just the total sink masses, the ``slow'' accretion model is systematically much more top-heavy in Figs.~\ref{fig:imf.toplevel}-\ref{fig:imf.system.vs.sink}, whether we consider just recently-formed sinks or all sinks formed since the highest resolution was reached. If everything were perfectly identical in these simulations on all scales, these total masses would be the same (the only difference would be in the sub-grid mass breakdown). However we stress that in every measureable {\em global} QAD/CQM property we considered in \paperone\ or \papertwo, this choice of models make no difference: the primary difference appears to be restricted to the {\em stars} themselves, and must be manifesting on scales of order the individual collapsing cores.

\begin{figure*}
	\centering\includegraphics[width=0.95\textwidth]{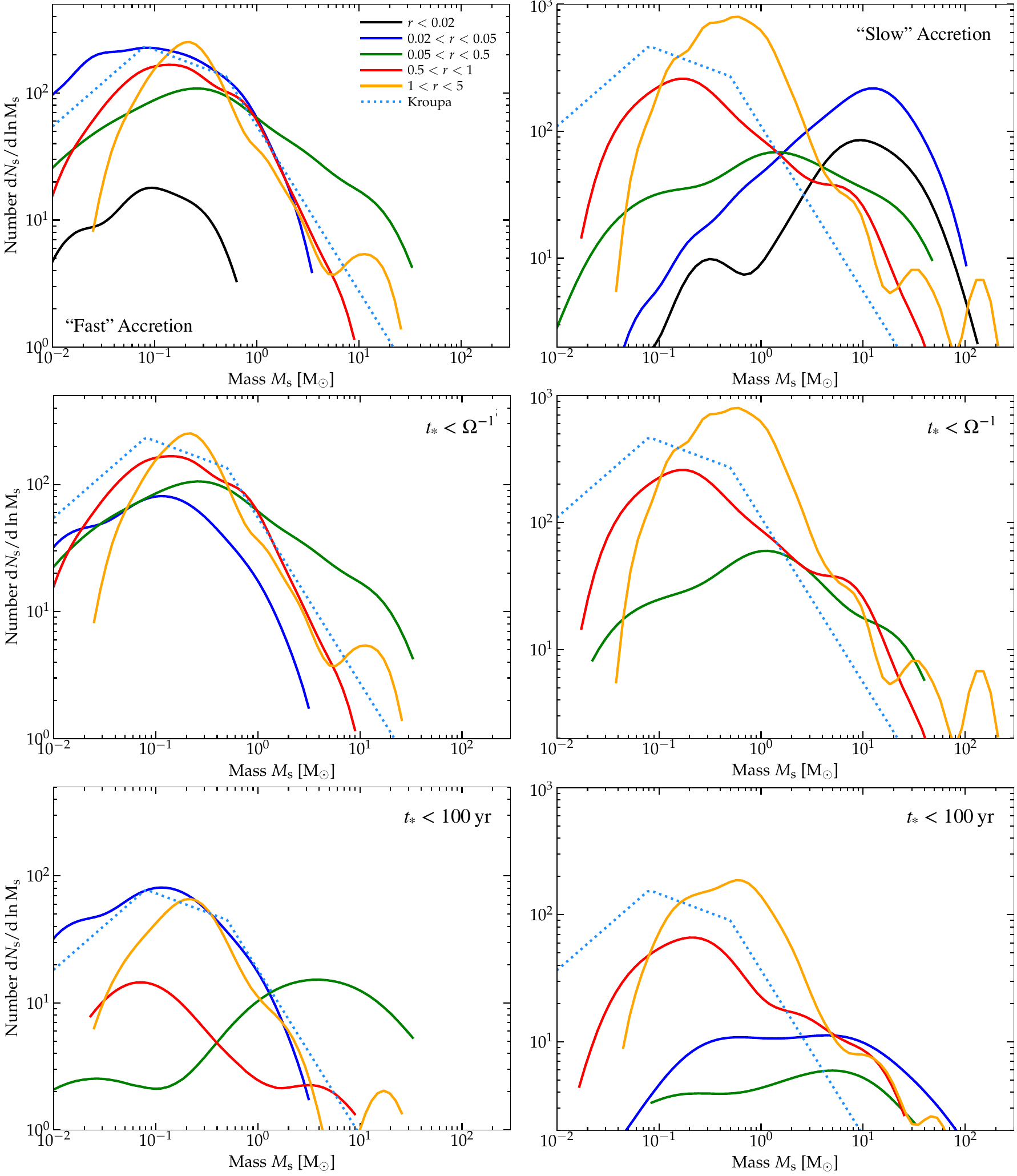} 
	\caption{IMF for the ``fast'' ({\em left}) and ``slow'' ({\em right}) PSD accretion models as Fig.~\ref{fig:imf.toplevel}, but dividing stars into different distances $r$ (in pc) from the SMBH (\S~\ref{sec:imf:distance}). We compare all stars at the end of the simulation ({\em top}), and young stars with time since formation $t_{\ast}<t_{\rm dyn} = \Omega^{-1}(r)$ ({\em middle}) or $t_{\ast}<100\,{\rm yr}$ ({\em bottom}).  the relative normalization of different curves reflects the different numbers of stars (we do not normalize by $N_{s}^{-1}$), so we can compare how many stars form at different times and places. At radii approaching the outer CQM/BHROI, $r\gtrsim$\,pc, where tidal forces from the SMBH become small compared to self-gravity, the IMF approaches the ``universal'' IMF form shown, especially in the Salpeter slope (even though ISM conditions still differ; \S~\ref{sec:ism.diff}). At intermediate radii, particularly the outer QAD around $\sim 0.1-0.5\,$pc, the IMF of young stars becomes notably more top-heavy. At the smallest radii ($r \ll 0.1$\,pc) whether it becomes more top-heavy or bottom-heavy depends on the PSD accretion model assumptions. At radii $r\lesssim 0.01\,$pc there is essentially no measurable star formation in the $\sim 10^{5}$ dynamical times simulated here. The IMF of stars at a given present-day position, including all stars, can be influenced by stars not formed at those radii -- e.g.\ the massive stars at small radii in the ``slow'' run primarily formed in the top-heavy outer-QAD ($\sim 0.1-0.5\,$pc) region, but on highly eccentric orbits and are now plunging through the QAD.
	\label{fig:imf.vs.radius}}
\end{figure*}

This suggests that, while it plays a relatively weak role on large scales, feedback from (proto)stars still plays an important role in regulating the total sink masses. In the ``slow'' accretion limit, $t_{\rm acc} \gg t_{\rm dyn}$, when mass is accreted into the sink it simply sits in the ``reservoir,'' doing nothing in our simulations, as all of the feedback processes we model (protostellar jets, main-sequence mass-loss, radiation, supernovae) depend on some combination of the mass of and mass accretion rate onto the actual sub-grid {\em star}, not the reservoir mass. In the ``fast'' accretion limit, the usual assumptions of efficient feedback (e.g.\ that an order-unity fraction of the swallowed mass is almost-instantaneously returned to the ISM in the form of energetic protostellar jets) hold. 

So given the strong effects of stellar feedback well-studied under more ``typical'' conditions, especially their known ability to regulate the massive end of the IMF \citep{guszejnov:2020.starforge.jets,guszejnov:environment.feedback.starforge.imf,grudic:2022.sf.fullstarforge.imf} it is not surprising that we see a less top-heavy IMF (suppression of the most massive system masses) in the ``fast'' accretion model. The key is that since the primary formation/accretion period for the stars is extremely short-lived or rapid in terms of physical years (\S~\ref{sec:accretion}), the accretion onto the (proto)star must be correspondingly fast to enable any effect of (proto)stellar feedback.

\subsubsection{IMF versus Distance from the SMBH}
\label{sec:imf:distance}

Fig.~\ref{fig:imf.vs.radius} plots the IMF versus ``birth'' location (distance from the SMBH, i.e.\ initial $R_{\rm initial}$ in the QAD/CQM). Note here we have to be careful to separate the birth location from the instantaneous location of the star, for the reasons in \S~\ref{sec:wandering} (stars move significantly in their orbits over the simulation duration on these small scales). So for both the ``fast'' and ``slow'' accretion models, we compare the IMFs at different radii for: (1) all sinks formed since the simulations reach their maximum resolution, (2) ``dynamically young'' sinks which formed less than one dynamical time ago ($t_{\ast} < \Omega^{-1}$, with $\Omega$ measured at the present radius $r$ of the sink), and (3) young sinks with absolute formation age $t_{\ast} < 100\,$yr. 

At $R_{\rm initial} \gtrsim0.5-1\,$pc, the IMF increasingly becomes ``normal'' or only slightly top-heavy. In particular, for {\em both} the ``fast'' and ``slow'' accretion cases, the high-mass slope appears to be rapidly converging to Salpeter-like with increasing distance from the SMBH. Interestingly, for both, at $0.5<r<1$\,pc, the IMF peak mass agrees very well with the observed individual-star IMF, while at somewhat larger radii ($\sim 1-5\,$pc) it becomes mildly top-heavy (shifting by a factor of $\sim 2-3$ to higher masses). It is possible this reflects real trends with e.g.\ the opacity limit or the amount of self-shielding to the intense galactic radiation field, but as discussed in \S~\ref{sec:resolution}, this is precisely the region of the disk where resolution concerns are maximized, and (owing to computational expense), we cannot run at our highest resolution for very long (the dynamical time at $5\,$pc reaches $\Omega^{-1} \sim 5\times10^{4}\,$yr), so it is also not clear if the IMF has actually reached steady-state at these very largest radii. As a result, we urge against over-interpreting the any apparent trend going from $\sim 0.5-1\,$pc to $\sim 1-5\,$pc. 

In any case, with these caveats in mind, it is clear that on scales $\gtrsim 0.5\,$pc well outside the QAD (but still in the CQM), the IMFs approach something more like the ``universal'' IMF. This is noteworthy because conditions at these radii (\S~\ref{sec:ism.diff}) are still quantitatively radically different from the Solar neighbood ISM, but the overall filamentary fragmentation, turbulence, and collapse process appears similar to ``scaled up'' versions of nearby SF (\S~\ref{sec:where:filament}). We discuss the implications in \S~\ref{sec:imf:physics}.

From $R_{\rm initial} \sim 0.05-0.5\,$pc (the location of the outer QAD) on the other hand, the ``top-heavy'' behavior of the IMFs is quite dramatic. Even in the ``fast'' accretion case, we see that (1) the high-mass slope is quite shallow, between $\sim 1.6$ for the dynamically-young stars and $\sim 0$ for the very youngest ($<100\,$yr-old) stars (though we caution the statistics are poor for that latter category so this is driven by just a few sinks). And the turnover mass is also shifted by a factor of a few to higher masses, despite the mass resolution being finer and duration of the simulation much longer. It is clear that especially for the ``fast'' accretion model, the ``overall'' top-heavy behavior of the total IMFs seen in Figs.~\ref{fig:imf.toplevel}-\ref{fig:imf.system.vs.sink} is driven in large part by the stars forming in the outer QAD (mostly around $\sim 0.1-0.3\,$pc). 

Interestingly, if we restrict to young stars forming at the radii discussed so far ($\gtrsim 0.05\,$pc), the differences between the ``slow'' and ``fast'' accretion models are not nearly as dramatic as they appeared in Fig.~\ref{fig:imf.toplevel}. It is still clearly the case that the ``slow'' accretion runs produce a more-top-heavy IMF at each radius, but it suggests that at least {\em some} of the most extreme time-and-location-integrated difference seen in Fig.~\ref{fig:imf.toplevel} owes not to the local physics of the IMF at a given point, but to the question of how many stars (relatively speaking) formed at a given location (since the IMF is quite sensitive to location). 

At slightly smaller radii, $0.02 \lesssim R_{\rm initial} \lesssim 0.05$\,pc, we see a more qualitative divergence in the IMFs. In the ``fast'' model, the IMF becomes less top-heavy again, with very few massive stars. But in the ``slow'' model, it remains comparably top-heavy to the IMF at slightly larger radii $0.05-0.5$\,pc. While the ``more top heavy'' qualitative character of the ``slow'' model is expected, the more dramatic apparent difference here may have to do as much with accidents of exactly where flips in the polarity of ${\bf B}$ occur (see \S~\ref{sec:how}). It is worth noting that almost all the sinks at this radius form towards the very outer edge of the ``bin'' (at more like $\sim 0.04-0.05\,$pc). But it may have interesting implications for the physics driving the IMF in each case. 

At $R_{\rm initial} \lesssim 0.03$\,pc, there is essentially no sink formation. Sinks can exist at these radii, if we consider ``all'' stars, but we see zero very young sinks here, because sink formation is strongly suppressed by the strong toroidal magnetic fields and rapidly-rising thermal Toomre $Q$ at radii interior to this (see Fig.~\ref{fig:profile.dynamics} \&\ \S~\ref{sec:where:disk}-\ref{sec:how}). We therefore cannot define a true ``initial'' mass function here. Rather, stars cross into these radii from larger formation radii. In the ``slow'' run, this leads to the ``apparent'' IMF (simply taking all stars, remembering that they all have ages $<10^{4}\,$yr, which can be extremely large compared to the dynamical time at $\lesssim 0.03\,$pc) being extremely top-heavy. But as we discuss below, this must reflect more which stars make it into these radii (and we similarly see the all-stars mass function at $0.02<r<0.05\,$pc being more top-heavy than the young-star IMF at the same radii when stars actually form).

\begin{figure*}
	\centering\includegraphics[width=0.77\textwidth]{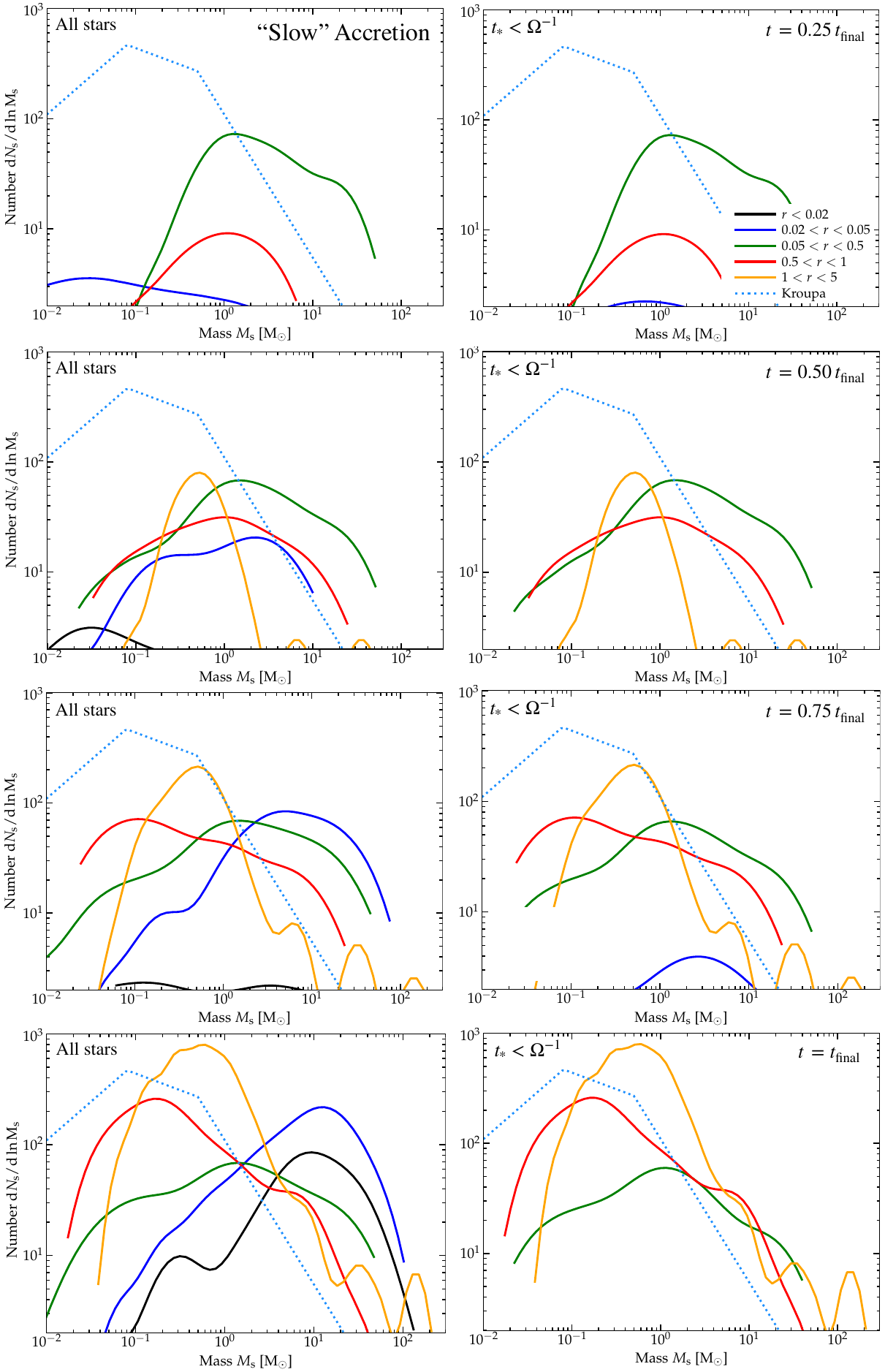} 
	\caption{IMF for the ``slow'' PSD accretion model versus distance from the SMBH (as Fig.~\ref{fig:imf.vs.radius}), comparing all stars formed after reaching the highest refinement level ({\em left}) and dynamically young stars with $t_{\ast}<t_{\rm dyn}(r) = \Omega^{-1}(r)$ ({\em right}), at four different times (early-to-late from {\em top} to {\em bottom}). If we {\em only} consider dynamically-young stars -- i.e.\ stars close to their ``birth sites'' -- the IMF at each $r$ is quite stable in time. Per \S~\ref{sec:imf:time}, the global evolution of the ``all stars'' IMF owes to two effects: (1) stars moving from their birth radii (the dominant effect at $r\lesssim 0.05\,{\rm pc}$ and reason for the existence of {\em any} stars at $r\lesssim 0.02\,{\rm pc}$); and (2) change in the relative number of stars forming at each radius with times (owing to e.g.\ the global mass flow, advection of the sign flips in $B_{\phi}$, etc.). Results are similar for the ``fast'' case.
	\label{fig:imf.time.evol}}
\end{figure*}

\subsubsection{Evolution over Time}
\label{sec:imf:time}

Fig.~\ref{fig:imf.time.evol} plots the IMFs versus time in the simulations. We focus on the ``slow'' model, since in Fig.~\ref{fig:imf.toplevel} this appears to show a notable difference between the IMF integrated over all time since the simulation began versus that of just young sinks (though of course there has been no time to have objects evolve off the main sequence). All our conclusions qualitatively cross-apply to the ``fast'' model case, but the differences there are much smaller. 

Specifically, in Fig.~\ref{fig:imf.time.evol} we plot both the MF (versus radius of formation as Fig.~\ref{fig:imf.vs.radius}) of all sinks formed since the simulation reaches maximum refinement, and just dynamically young sinks with $t_{\ast} < \Omega^{-1}$ at their radii, at four equally-spaced times leading to the final time (what we plotted in Fig.~\ref{fig:imf.vs.radius}). Note that at much earlier times than the earliest already plotted, there are simply not enough sinks to meaningfully sample the MFs. 

If we first just focus on the dynamically-young stars, the real ``IMF'' here, then we see that there is not actually much significant evolution in the MF of sinks just formed {\em at a given radius}. There are some subtle effects, but most of these are within the sampling/Poisson errors (binning by both time and radius and age leads to only a few sinks in several of the bins). For example, the IMF at $0.5<r<1$ pc may become slightly less top-heavy after the earlier times, as the CQM on these radii relaxes somewhat and comes more into steady-state after refinement is completed, but the effect is modest. Overall, this is reasonable: as shown in \paperone\ and \papertwo, the global QAD and CQM properties are in steady-state for the duration of our simulations. 

Interestingly, however, if we examine the MF of all sinks versus time, there appears to be more dramatic evolution of the MFs at $r < 0.05\,$pc, in particular. These become much more top-heavy as time goes by. Since especially at $r<0.02\,$pc this is just a few sinks, we have checked that this evolution is not driven by some sort of late accretion episodes onto these stars (see Fig.~\ref{fig:dt.accretion} for explicit illustration of this). This is consistent with what we showed in \S~\ref{sec:speed} and Fig.~\ref{fig:dt.accretion}, that these stars accrete little after their dynamically-young period. 

Rather, we can see that this trend comes entirely from ``new'' stars (stars which had not previously been at $r<0.02\,$pc) appearing at these radii despite forming at larger radii. Note that at just larger radii, $0.05<r<0.5\,$pc, the IMF is quite top-heavy, so almost any movement of stars from that radius will make the MF at smaller radii more top-heavy. As noted in \S~\ref{sec:wandering} these stars are not ``sinking'' via something like dynamical friction or Type I migration (in the sense of systematically migrating in on slowly-decaying nearly-circular midplane orbits), as the timescale for this to occur is much larger than the Hubble time. Rather, direct examination of their orbital parameters confirms the stars of interest are simply ``passing through'' the inner disk on highly-eccentric, often inclined orbits (see Figs.~\ref{fig:edgeon.oldstars.and.ejecta} \&\ \ref{fig:stellar.velocities}).

\subsubsection{What Properties do the ``Reservoirs'' Have, and Should They Accrete Rapidly, or Fragment?}
\label{sec:imf:frag}

Given the difference in \S~\ref{sec:imf:accmodel} between the ``fast'' vs.\ ``slow'' sub-grid accretion models for our sink particles, it is natural to ask what ``should'' occur, and our simulations can provide some insight (although fundamentally, the goal here it to motivate more detailed future work, as described below). If accretion is fast ($t_{\rm acc} \lesssim t_{\rm dyn}$), then the mass will be accreted from the sink radius (where it is gravitationally captured by the sink in a resolved sense) onto the protostar rapidly, and there will not be much reservoir or ``PSD'' to speak of, but if accretion is ``slow'' ($t_{\rm acc} \gg t_{\rm dyn}$) then mass piles up in the reservoir by definition. One might imagine this occurring if, for example, gas accreted into quasi-circular orbits which form some kind of $\alpha$-disk around the true protostar, accreting slowly onto said protostar with a very low $\alpha \ll 1$. As shown in Fig.~\ref{fig:imf.system.vs.sink}, if this occurred at the rates assumed in our ``slow'' accretion scenario, it would imply a buildup of a reservoir or ``PSD'' mass $M_{\rm PSD} \gg M_{\ast}$. So if this really were in a PSD and accretion were ``slow,'' it seems likely such a system would fragment catastrophically into close multiples. 

To gain some insight into what is reasonable, recall (per \S~\ref{sec:ism.diff:tidal}) that the dynamical time in the CQM is $t_{\rm dyn}^{\rm CQM} = 1/\Omega = \sqrt{R^{3}/G\,M_{\rm BH}} \sim 5000\,{\rm yr}\,R_{\rm pc}^{3/2}\,M_{\rm BH,\,7}^{-1/2}$ ($M_{\rm BH,\,7} \equiv M_{\rm BH}/10^{7}\,{\rm M_{\odot}}$, $R_{\rm pc} \equiv R/{\rm pc}$), and the tidal radius of a PSD+star system with system mass $M_{\rm s} \equiv M_{\ast} + M_{\rm PSD} = M_{\rm s,\,\odot}\,{\rm M_{\odot}}$ is $r_{\rm t} = R\,(M_{\rm s}/4\,M_{\rm BH})^{1/3} \sim 1000\,{\rm au}\,M_{\rm s,\,\odot}^{1/3}\,M_{7}^{-1/3}\,R_{\rm pc}$, which gives a maximum PSD dynamical time of $t_{\rm dyn}^{\rm s,\,max} \equiv  \sqrt{r_{\rm t}^{3}/G\,M_{\rm s}} = 1/2\,\Omega \sim 2500\,{\rm yr}\,R_{\rm pc}^{3/2}\,M_{\rm BH,\,7}^{-1/2}$. Given the accretion criteria adopted in our simulations (see \S~\ref{sec:methods:ov}), it is implicitly required that any accreted gas onto a a protostellar sink would have a circularization radius less than $r_{\rm t}$ (since it must be bound to the sink, have apocentric radius inside the sink radius, etc.). So the key question is whether or not accretion in the PSDs is ``dynamical'' ($t_{\rm acc} \sim t_{\rm dyn}^{\rm s,\,max}$) or ``secular'' ($t_{\rm acc} \gg t_{\rm dyn}^{\rm s,\,max}$). In gravitationally-unstable, rapidly cooling, or strongly-magnetized PSDs, the former (dynamical accretion) is usually true in idealized simulations \citep{gammie:2001.cooling.in.keplerian.disks,
deng:gravito.turb.frag.convergence.gizmo.methods,deng:2020.global.magnetized.protoplanetary.disk.sims.gravito.turb.leads.to.large.B.saturation.vs.mri,riols:2016.mhd.ppd.gravitoturb,kratter:grav.instab.review}; but in a traditional weakly-turbulent \citet{shakurasunyaev73}-like $\alpha$-disk PSD with $\alpha \ll 1$, the accretion is secular with $t_{\rm acc}/t_{\rm dyn} \sim (c_{s}^{\rm PSD}/V_{c}^{\rm PSD})^{-2}\,\alpha^{-1} \gg 1$. 

It is notable that our numerical sink accretion radii $r_{\rm sink} \sim 80-100\,$au are within $r_{\rm sink} < r_{\rm t}$ so long as $R_{\rm pc} \gtrsim 0.1\,M_{\rm s,\,\odot}^{-1/3}$ (and we show this explicitly below when we discuss multiplicity), and our mass resolution is such that the more massive PSDs would have as many as $\sim 10^{4}$ cells. So in principle, we should be able to marginally resolve at least the formation of these PSDs or ``stalled'' accretion reservoirs if the accretion were really ``slow,'' at least at larger radii $\sim 1\,$pc. But we find that at any given time, it is very rare to find any systems with extended resolved PSDs or quasi-hydrostatic accreting gas ``reservoirs,'' in {\em either} the ``fast'' or ``slow'' $t_{\rm acc}$ simulations. We see none in the few stars that form in our higher-resolution test in \S~\ref{sec:resolution}. And we show below that multiplicity is very strongly suppressed in the environments here. This, coupled to the sink growth histories in Fig.~\ref{fig:dt.accretion}, suggests that at least on semi-resolved scales from $\sim 100-1000$\,au, the accretion is indeed ``dynamical'' given the physics in our simulations,\footnote{This does not necessarily mean the accretion would be dynamical throughout the entire life of the star and/or reservoir/PSD, as we are really only interested in and able to follow the duration of the initial phase of rapid stellar growth. A more depleted compact PSD or reservoir, with mass much less than the star could of course persist much longer with much lower accretion rates, more akin to typically observed PSDs in the Solar neighborhood with $\gtrsim 10^{5}-10^{6}\,$yr ages that have depleted to $M_{\rm PSD} \ll M_{\ast}$.} and that massive PSDs never really build up in the first place. In contrast, if accretion were ``slow'' on resolved scales we would expect to see massive PSDs in both simulations, but with an over-depleted ``hole'' in the center in our ``fast accretion'' simulations.

That should not, perhaps, be surprising, for at least three reasons. First, as noted in \S~\ref{sec:ism.diff}, the CQM is extremely strongly-magnetized and well-ionized, with the colder gas accreting onto protostars having typical $\beta$ as low as $\sim 10^{-4}$. Magnetic braking should therefore produce efficient torques and prevent PSD formation, far moreso than in ``typical'' (less-strongly magnetized) GMC conditions. Second, the early stages of even pure-hydrodynamic collapse follow more closely a \citet{shu:isothermal.sphere.collapse}-type quasi-spherical, quasi-isothermal collapse with $\dot{M}_{\ast} \sim c_{s}^{3}/G$, and $c_{s}$ here is much larger than in the LISM (\S~\ref{sec:ism.diff}) producing larger accretion rates by a factor of $\sim 10^{4}-10^{6}$ compared to a $\sim 10\,$K standard cold molecular medium in the Solar neighborhood. Third, even if one tried to form a PSD,  the massive PSDs of greatest interest here would likely be highly unstable: for a PSD with most of $M_{\rm s}$ in $M_{\rm PSD}$, with temperature in the PSD $T_{\rm PSD,\,gas}$ and size equal to the tidal radius $r_{t}$, we would have $Q \sim 0.4\,(T_{\rm PSD,\,gas}/1000\,{\rm K})^{1/2}\,R_{\rm pc}^{1/2}\,(M_{\rm s}/100\,{\rm M_{\odot}})^{-1/3}\,M_{\rm bh,\,7}^{-1/6}$, with a cooling time at $r_{\rm t}$ of $t_{\rm cool} \ll \Omega^{-1} \sim t_{\rm dyn,\,PSD}$. So the PSDs would be gravitoturbulently unstable without magnetic fields \citep{gammie:2001.cooling.in.keplerian.disks,forgan:2017.mhd.gravitoturb.sims,riols:2018.mhd.sims.ppd.gravitoturb.fx.on.mri}.

It is important to note that all of these properties strongly differentiate the circum-stellar PSDs around stars embedded in the QAD from circum-planetary (or satellitesimal) disks around planets embedded in a ``typical'' circum-stellar disk, a point we discuss further in \S~\ref{sec:sf.qad.planets.psd}.

Regarding the strong magnetization, one might worry that non-ideal MHD effects should lead to less efficient angular momentum transport (as commonly argued for Solar-neighborhood PSDs). However, we note that (unlike STARFORGE simulations of more typical Solar-neighborhood GMCs), we also do not see prominent PSDs on these scales even in our simulations {\em without} magnetic fields, discussed below. So it cannot be a purely magnetic effect. And in the ``diffuse'' CQM (outside of PSDs) it is easy to verify that non-ideal MHD effects are irrelevant: as shown in detail in \paperone, and discussed in \S~\ref{sec:ism.diff}, given the high temperatures ($\gtrsim 1000\,$K) of even molecular gas, extreme irradiation and cosmic ray background in the vicinity of a bright quasar and nuclear galactic starburst, the ionization fractions more closely resemble the warm neutral medium (WNM) and warm ionized medium (WIM) (with $\sim 1-100\%$ ionization fraction) so ideal MHD is actually an excellent approximation. Moreover, as discussed in \paperone, we have experimented with turning on and off different non-ideal MHD terms. Ohmic resistivity and ambipolar diffusion have negligible effects here: the former (Ohmic) because we never reach low enough ionized fractions and high enough densities for it to dominate among the non-ideal effects, the latter (ambipolar) because even where it is the ``most important'' non-ideal MHD effect, the relevant ambipolar diffusion timescales are vastly longer than the effective turbulent diffusivity timescales (given the trans-\Alf{ic} and highly super-sonic turbulence) on resolved scales here \citep[consistent with a number of other recent simulations of even more typical Solar-neighborhood GMCs; see][]{mac-low:2004.turb.sf.review,vazquez.semadeni:2011.weak.fx.ambipolar.diffusion.gmc.sf,chen:2014.core.collapse.with.without.ambipolar.similar,wurster:2021.ambipolar.small.fx.starform.hall.fx.larger,sadanari:2022.ambipolar.diffusion.first.stars.fx.small}. 

The Hall effect is slightly more complicated. We do see some places where Hall MHD could in principle be important (at least in the sense that the fastest Hall timescales -- those associated with fast whistler waves -- become shorter than other resolved timescales like \Alf\ or turbulent crossing times at our resolution). However this only occurs in a very small subset of the gas in the simulation, which has extremely small ion fractions $x_{i} \sim n_{\rm ion}/n_{\rm neutral} \sim 10^{-17}-10^{-15}$ (vastly lower than the diffuse CQM). But there are several numerical challenges involved here: (1) gas only ever appears to reach these conditions on extremely small scales of order one resolution element around sink particles (at the very highest densities before being accreted), so coherent Hall MHD effects are not well-resolved; (2) this usually involves very thermally-cold cells dropping to around the temperature ``floor'' set by the CMB and other physics, but these cells always have plasma $\beta$ extremely small (often $\ll 10^{-4}$), so accurately integrating their thermal evolution is extremely challenging (very small integration errors in their kinetic/magnetic energies can easily dominate over the physical thermal energy); (3) these fast Whistler waves impose a very short timestep requirement, so in our default simulations we choose to ignore this effect as it limits our simulation runtime; (4) it is well known that the Hall effect is very sensitive to the thermo-chemistry and especially dust properties of the PSDs \citep{bai:2017.hall.magnetic.transport.ppds,zhao:2020.hall.fx.ppds,lee:2021.non.ideal.mhd.fx.planet.disks,tsukamoto:hall.mhd.fx.ppd.form}, which are very different here compared to ``typical'' PSDs in the Solar neighborhood, so it is by no means clear if our chemical network is totally adequate for these extremes; and (5) the implied Hall currents for this to be important are often superthermal, so traditional scalings for the Hall coefficient break down \citep{hopkins:2024.hall.coeff.in.neutral.limit.superthermal.drifts}. So while it is conceivable that these effects could be important in the most-dense PSDs which form here, following Hall effects believably clearly requires much more highly-resolved simulations.

We conclude that our ``fast'' sub-grid accretion model is probably more reasonable than the ``slow'' model. But saying anything more requires future work. In particular, this would be better studied by idealized simulations of a collapsing core in the extreme environments here. Critical for this problem is to include (1) the extremely strong tidal/shear background (analytically as in a shearing-box simulation), (2) the very strong fluxes of mass (with $\dot{M}$ reaching $\gtrsim 1\,{\rm M_{\odot}\,yr^{-1}}$ in the rapid collapse phase), magnetic flux, and thermal energy from the ambient warm ionized medium, (3) the much stronger local interstellar radiation field and cosmic ray background, and (4) some model for dust sublimation.

\begin{figure*}
	\centering\includegraphics[width=0.98\textwidth]{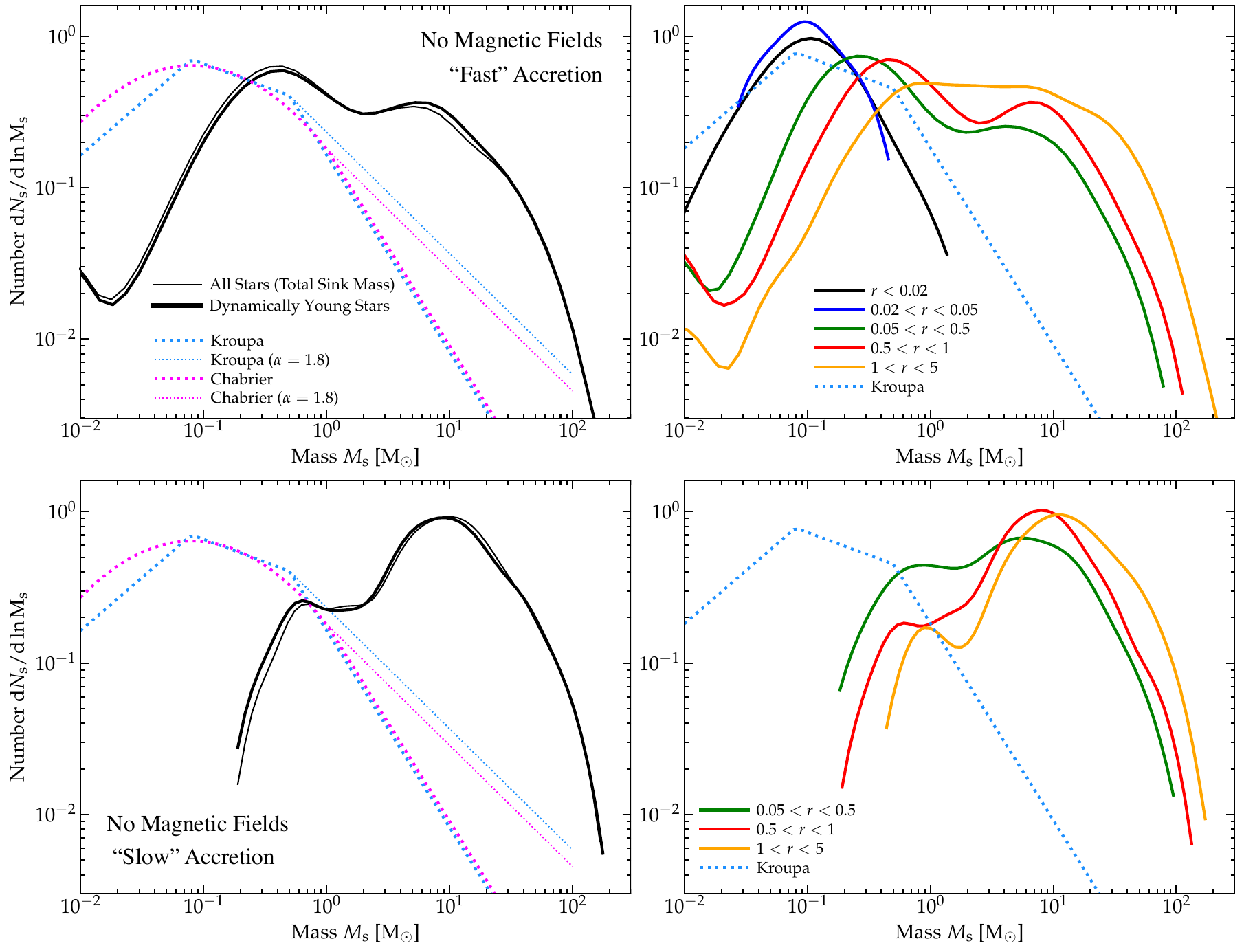} 
	\caption{IMF for the ``fast'' ({\em top}) and ``slow'' ({\em bottom}) PSD accretion model for all stars interior to the BHROI as Fig.~\ref{fig:imf.toplevel} ({\em left}) and versus radius as Fig.~\ref{fig:imf.vs.radius} ({\em right}), but in simulations where we repeat our hyper-refinement experiment keeping all other physics identical but removing magnetic fields from the initial conditions per \S~\ref{sec:imf:bfields}. These are normalized by $N_{s}^{-1}$: the absolute SFR and number of stars is $\gg 100$ times larger at $r\ll 1$\,pc compared to the runs with magnetic fields. Per \paperone\ and expectations from \S~\ref{sec:ism.diff}, artificially removing magnetic fields qualitatively changes the properties of the QAD+CQM, leading to runaway fragmentation on sub-pc scales, orders-of-magnitude higher SFRs, and a much smaller, thinner, less-massive QAD with a factor $\sim 500$ lower accretion rate onto the SMBH. But despite the {\em enhancement} of global fragmentation and star formation and lower gas masses in the QAD, the predicted IMF actually becomes much more top-heavy absent magnetic fields (in both PSD accretion-rate scenarios). Moreover, the predicted IMFs become monotonically {\em more} top-heavy at larger radii $r$ interior to the BHROI (perhaps simply limited by the small QAD mass as $r\rightarrow 0$ in these simulations). This demonstrates the critical role of magnetic fields suppressing not just the global fragmentation/SFR (discussed in \paperone), but also the masses of individual cores and their ability to accrete.
	\label{fig:imf.vs.mhd}}
\end{figure*}

\subsubsection{Dependence on Magnetic Fields}
\label{sec:imf:bfields}

In \paperone\ \&\ \papertwo, we compared a re-run of this simulation without magnetic fields, and showed that this produces radically different global properties. In particular, suppression of SF and gravitoturbulent fragmentation in the QAD depends strongly on magnetic fields. In the full-physics simulation, magnetic pressure dominates the CQM and QAD pressure, and so removing magnetic fields leads to a much thinner, less massive, smaller-spatial-scale, rapidly gas-depleted QAD, with the SFR at CQM scales larger by factors $\sim 10^{3}-10^{5}$, and the accretion rates onto the SMBH decreasing by factors of $\sim 10^{2}-10^{3}$. So clearly magnetic fields play a crucial role in the global dynamics on these scales.

In Fig.~\ref{fig:imf.vs.mhd}, we show how removing MHD influences the predicted IMF in these simulations. Even on timescales as short as a few hundred years after the highest refinement level is reached (shorter than the duration of our full-physics run at its highest resolution), we see that absent magnetic fields, the IMF ``runs away'' to much more top-heavy values. Even with the ``fast'' PSD accretion model (which produces a mildly top-heavy IMF in the full-physics case), the IMF absent magnetic fields shifts to resemble a second lognormal peak at $\gtrsim 10\,M_{\odot}$. Also important, without MHD the IMF becomes similarly top-heavy at all radii where we can predict it (where we employ full STARFORGE physics). Recall, in the full-physics case, the IMF becomes progressively less top-heavy at larger radii, resembling a ``typical'' IMF in the more ambient ISM at $R \gg 1\,$pc. Absent MHD, even at $>\,$pc scales, the IMF appears to be getting systematically more top-heavy at larger-$r$.

The fact that magnetic fields in our simulations help to suppress global fragmentation and star formation rates overall is qualitatively consistent (albeit more dramatic, owing to much stronger fields) with most previous studies of star formation in Solar-neighborhood-like ISM conditions \citep[see][and references therein]{krumholz.federrath:2019.review.bfields.sf}. However, studies of the IMF in LISM conditions (including those with the same STARFORGE physics used here) have generally concluded that magnetic fields have a relatively weak net effect on the IMF of the stars that do form \citep[see][]{myers:2014.orion.imf.with.without.magnetic.fields,lee:2019.imf.magnetic.field.fx,krumholz.federrath:2019.review.bfields.sf,sharda:pop3.imf.with.bfields}, with the sign of the net effect varying as magnetic fields can both suppress fragmentation (producing larger ``cores'') and also accretion (suppressing growth after initial collapse). But our results are more similar to \citet{hocuk:2012.b.fields.in.circumnuclear.smbh.imf.star.form}, who considered idealized simulations of a $\sim 800\,M_{\odot}$ cloud at a distance of $\sim 10$\,pc from a $\sim 10^{7}\,M_{\odot}$ BH,\footnote{We do caution that while the tidal environment considered is similar, the simulations in \citet{hocuk:2012.b.fields.in.circumnuclear.smbh.imf.star.form} differ fairly radically from those here (\S~\ref{sec:ism.diff}) in that their simulations considered orders-of-magnitude smaller mass, density, column density, temperature, and magnetic field values.} and showed adding strong magnetic fields shifted their predicted turnover mass from $\sim 1\,M_{\odot}$ to $\sim 0.1\,M_{\odot}$ (see also \citealt{li:2010.imf.lower.mass.without.magnetic.fields}). Those authors noted that under these conditions, the magnetic fields strongly suppress accretion, and can even lower the thermal Jeans mass while also suppressing fragmentation by restricting fragmentation to a limited set of more extreme over-dense regions (with higher $\rho$ so lower $M_{\rm Jeans}$) -- both effects we also see and describe above. The more extreme difference we see between our magnetized and non-magnetized runs likely owes to the more extreme magnetic field strengths. 

Even without magnetic fields, as noted above, existing hydro-only QAD simulations from \S~\ref{sec:intro} \citep{nayakshin:sfr.in.clumps.vs.coolingrate,bonnell.rice:2008.gmc.infall.smbh.sf.sims,hobbs:2009.mw.nucdisk.sim,hocuk.spaans:2011.radiated.clouds.near.galactic.center.quasars.imf,alig:2013.nuclear.disk.sims.idealized.hydro.only} and ours predict top-heavy IMFs but do not predict the sort of supermassive stars or runaway accretion argued for analytically by some \citep[e.g.][]{goodman.tan:2004.qso.disk.supermassive.stars}. Reasons for this are discussed in \S~\ref{sec:accretion}, \ref{sec:imf:time}, \&\ \ref{sec:imf:physics}, but in brief the situation is not akin to giant planet formation around a star, but more like the ISM as just discussed. Absent magnetic fields, if $Q_{\rm thermal} \ll 1$, the medium does fragment into Toomre-mass clumps as som analytic QAD SF models assume, but those clumps cannot contract homologously into single extremely massive stars, nor open gaps and experience runaway accretion -- that would require slow cooling $t_{\rm cool} \lesssim t_{\rm dyn}$ in a weakly-turbulent, thin, quasi-laminar QAD (with ``background'' QAD gas mass much larger than QAD stellar mass). Instead like GMCs, in any supersonically turbulent or efficiently-cooling system, fragmentation occurs from the maximally unstable scale down to the opacity limit/Larson core mass (Figs.~\ref{fig:imf.vs.mhd} \&\ \ref{fig:profile.fragmass}). Because stellar feedback is globally ineffective owing to the deep potential, efficient fragmentation means that the QAD stellar mass or $\Sigma_{\ast}$ at all radii resolved here becomes larger than the QAD gas mass or $\Sigma_{\rm gas}$ within a dynamical time (\paperone, \papertwo), and so accretion proceeds at orders-of-magnitude lower rates with the remaining gas via gravitational torques between stellar and gas patterns (as anticipated in \citealt{paczynski:1978.selfgrav.disk} and studied in detail in \citealt{levine2008:nuclear.zoom,hopkins:zoom.sims,hopkins:inflow.analytics,prieto:2016.zoomin.sims.to.fewpc.hydro.cosmo.highz}), with star formation moving inwards along with the gas, as described in \citet{daa:20.hyperrefinement.bh.growth}.

\begin{figure}
	\centering\includegraphics[width=0.95\columnwidth]{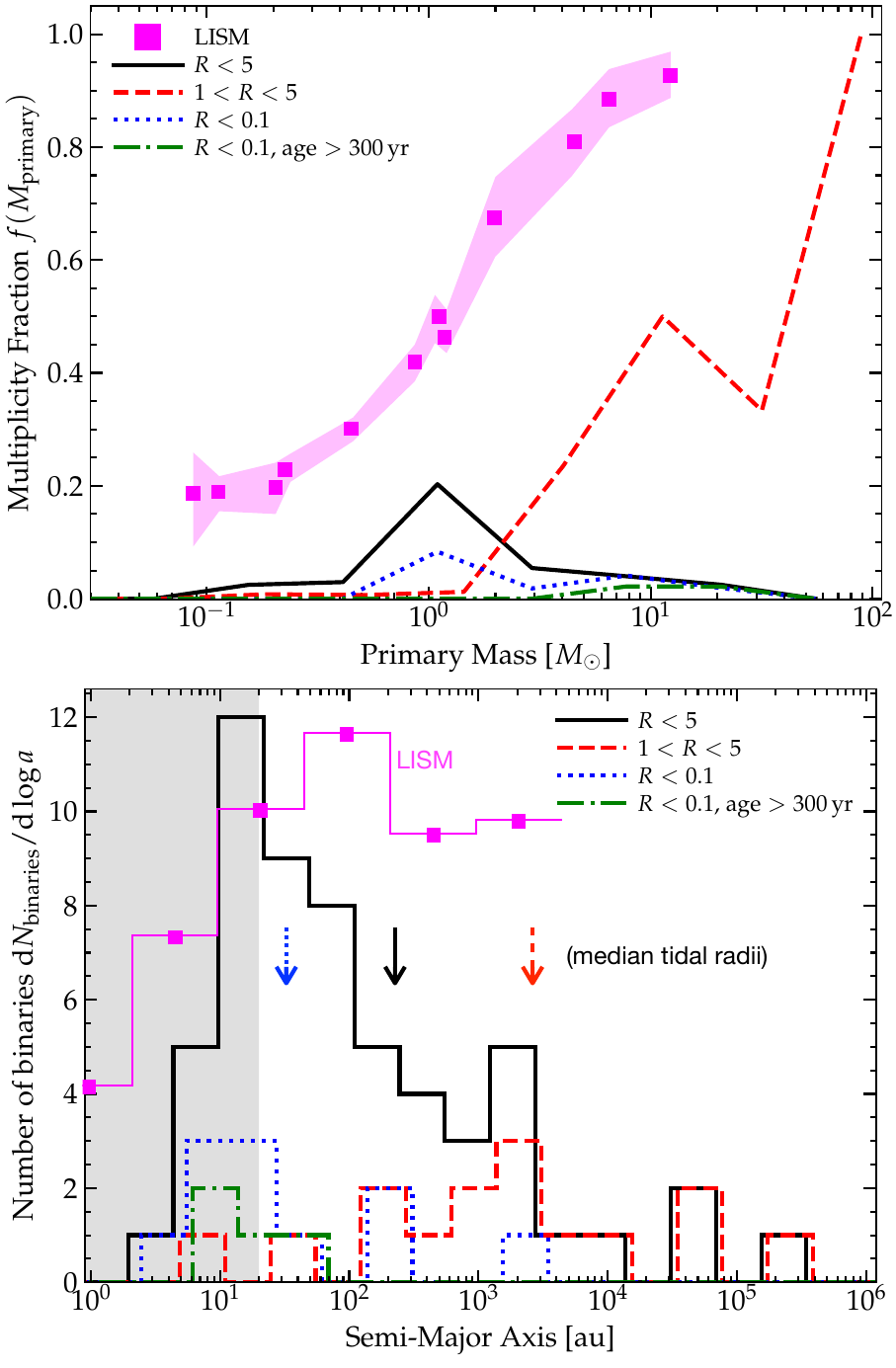} 
	\caption{Resolved multiplicity statistics (\S~\ref{sec:imf:multiplicity}). 
	{\em Top:} For the ``slow'' PSD accretion simulation, we plot the multiplicity fraction $f$ of stars versus primary sink mass $M_{1}$, including all mass ratios, selecting stars with formation distances from the SMBH $r$. We compare LISM measurements compiled in \citet{offner:2022.multiplicity.review}, similar to STARFORGE simulations of typical LISM GMCs \citep{guszejnov:starforge.environment.multiplicity}. 
	{\em Bottom:} Semi-major axis ($a$) distribution of the same multiples (given the low $f$, these are almost all binaries). Arrows indicate a typical tidal radius $r_{\rm t} = R\,(M_{\rm s}/4\,M_{\rm BH})^{1/3} \sim 500\,{\rm au}\,(M_{\rm s}/{\rm M_{\odot}})^{1/3}\,(R/{\rm pc})$ for a binary (of mass $M_{\rm s}$ equal to the median mass of the binaries in each line) in the SMBH potential. Note that the ``tails'' of stars with larger $a$ largely correspond to more-massive stars or stars near the outer edge of the radii ranges, for which $r_{\rm t}$ is larger. We plot LISM observations from \citet{raghavan:2010.period.axis.distrib.stars.lognormal} after completeness correction in \citet{moe:2017.binary.statistics.period.massratio.distribs}.
	Shaded range shows sink gravitational softening radii (harder binaries are un-resolved).
	At radii $\gtrsim\,$pc where the tidal effects of the SMBH become relatively weaker, multiplicity statistics begin to approach the LISM results (though still suppressed), but at smaller radii, multiplicity is very strongly suppressed. 
	The suppression occurs both at formation (fewer ``common core'' binaries form), and then via subsequent evolution/tidal disruption. We see slightly older stars, even $>300\,$yr or $\sim 3\,\Omega^{-1}$ at $r \lesssim 0.1\,$pc, have even lower multiplicity, and the semi-major axis distribution is truncated at the tidal radii as expected. Trends are similar for the ``fast'' accretion case.
	\label{fig:multiplicity}}
\end{figure}

\subsubsection{Multiplicity}
\label{sec:imf:multiplicity}

We see strongly suppressed multiplicity in the QAD environment. This is true whether we attempt to visually identify binaries or use more sophisticated linking algorithms such as those in \citet{guszejnov:starforge.environment.multiplicity}. We emphasize this is quite different from simulations with the same physics and code in ``typical'' Solar-neighborhood ISM GMCs \citep{bate:2012.rmhd.sims,lee:2019.binary.sim,guszejnov:starforge.environment.multiplicity,kuruwita:2023.binary.sim}. But it is not surprising: even if binaries can form, unequal-mass stellar binaries which have a separation comparable to or larger than the tidal radius $r_{\rm t} = R\,(M_{\rm s}/4\,M_{\rm BH})^{1/3} \sim 500\,{\rm au}\,(M_{\rm s}/{\rm M_{\odot}})^{1/3}\,(R/{\rm pc})$ will be tidally sheared apart in these extreme environments. And binary formation, at least ``common core'' binaries, should be suppressed in these environments by multiple effects: (1) the warm gas temperatures and strong magnetic fields raise the Jeans/magnetic critical mass as discussed above and in \S~\ref{sec:imf:physics} below; and (2) the strong magnetic fields and high ionization fraction lead to rapid accretion/magnetic braking and inhibit formation of PSDs (\S~\ref{sec:imf:frag}). 

Fig.~\ref{fig:multiplicity} shows this quantitatively: we plot the multiplicity fraction, defined using the algorithm to identify all bound multiples (including hierarchical multiples) from \citet{bate:2012.rmhd.sims} as applied in \citet{guszejnov:starforge.environment.multiplicity}.\footnote{Note that we do not apply the additional mass-ratio and age cuts considered in \citet{guszejnov:starforge.environment.multiplicity}. The age cuts of $\gtrsim$\,Myr would not make sense given the timescales of interest here; the mass-ratio cuts make little difference to the result but will slightly decrease the multiplicity fractions if we did apply them.} First, we show the multiple fraction (defined here simply as the fraction of stars in bound multiples),\footnote{Because the multiplicity is low, most of the multiples are simple binaries, not higher-order multiples, so it makes little difference here if we consider a binary fraction versus some higher-order multiplicity fraction statistics as discussed in \citet{guszejnov:starforge.environment.multiplicity}.} versus stellar mass, for stars at various BH-centric radii $R$ and with various ages since the initial sink formation time. For reference, we also plot the observational compilation in the LISM with which the simulations in \citet{guszejnov:starforge.environment.multiplicity} agreed. Recall, these used the same STARFORGE physics to simulate ``typical'' Solar neighborhood GMC/LISM environments, and there the multiplicity fraction rises from $\sim 0.5$ at $\sim 1\,M_{\odot}$ to $\sim 0.9$ at $>10\,M_{\odot}$, much larger than we see here. 

Clearly there is some radial trend where at larger radii, the simulations begin to approach the ``typical'' ISM, though still with a notable suppression, which again is expected from the simple tidal radius argument above. We also see clear evidence for this in the semi-major axis ($a$) distribution: at e.g.\ $R<0.1\,$pc, considering binaries $>300\,$yr old, the only surviving (small number) of examples have semi-major axes $a<100\,$au, more or less exactly the tidal radius predicted for this $R$ and typical stellar masses $\sim 10-30\,M_{\odot}$ in the simulation at those radii (this is the example with ``slow'' PSD accretion). Even at $R\sim 1$\,pc, the semi-major axis distribution is sharply suppressed at $a$ larger than a few thousand au, again roughly the tidal radius at these BH-centric radii. For comparison, the observed semi-major axis distribution in the LISM (for Solar-type and massive primaries) is flat up to at least $a \sim 10^{4}$\,au. 

So clearly some of the suppression of multiplicity owes to the strong tidal forces, but this is probably not all of it. If we simply take the Solar-neighborhood binary fractions and truncate them by eliminating any binaries with $a \gtrsim r_{\rm t}$ (as well as removing any which would be numerically-unresolved with $a \lesssim 20\,$au), we would still predict a ``surviving'' binary fraction of $\sim 20-30\%$ in the massive $\gtrsim 10\,M_{\odot}$ stars, for example. But the simulations predict a binary fraction $\lesssim 5-10\%$. This additional suppression of stellar multiplicity could occur if a substantial fraction of binaries at $\lesssim 20$au separation form through fragmentation at larger separations and then migrate inwards through gas or dynamical friction \citep{offner:2016.core.frag.spin.misalign,lee:2019.binary.sim,kuruwita:2023.binary.sim}. Here, such binaries would be tidally disrupted before migration could occur. Alternatively, the remaining suppression owes to the physics described above suppressing the incidence of fragmentation (both core and disk).

Briefly, we do not see any obvious unusual behavior in the mass ratio distribution, but given the combination of small statistics and resolution limitations (and incompleteness for observational mass-ratio distributions), this is not a particularly significant statement.

Of course, we cannot resolve compact binaries that could form in principle via fragmentation in some dense PSD or collapsing cores at scales below $\sim 10\,$au in these simulations. Those are particularly interesting for e.g.\ TDE and LIGO/LISA sources as noted in \S~\ref{sec:intro}. Predictions for such compact binaries will fundamentally require dedicated, much higher-resolution simulations of the formation of individual systems with the extreme boundary/initial conditions provided by the simulations here, as argued for in \S~\ref{sec:imf:frag}.

\subsection{Physics: Why Is the IMF Top-Heavy?}
\label{sec:imf:physics}

What physics drives the IMF to be top-heavy in the simulations?

\begin{figure}
	\centering\includegraphics[width=0.98\columnwidth]{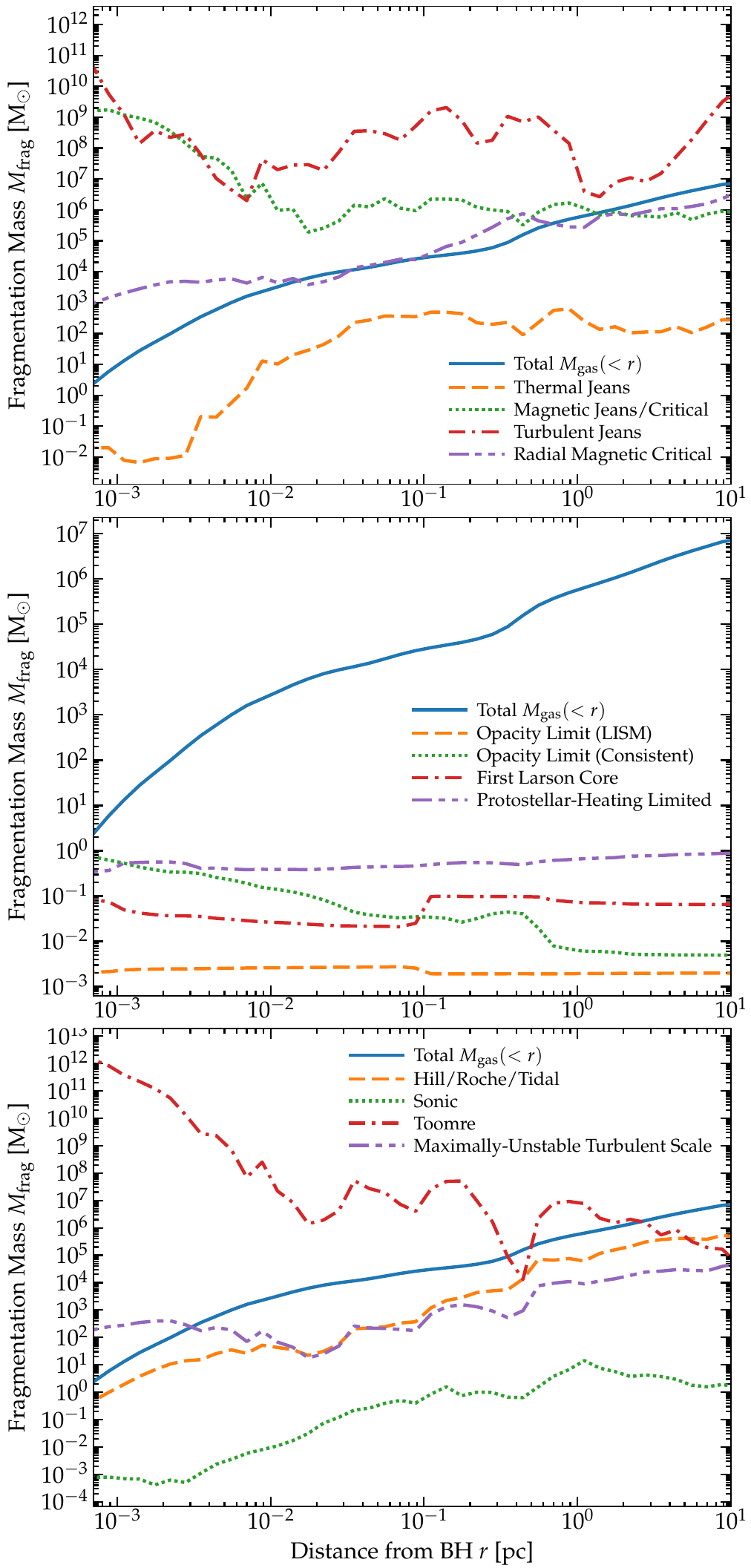} 
	\caption{Different characteristic mass scales versus distance to the SMBH (\S~\ref{sec:imf:physics}), defined as the mass-weighted median of said mass scale at the end of the ``fast'' PSD accretion simulation, in radial annuli. Using the mean or other weighting, or comparing different times, or the ``slow'' PSD simulation, makes little difference. We compare the total enclosed gas mass $M_{\rm gas}(<r)$ and a number of other mass scales which might set the IMF mass scale. None of these alone appears able to entirely explain the IMF trends seen in the simulations \S~\ref{sec:imf:physics:bad}. However some combination may be able to do so for a couple scenarios discussed in \S~\ref{sec:imf:physics:good:turbfrag.for.slow}-\ref{sec:imf:physics:good:feedback.for.fast}.
	\label{fig:profile.fragmass}}
\end{figure}

\subsubsection{Scalings Which Cannot Alone Explain the Simulation IMF Trends}
\label{sec:imf:physics:bad}

First, it is worth discussing many candidate explanations that can be ruled out as the sole explanation for the IMF variation we see. In Fig.~\ref{fig:profile.fragmass}, we show the typical values of many different characteristic mass scales, which one might imagine setting the IMF in principle. None of these exhibits the behaviors needed to fully explain the variation in IMF behavior we see, but we briefly discuss them in turn. To do so, it is helpful to divide them into a few different categories. 

{\bf (1) Jeans-Type Masses:} First, one can consider some Jeans-type length $\lambda_{J} \sim v_{0}/\sqrt{G\,\rho}$ and corresponding Jeans mass $M_{J} \sim (\lambda_{J}/2)^{3}\,(4\pi/3)\,\rho$, which have been argued to be proportional to the IMF mean or turnover mass \citep{bate:2005.imf.mass.jeansmass}. One can then generalize the classic {\bf thermal Jeans} $v_{0}\rightarrow c_{s}$ mass, to a {\bf magnetic Jeans} $v_{0}\rightarrow v_{A}$, or {\bf turbulent Jeans} $v_{0} \rightarrow \langle \delta v_{\rm turb}^{2}(\lambda_{J})\rangle^{1/2}$ mass. We see that the magnetic and turbulent Jeans masses are vastly too-large at all radii to represent some mean mass in the IMF -- indeed, they exceed the total disk mass $M_{\rm gas}(<r)$, which means such collapse would not generically occur. And the thermal Jeans mass is much too high at large $r$ and {\em increases} with $r$ (the opposite of the IMF characteristic mass). We can also define the {\bf magnetic critical mass} $M_{B} \sim 0.1\,\Phi_{B}/G^{1/2}$. If we assume $\Phi_{B} \sim |{\bf B}|\,\pi\,R^{2}$,\footnote{For the actual plot we calculate a more appropriate effective $\int {\bf B}\cdot {\rm d} {\bf A}$ with the projection area defined as the normal to the mean ${\bf B}$ field in cylindrical annuli.} then $M_{B} \sim M_{\rm gas}\,(M_{J,\,B}/M_{\rm gas})^{1/3}$, so this is also much too large ($>M_{\rm gas}$) at all radii of interest, and it (like $M_{J,\,T}$) increases with $r$. It is also well-known that these hypotheses would predict vastly more-dramatic IMF variation in the Solar neighborhood compared to either that observed or seen in simulations with the full physics (e.g.\ stellar feedback) here \citep{guszejnov:imf.var.mw,guszejnov:2019.imf.variation.vs.galaxy.props.not.variable}.

{\bf (2) Opacity/Heating-Limited Masses:} We can also consider the classic opacity limit \citep{rees:1976.opacity.limit} and related arguments, where a critical mass scale is set by some point where heating balances cooling and effective cooling times become long compared to collapse times, preventing fragmentation. If we just take the naive scalings for the LISM, assuming opacities are dominated by dust only and that compression dominates heating, as in \citet{jappsen:2005.imf.scale.thermalphysics,bate:2009.rad.fb.imf,lee:2018.stellar.imf.larson.core}, we obtain the {\bf opacity limit (LISM)} mass. As shown in \citet{grudic.hopkins:2023.opacity.limit.sf.scalings}, one can more carefully generalize this to cases with arbitrary gas-phase plus dust-phase opacities and for the limit where external radiative heating dominates over compressional heating -- using these scalings (e.g.\ Eqs.~19 \&\ 21 therein) with the exact compressional and radiative heating rates and opacities taken directly from the simulations gives the {\bf opacity limit (consistent)} mass. One can then scale to a {\bf first Larson core} mass as shown therein (Eq.~23 therein) assuming quasi-adiabatic collapse past the limit of inefficient cooling \citep{lee:2018.stellar.imf.larson.core}. Alternatively, but related in a physical sense, some models predict that the IMF turnover mass is strictly self-regulated by {\bf protostellar feedback} from radiation \citep{krumholz:2011.cooling.lum.regulates.imf.mass,guszejnov.2015:feedback.imf.invariance}. Alone, these models predict an extremely {\em weak} dependence of the turnover mass on local conditions such as $T$ and $\rho$, as e.g.\ $M_{\rm PSFB} \propto (\rho\,T)^{-1/18}$. Most of these masses scale far too weakly with radius $r$ to explain the IMFs in our simulations, moreover they are much smaller than the mass scales where we see interesting IMF variation. The one potentially interesting case is the consistent opacity limit mass -- this rises at small radii owing to the rapidly-rising gas temperatures and gas-phase opacities (even though the dust is sublimated, the gas opacities can be quite large here), combined with the stronger sensitivity to conditions in the radiation-dominated regime as derived in \citet{grudic.hopkins:2023.opacity.limit.sf.scalings}. But given the (again very low) mass scales of (e.g.\ $\sim 0.03\,{\rm M_{\odot}}$ at $r\sim 0.1\,$pc), this is likely at most relevant for explaining the weak variation in the very low-mass end of the IMF.

{\bf (3) Global Fragmentation Mass Scales:} There are some characteristic global mass scales in fragmentation cascades. One is the {\bf total mass} $M_{\rm gas}(<r)$. This and similarly any {\bf ``tidally limited'' mass}, e.g.\ the classic {\bf Roche} or {\bf Hill} mass (making the analogy of star formation in the QAD around the SMBH to planet formation in a PSD around a single star),\footnote{We show the Roche/Hill mass rather than isolation mass, because the latter can only be self-consistently defined if the tidal radius of stellar ``seeds'' exceeds the disk scale-height $r_{t} \sim (M_{\ast}/4\,M_{\rm BH})^{1/3} R \gtrsim H$ \citep{lin.papaloizou:1986.migration} which would require stars of mass $\gtrsim 10^{6}\,M_{\odot}$ so is not possible here (owing to the large disk thickness). If we simply ignore this and take the usual scaling anyways, $M_{\rm iso} \sim (4\pi r^{2} \Sigma_{\rm gas} f_{\rm H})^{3/2}/(3 M_{\rm BH})^{1/2}$, we obtain $M_{\rm iso} \sim (0.001,\, 1000,\, 10^{7}) {\rm M_{\odot}}$ at radii $R\sim(0.001,\, 0.05,\,10){\rm pc}$, so this predicts qualitatively opposite and orders-of-magnitude discrepant trends from the simulations.} $M_{\rm Hill} \sim \Sigma_{\rm gas}\,H^{2}$, all scale in the opposite manner desired, increasing monotonically with radius $r$. Alternatively the {\bf Toomre mass} scales as the Hill/Roche/Tidal mass multiplied by $\sim Q^{2}$ -- this makes it vastly larger than $M_{\rm gas}(<r)$ and the characteristic IMF scales. Unsurprisingly, on large scales these have more to do with the mass scales of GMCs/star-forming cloud complexes than individual stars. In a turbulent cascade, there are two additional commonly-cited mass scales. First is the {\bf ``sonic mass''} or {\bf turbulent Bonner-Ebert sphere mass} $M_{\rm sonic} \sim c_{s}^{2}\,R_{\rm sonic}/G$ (where $R_{\rm sonic}$ is the scale where the turbulence becomes sub-sonic; \citealt{padoan:2002.density.pdf,hopkins:excursion.imf,hopkins:excursion.clustering,hopkins:excursion.imf.variation,guszejnov:cmf.imf}) -- this is fairly constant around $\sim 1\,M_{\odot}$ at large scales but again increases with $r$ from $r\sim 0.01-1\,$pc. Second is the {\bf ``maximally-unstable turbulent scale''} defined by the first-crossing distribution \citep{hopkins:excursion.ism,hopkins:frag.theory} (Eq.~44, combined with 16 \&\ 24 in \citealt{hopkins:frag.theory}), analogous to the Hill/tidal scale but in an inhomogeneous turbulent medium (and identical to these and the ``turbulent Toomre mass'' in a supersonically-turbulent, $Q_{\rm turb} \sim 1$, weakly-magnetized medium). Again this increases with $r$ (and at the largest radii is much too large to represent the IMF). 

{\bf (4) Accretion-Based Models:} We have extensively discussed (\S~\ref{sec:speed}, \ref{sec:wandering}, \ref{sec:accretion}, \ref{sec:imf:bfields}) how accretion is {\em strongly} suppressed in these stars, beyond their initial collapse masses. So they are accreting {\em less}, not more, from their environments than comparably massive stars in typical GMCs. Thus any explanation for the IMF relying on significant accretion beyond the initial collapsing core (e.g.\ {\bf competitive accretion}) is immediately ruled out.

{\bf (5) Metallicity-Driven or Strictly Cooling-Rate-Driven Models:} Various models have proposed heuristically that the IMF could depend on some simple power-law function of metallicity, or otherwise on the ratio of cooling to dynamical time \citep{cappellari:2012.imf.variation,spiniello:2012.bottom.heavy.imf.massive.gal,narayanan:2013.imf.evol.jeans.cooling,martin.navarro:2015.imf.vs.metals.galaxies}, although more detailed assessments such as \citet{offner:2014.imf.review} have noted that observations require any such dependence be weak and/or highly indirect (as in e.g.\ the opacity limit, which can depend on metallicity, but very weakly, and depends on other parameters as well). However, as shown in \paperone, the entire galactic nuclear region for which we model the IMF is quite chemically homogeneous here: there is no appreciable difference in abundances, nor in the ratio $t_{\rm cool}/t_{\rm dyn} \ll 1$, over the dynamic range where we see strong IMF variation.

\subsubsection{A Potential Explanation for the IMF in the ``Slow'' Accretion Limit: (Gravito)Turbulent Fragmentation in a Quasi-Stable, Quasi-Keplerian, Trans-\Alf{ic} QAD}
\label{sec:imf:physics:good:turbfrag.for.slow}

One possible explanation for the behavior of the IMF in our ``slow'' accretion simulations is suggested by the remarkable similarity between the IMF shape where it is top heavy, and the IMF of {\em planets} formed by gravito-turbulent fragmentation in a quasi-Keplerian disk in \citet{hopkins:2013.turb.planet.direct.collapse} (after, of course, scaling the absolute units of the disk to the QAD here). That study considered turbulent fragmentation (generalizing models for star formation in the ISM from e.g.\ \citealt{hennebelle:2009.imf.variation,hopkins:excursion.imf}), in a Keplerian disk, with a quasi-isothermal EOS ($t_{\rm cool} \ll t_{\rm dyn}$), modest {\rm thermal} Toomre $0.1 \lesssim Q \lesssim 100$, and modestly compressible turbulence (compressive Mach number of $\sim 1$ producing $\mathcal{O}(1)$ local density fluctuations in shocks). While subsequent observations have argued that these assumptions probably do not apply to most LISM PSDs \citep{flaherty:2015.weak.ppd.turb,flaherty2017.weak.ppd.turb,2018ApJ...869L..41A}, they are an excellent description of the QAD. Moreover, unlike in LISM GMC star formation, we can safely ignore the effects of stellar feedback both locally and globally regulating SF in this particular simulation.

Essentially, what happens in this model (and can be more generally derived from the models in \citet{hopkins:frag.theory}), is as follows. As discussed above (\S~\ref{sec:imf:physics:bad}) there are two key turbulent fragmentation scales: the sonic mass $M_{\rm sonic}$ and maximally-unstable turbulent mass $M_{\rm max}$, which define the potential dynamic range of the turbulent fragmentation cascade. When ``effective'' compressive Mach number $\mathcal{M}_{c,\,\rm eff}$ (defined in practice as directly related to the magnitude of rms density fluctuations generated by turbulence, $\delta \rho/\rho \sim \mathcal{M}_{c,\,\rm eff}$) is large ($\mathcal{M}_{c,\,\rm eff} \gg 1$), then the entire fragmentation cascade is realized: objects form initially at the first-crossing $M_{\rm max}$ scale and hierarchically fragment rapidly into progressively smaller sub-structures (e.g.\ clouds/clumps/cores, etc.) producing scale-free mass functions and correlation functions all the way down to the last-crossing/sonic scale $M_{\rm sonic}$ (see e.g.\ \citealt{guszejnov:universal.scalings}), with the dynamic range between the two scaling as $\sim \mathcal{M}_{c,\,\rm eff}^{4}/Q$. In the diffuse ISM, this can qualitatively explain the MFs of GMCs/clumps/cores and the stellar IMF down to the mean/turnover mass $\sim M_{\rm sonic}$ (see references above). But when $\mathcal{M}_{c,\,\rm eff}$ becomes $\lesssim 1$ with $Q_{\rm thermal}+Q_{\rm magnetic} \gtrsim 1$, as considered in \citet{hopkins:2013.turb.planet.direct.collapse} the IMF becomes strongly-peaked around $M_{\rm max}$, because essentially all mass scales are formally ``stable'' if the disk were laminar, but turbulent fluctuations mean there is a small (but non-vanishing) probability of a sufficiently overdense fluctuation which can collapse and that probability is (by definition) maximized at $M_{\rm max}$ (it is generically more difficult to generate a fluctuation with sufficient mass to collapse on progressively smaller scales). So in practice, we might expect the mean sink mass to ``jump'' at this critical point from $M_{\rm sonic} \sim 1-3\,{\rm M_{\odot}}$ at $r \gtrsim \,$pc to $M_{\rm max} \sim 30-100\,{\rm M_{\odot}}$ in Fig.~\ref{fig:profile.fragmass} at $r\ll \,$pc, which agrees at least crudely with the sharp change in behavior seen at the outer QAD (Fig.~\ref{fig:imf.vs.radius}). 

This transition occurs fairly abruptly because of the same physics discussed in \S~\ref{sec:how}. At radii outside the QAD in the CQM, the turbulence is super-\Alf{ic}, highly super-sonic, and the magnetic fields are isotropically tangled (see \paperone\ \&\ \papertwo). In this regime, as discussed at length in \citet{hopkins:2013.turb.planet.direct.collapse} and \citet{hopkins:frag.theory} and shown in idealized numerical simulations in \citet{kowal:2007.log.density.turb.spectra,lemaster:2009.density.pdf.turb.review,kritsuk:2011.mhd.turb.comparison,molina:2012.mhd.mach.dispersion.relation}, magnetic fields only modify the effective compressive Mach number by a geometric pre-factor $\mathcal{M}_{c,\,\rm eff} \rightarrow b\,\mathcal{M}_{\rm sonic}/\sqrt{3} $ (with $b \sim 1/2$) as $\beta\rightarrow 0$, because the fields cannot resist compressive motions along the field lines. But in the QAD, the turbulence rapidly drops to sub-\Alf{ic} (\paperone) and the field becomes dominated by the mean toroidal field, which strongly suppresses the compressibility of the gas to Toomre-style modes (suppressing $\mathcal{M}_{c,\,\rm eff}$ by a factor $\sim c_{s}/v_{A} \sim \beta^{1/2}$ in the perpendicular radial/vertical directions; see \citealt{hopkins:frag.theory} \S~6 for a theoretical discussion or \citealt{beattie:2021.turb.intermittency.mhd.subalfvenic} for simulation examples), while of course strong shear in the QAD prevents azimuthal collapse (suppressing $\mathcal{M}_{c,\,\rm eff}$ in that direction by a factor $\sim c_{s} / \kappa\,H \sim c_{s}/v_{A}$ for the magnetically-supported disks here, see \citealt{hopkins:2013.turb.planet.direct.collapse}).\footnote{Another way of saying this is that the sonic mass/length defined purely in terms of the sonic Mach number $\mathcal{M}_{s}$ is no longer relevant in the QAD, because in turbulent fragmentation models this is just a proxy for the scale below which large compressions are rare/difficult, so the ``effective'' sonic mass jumps to a value $\gtrsim M_{\rm max}$.} So $\mathcal{M}_{c,\,\rm eff}$ effectively drops to $\sim \mathcal{M}_{A} \lesssim 1$, and the IMF changes abruptly. For the reasons in \S~\ref{sec:wandering}-\ref{sec:accretion}, stars accrete little after their formation, so the IMF is ``frozen'' after initial collapse.

\subsubsection{A Potential Explanation for the IMF in the ``Fast'' Accretion Limit: Feedback-Regulated Stellar Masses}
\label{sec:imf:physics:good:feedback.for.fast}

In the ``fast'' accretion model, the behavior of the IMF is somewhat more subtle. As discussed in \S~\ref{sec:imf:accmodel}, the key difference is that in this case, stellar feedback -- primarily in the form of radiation from accretion+contraction luminosity and protostellar jets, which scale with the sub-grid mass and accretion rate onto the ``star'' within the sink -- is able to act because mass can transfer from sink to ``star'' rapidly compared to the local QAD/CQM dynamical time.

At large radii in the CQM, the physics is probably similar to the LISM. The sonic mass may play a role in setting the ``turnover'' mass \citep{guszejnov:2020.mhd.turb.isothermal.imf.cannot.solve}, but this is modified by feedback: massive stars in particular self-regulate via feedback making the high-mass slope less steep \citep{grudic:2022.sf.fullstarforge.imf}, and as many studies have shown protostellar jets shift the IMF peak mass a factor of a few\footnote{We stress that this is not ``put in by hand'': the jet mass-loading is just $\sim 30\%$ of the accreted mass \citep{grudic:starforge.methods}. But as shown in these studies the jets efficiently entrain material which would otherwise accrete onto the protostar. Here, while we showed the jets have little effect on the {\em global} properties of the QAD/CQM, they by definition can unbind a significant mass from accreting onto the protostar since they are assigned a launch velocity comparable to the escape velocity from the surface of the star, which of course must be larger than the escape velocity from the local potential perturbation caused by the star itself on larger (resolved) scales at $\sim 10-1000\,$au distances (even if it is less than the escape velocity from the global potential of the SMBH, so the outflows are not unbound from the global QAD/CQM).} lower \citep{guszejnov:2020.starforge.jets,guszejnov:environment.feedback.starforge.imf,lebruilly:2023.jets.needed.for.reasonable.imf}, naturally explaining the difference in the IMF shapes and peak masses between the ``fast'' and ``slow'' accretion models (Figs.~\ref{fig:imf.toplevel}-\ref{fig:imf.vs.radius}). 
Any modest residual difference between the IMF shape at $\gtrsim$\,pc scales and the ``universal'' IMF can be largely explained by the suppression of fragmentation on smaller scales (\S~\ref{sec:imf:multiplicity}), so the predicted IMF is more like the system IMF (Fig.~\ref{fig:imf.system.vs.sink}) -- but we caution that we need higher resolution to make that a more rigorous and quantitative statement (\S~\ref{sec:resolution}). We do not see much evidence at the lowest masses for a role of a radially-dependent opacity limit as in Fig.~\ref{fig:profile.fragmass} (if anything, the ``cutoff'' mass of the IMF may be slightly larger at the largest $r$, opposite this expectation), but if the ``first Larson core mass'' is more relevant than the ``opacity limit mass'' the weak evolution in Fig.~\ref{fig:profile.fragmass} is mostly cancelled out and it would explain the weak evolution in the lower-limit or ``cutoff'' mass at the lowest resolved masses.

But clearly, something still happens around the outer QAD (see Fig.~\ref{fig:imf.vs.radius}). At that specific radius, the IMF is comparably top-heavy and shallow in slope to the ``slow accretion'' simulation, though with perhaps the peak mass shifted down by a factor of a few. The shift between ``fast'' and ``slow'' models again can be attributed to feedback. But the stark ``jump'' to a more top-heavy shape of the IMF in the outer QAD likely has a similar origin to that discussed in \S~\ref{sec:imf:physics:good:turbfrag.for.slow} -- this is a special environment where the effective compressibility of the gas is strongly suppressed. This overall suppresses star formation, but means that where stars do form, they are most likely to be at the largest (most unstable) scales. This is further supported by the fact that we see the sites of star formation in the outer QAD are the same for the the ``fast'' and ``slow'' cases (\S~\ref{sec:how}). 

The one behavior in the ``fast'' run which is somewhat more difficult to explain with the arguments above is the apparent reversion to a ``normal'' IMF at slightly smaller radii $0.02<r<0.05\,$pc, just before SF is completely suppressed at smaller radii in the QAD (see Fig.~\ref{fig:imf.vs.radius}). But as discussed in \S~\ref{sec:imf:distance}, the statistics for this region are quite sparse (there are just a few stars forming here) so it is difficult to draw a robust conclusion. But this appears to be worth future investigation.

\subsection{Is Star Formation in the QAD Analogous to Planet formation in a ``Typical'' PSD?}
\label{sec:sf.qad.planets.psd}

It is common to draw the analogy between star formation in a QAD and planet formation in ``typical'' PSDs (at least planet formation via gravitational instability, as opposed to solid core formation), and indeed we did this to some extent in \S~\ref{sec:imf:physics}. As argued there, some parts of that analogy are reasonable -- after all, gravity is strictly scale-free, and the mass ratios are conceivably similar in the QAD-star and PSD-planet cases. And indeed some key aspects of gravito-turbulent fragmentation appear to cross-apply, at least with a {\em careful} accounting for the correct ``effective'' compressive Mach numbers (after accounting for the effects of magnetic fields and radiation explicitly), as discussed in \S~\ref{sec:imf:physics:good:turbfrag.for.slow}.

However, most of our discussion above should make it clear that the analogy should not be pushed too far. Much of the physics which plays a crucial role (e.g.\ the thermo-chemistry) is not scale-free. And much of the physics that is scale-free in principle (e.g.\ ideal MHD) occupies a qualitatively different dimensionless parameter space (see \S~\ref{sec:ism.diff}). Most obviously, the plasma $\beta$ in the QADs here is $\beta_{\rm QAD} \sim 10^{-4}$, whereas in PSDs it is generally assumed to be $\gg 1$ and in standard models for MRI-active PSDs, the effective viscosity parameter $\alpha \sim \beta_{\rm PSD}^{-1}$ implying $\beta_{\rm PSD} \gtrsim 10^{4}-10^{6}$ in observations (see \citealt{hughes:2011.turb.protoplanetary.disk.obs,flaherty:2015.weak.ppd.turb,flaherty2017.weak.ppd.turb,2018ApJ...869L..41A}). Moreover in the QAD, the magnetic fields are extremely well-coupled to the gas and well in the ideal limit, with ionization fractions (given the warm temperatures and extreme radiation environments) rising from $f_{\rm ion} \sim 0.01$ in the outskirts to $\sim 1$ in the inner QAD. In contrast typical PSDs are expected to have $f_{\rm ion} \ll 10^{-15}$ throughout much of the PSD, making magnetic fields as a whole far less dynamically important and making Hall MHD and Ohmic resistivity prominent \citep{wardle.2007:mhd.protoplanetary.disk.review,keith:2014.planetary.disk.ionization.model,tsukamoto.2015:ohmic.ambipolar.firstcores,bai:2017.mhd.disk.winds,zhao:2020.hall.fx.ppds,lee:2021.non.ideal.mhd.fx.planet.disks}. As discussed in \S~\ref{sec:ism.diff}, \ref{sec:how}, \ref{sec:imf:bfields} these magnetic fields also elevate $Q$ and strongly suppress gravitoturbulence and fragmentation.

Similarly, the QADs here are vastly more strongly turbulent even in a dimensionless sense compared to observed PSDs. Typical sonic Mach numbers in the QAD are $\mathcal{M}_{s} \sim 30-100$, and even the (much lower) \Alf\ Mach numbers are $\mathcal{M}_{A} \sim 1$, while for PSDs inferred Mach numbers are $\mathcal{M}_{s} \sim 0.001 - 0.01$ \citep{hughes:2011.turb.protoplanetary.disk.obs,flaherty:2015.weak.ppd.turb,flaherty2017.weak.ppd.turb,2018ApJ...869L..41A}. Both turbulence and magnetic support make the QADs much thicker with $H/R \sim 0.1-1$, which has important consequences for star-gas drift velocities (\S~\ref{sec:wandering}), QAD midplane densities (Fig.~\ref{fig:profile.dynamics}), accretion rates, migration times, and many other processes.

Related to this, the cooling physics is qualitatively different, not just in the dominant cooling channels and opacities (though these are fundamentally different, with dust sublimated in the QAD and orders-of-magnitude stronger radiation fields and cosmic ray densities), but in the dimensionless ratio of cooling time $t_{\rm cool}$ to dynamical time $t_{\rm dyn}$. For a PSD, except perhaps for a very short transient period when they first form, one generically expects (and again, observations appear to support) $t_{\rm cool} \gg t_{\rm dyn}$, while for the QADs here cooling is efficient with $t_{\rm cool} \ll t_{\rm dyn}$. This also means that the QADs can be multi-phase and feature radiative (quasi-isothermal) shocks, and qualitatively different behaviors upon contraction. Yet despite the much faster cooling in this sense in the QADs, the absolute temperatures of the QAD are significantly warmer owing to the strong accretion heating and extreme environment. Those higher QAD temperatures and accretion rates also mean that the QADs feature orders-of-magnitude higher ratios of radiation pressure to thermal gas pressure: in typical PSDs the radiation-to-gas-thermal pressure ratios are $\sim 10^{-7}$ \citep[e.g.][]{chiang:2010.planetesimal.formation.review}, while in the QADs here they exceed unity throughout much of the QAD (and even in the outermost QAD never drop much below $\sim 1$). 
 
And obviously, even if dust were not sublimated in the QADs, the Stokes numbers are orders-of-magnitude different and efficient coagulation will not occur as it does in PSDs, so there is also no direct analogy for planetesimal formation, pebble accretion, the streaming instability (though other resonant drag instabilities do exist in QAD environments, they behave in qualitatively different ways; see \citealt{squire.hopkins:RDI,squire:rdi.ppd,soliman:agn.variability.from.dust.instabilities}), or core accretion.

These differences will also strongly modify the dynamics of the extant stars and black holes in the QAD, via their coupling to the gaseous disk (for one example, see \S~\ref{sec:wandering}). But even for a pure $N$-body disk, i.e.\ after the gas disk is fully consumed or expelled, there are some subtle but important distinctions. In a planetary disk, the deviations from Keplerian orbits are, to leading order, dominated by an $\mathcal{O}(1)$ number $N$ of large bodies (e.g.\ Jupiter), while for a QAD (assuming the gas has been removed) they are dominated by $N\gg 1$ broadly similar-mass stars and/or compact objects. Moreover for the planetary disk, the total mass of orbiting bodies is generally much less than the mass of the host star, while it is basically guaranteed that in the outer CQM (approaching the BHROI), the total enclosed mass of stars becomes comparable to the SMBH mass. And as noted above the relic stellar disks here are thick with $H/R \sim 0.1-1$, not thin. These differences will strongly influence and often qualitatively change the outcomes of phenomena such as migration, resonant relaxation, propagation of $m=1$ modes and other perturbations, and resonance and trojan trapping.

\subsection{Future Work: Generalizing to Other Environments}
\label{sec:generalizations}

Various recent simulations and observational arguments summarized in \S~\ref{sec:intro} suggest that qualitatively similar, strongly-magnetized QADs may be the norm for high accretion rate AGN (with Eddington ratios $\gtrsim 0.01$, see \citealt{hopkins:multiphase.mag.dom.disks}). This suggests that many of our qualitative conclusions might apply across a broad range of systems. It is more challenging to make quantitative predictions for e.g.\ the characteristic IMF turnover mass and shape, given our single case study. Even within our specific ICs, we have stressed this depends strongly on the uncertain PSD-scale accretion physics (\S~\ref{sec:imf:accmodel}). Still, in future work, it would be useful to explore analytic models, taking either the ``fast'' or ``slow'' regimes. For example, taking the analytic model discussed in  \S~\ref{sec:imf:physics:good:turbfrag.for.slow} \citep{hopkins:2013.turb.planet.direct.collapse}, for the IMF as a function of QAD properties, where magnetic fields suppress fragmentation strongly, coupled to the analytic model for the salient disk structural properties from \citet{hopkins:superzoom.analytic}. But even then, as noted in \S~\ref{sec:imf:physics:good:turbfrag.for.slow}, this model only applies interior to radii where the gas and magnetic fields circularize (building a strong mean toroidal field) and the QAD has sufficiently large, but not {\em too} large, Toomre $Q$ and optical depth to its own cooling radiation. At larger radii (where fields are super-\Alf{ically} turbulent, $Q$ is relatively small, and the gas is more optically thin) star formation proceeds more akin to the ISM (e.g.\ \S~\ref{sec:where:filament}, \ref{sec:how}), while at smaller radii star formation is suppressed completely. So one would further need to couple these to analytic models for the opacities and thermochemical properties (e.g.\ \citealt{hopkins:multiphase.mag.dom.disks}) as well as models for the magnetic field and gas geometry/circularization at radii $\sim 0.01-10\,$pc interior to the BHROI. It will clearly be helpful to these efforts to study star formation in other similar hyper-refinement simulations studying different accretion rates, BH masses, and environments.

\section{Conclusions}
\label{sec:conclusions}

We have studied the stellar IMF in the outer regions of quasar accretion disks and the circum-quasar medium (QADs and CQM), where the SMBH dominates the potential and star formation is suppressed but still can occur at some level. For the first time, we consider this with a combination of detailed physics, which have been shown (in previous STARFORGE papers) to reproduce a plausible observed IMF in LISM/``typical'' Galactic GMC conditions, including: non-equilibrium ion/atomic/molecular/dust thermo-chemistry coupled to explicit multi-band radiation transport capable of handling both optically thick and thin regimes, magnetic fields, self-gravity, and star formation and accretion with detailed proto-stellar and main-sequence stellar evolutionary models enabling feedback in the form of (proto)stellar jets, winds/mass-loss, radiation, and supernovae. We also for the first time apply this to simulations which self-consistently form a QAD around a SMBH from cosmological initial conditions, so we do not have to make an ad-hoc assumption for critical properties such as boundary/initial conditions, magnetic field strengths/geometries, etc. Our important conclusions include:

\begin{enumerate}

\item{\bf Star formation is possible in the CQM \&\ outer QAD}, though it is strongly suppressed on sub-pc scales around the quasar by magnetic fields and optical depth effects (\paperone), and almost completely vanishes on $< 0.01\,$pc scales. With the physics here the SFR is much less than the inflow rates to the SMBH so stars and SF do not appreciably influence the structure or global dynamics of the QAD or quasar accretion. It is possible that much later, very low-level accretion could be driven by stellar mass-loss after the QAD is completely consumed or blown away, but this would necessarily involve orders-of-magnitude lower accretion rates. 

\item{\bf The sites of star formation are ``special.''} There are two ``zones'' of star formation with different detailed structure. In the CQM inside the BHROI but outside the circularized QAD, star formation occurs in fragmenting filamentary inflows enhanced by strong super-sonic transverse tidal compression/shocks, producing a population of stars on highly radial/eccentric orbits. Inside where the QAD circularizes (the outer QAD) star formation occurs via gravito-turbulence in special regions where toroidal magnetic field lines stretched from the CQM reverse sign, hence where magnetic pressure is locally unable to fully-stabilize the QAD against local collapse.

\item{\bf The IMF is likely top-heavy at ``intermediate'' radii $\sim 0.1\,$pc} between the innermost CQM and outermost QAD. At larger radii approaching the BHROI, the system IMF approaches the ``typical'' or ``universal'' IMF form (despite the fact that the physical conditions at the BHROI are still orders-of-magnitude different from typical Solar-neighborhood GMCs). At very small radii $\ll 0.01\,$pc essentially no star formation occurs. The variation of IMF versus radius within the QAD (but outside this radius) depends on the unresolved sub-grid assumptions regarding efficiency of accretion from PSDs onto stars in these extreme conditions, but the IMF can at some point become more bottom-heavy at small radii simply because the total enclosed QAD gas mass (and corresponding Hill mass) at some $R$ become very small ($\ll 10\,M_{\odot}$) and so massive star formation is not possible.

\item{\bf {\em How} ``top-heavy'' the IMF actually is, and perhaps whether close binaries and/or planets can form, depends sensitively on unresolved  accretion from sink radii to (proto)stars}. Even in our extremely high-resolution simulations, we cannot resolve the interior structure of accretion flows/PSDs around newly-forming protostars interior to the numerical sink radii. Typical assumptions motivated by PSDs in the Solar-neighborhood ISM naively predict timescales for accretion over this range of scales which are many thousands of global orbital times in the QAD -- in other words, essentially no accretion onto the (proto)star. But this is almost certainly unphysical, as the accretion ``reservoir,'' even if it could form a PSD (which seems unlikely) must be far more compact, dense, warm, and strongly magnetized if they form in the QAD owing to the extreme environment. We show that assuming the accretion is ``fast'' (occurring on a few sink dynamical times) leads to a far less extremely top-heavy system IMF. This clearly motivates dedicated simulations of {\em individual} core and (proto)star formation which can resolve scales $\ll$\,au, with initial/boundary conditions motivated by the simulations here.

\item{\bf Accretion beyond the initial core is negligible}. Beyond the initial tightly-bound core collapse, stars accrete a negligible amount of gas from the CQM/QAD, even integrating their trajectories over $\sim 10^{5}$ QAD dynamical times during which the global QAD properties (e.g.\ the hypothetical mass supply available to the star) are constant. If anything, the IMF becomes slightly {\em less} top-heavy with time. All the initial core accretion occurs on $\lesssim 1$ initial clump dynamical time. The accretion is strongly suppressed, even in the absence of any (proto)stellar feedback, by the strong magnetic fields, turbulence, radiation, warm gas pressures, and non-trivial orbit structure (e.g.\ eccentric and spiral modes in the QAD as well as stellar-gas orbit separation as detailed below). This is qualitatively different from simulations with the same code and physics in Solar-neighborhood GMCs, where the most massive stars become so by sustaining accretion over $\gtrsim$\,Myr timescales \citep{grudic:2022.sf.fullstarforge.imf,guszejnov:2022.starforge.cluster.assembly}.

\item{\bf Stars almost immediately dynamically decouple from gas after formation}, and follow orbits which are fundamentally distinct from the surrounding gas, with typical systematic relative velocities between the stars and the immediately surrounding ambient gas between $\sim 0.1-1$ times the Keplerian speed, i.e.\ $\gtrsim 100-1000\,{\rm km\,s^{-1}}$ at the radii of interest. This is expected and indeed inevitable, for fundamental dynamical reasons in any disk which has non-negligible eccentricity, warps, precession, pressure/magnetic/radiation/turbulent support, sinking of clumps and/or stars, etc. As a consequence, stars and gas are not co-orbital, which has important effects on their post-initial collapse evolution, helping to suppress accretion by orders of magnitude, and leaving ejecta/jet material behind on cometary trails rather than allowing it to coherently locally regulate the structure of the gas.

\item{\bf Multiplicity on $\gg 10\,$au scales is strongly suppressed}, owing to a combination of inefficient fragmentation (the same physics that pushes the IMF to be top-heavy) and the strong tidal environment disrupting any binaries that are not very compact/hard. Whether compact binaries form is un-resolved here and probably depends on the same uncertain physics of the un-resolved PSD accretion/structure described above.

\item{\bf Magnetic fields play a crucial role.} The magnetic fields in the CQM and QAD are extremely strong (plasma $\beta \sim 10^{-4}-10^{-2}$, field strengths $\gg$ Gauss) and the ionization fractions extremely high ($\gtrsim 0.01$, given the warm gas temperatures and extreme radiation environments) -- hence the magnetic fields much more strongly-coupled -- compared to typical Galactic GMCs and PSDs. We showed in \paperone\ and \papertwo\ that the properties of the CQM and QAD are fundamentally different without magnetic fields, including (notably) that the system experiences runaway catastrophic fragmentation on these scales if we (unphysically) remove said fields. So obviously they shape the initial conditions in a key manner at all radii we consider (not just in the QAD). We show here that the ensuing IMF absent magnetic fields is vastly more top-heavy than the (already top-heavy) IMF with magnetic fields, and becomes even more top-heavy with increasing distance from the SMBH, so magnetic fields are also playing a key role regulating the masses of these systems.

\item{\bf Many proposed explanations/scalings for the IMF mass are ruled out} at least as an explanation for the simulations here. For example, simply assuming the IMF turnover mass should scale with e.g.\ a thermal Jeans/Toomre/magnetic critical/Hill/Roche/tidal/first Larson core/opacity limit mass, gives the qualitatively incorrect scalings compared to the behavior we see. Likewise explanations predicated on competitive accretion are not viable given the lack of accretion noted above. We argue that the IMF shape in simulations with slow PSD accretion (hence weak stellar feedback) are potentially explained by analytic models for gravito-turbulent fragmentation in a disk with trans-\Alf{ic} turbulence and efficient cooling from \citet{hopkins:2013.turb.planet.direct.collapse}; while in simulations with fast PSD accretion (hence more efficient stellar feedback) are better explained by a feedback-regulated model as in \citet{guszejnov:environment.feedback.starforge.imf}.

\end{enumerate}

There are many obvious interesting questions to explore in future work, including more detailed exploration of properties such as multiplicity and stellar correlation functions, and pushing further out in distance from the SMBH at sufficiently high resolution to model the IMF in galactic nuclear starburst environments. It would also of course be interesting to explore similar simulations zooming in on different times and in different galaxies accretion phases, to explore different conditions, such as star formation around SMBHs more akin to Sgr A$^{\ast}$ (at much later cosmic times with vastly lower gas supply to the nucleus even during the presumptive star-forming episode). Indeed, although it is intriguing to note that the predicted IMFs here are top heavy at a similar spatial scale around the SMBH to where the Galactic nucleus is also observed to have a top-heavy IMF (and the IMF shapes are potentially similar to those observational claims), we strongly caution that we are simulating a vastly different environment from the Galactic nucleus. The simulations here or others like them could also be used to explore predictions and consequences for LIGO sources and TDEs (evolving the stars and their relics dynamically in more idealized simulations or analytic calculations, using their formation properties here as initial conditions). And of course, as we have noted above, it would be especially interesting, and important for informing future generations of circum-SMBH star formation simulations, to consider much higher-resolution/smaller-scale simulations of individual cores/accretion flows/PSDs and formation of individual massive stars with initial/boundary conditions taken from the simulations here, to inform the crucial uncertain sub-grid physics herein.

\begin{acknowledgements}
We thank the anonymous referee for a number of insightful suggestions. Support for PFH was provided by NSF Research Grants 1911233, 20009234, 2108318, NSF CAREER grant 1455342, NASA grants 80NSSC18K0562, HST-AR-15800. Numerical calculations were run on the Caltech compute cluster ``Wheeler,'' allocations AST21010 and AST20016 supported by the NSF and TACC, and NASA HEC SMD-16-7592. Support for MYG, KK, and DG was provided by NASA through the NASA Hubble Fellowship grants \#HST-HF2-51479, 51506, 51510 awarded  by  the  Space  Telescope  Science  Institute,  which  is  operated  by  the   Association  of  Universities  for  Research  in  Astronomy,  Inc.,  for  NASA,  under  contract NAS5-26555. 
\end{acknowledgements}

\bibliographystyle{mn2e}
\bibliography{ms_extracted}

\begin{appendix}

\section{Potential Effects of Resolution} 
\label{sec:resolution}

\begin{figure*}
	\centering\includegraphics[width=0.98\textwidth]{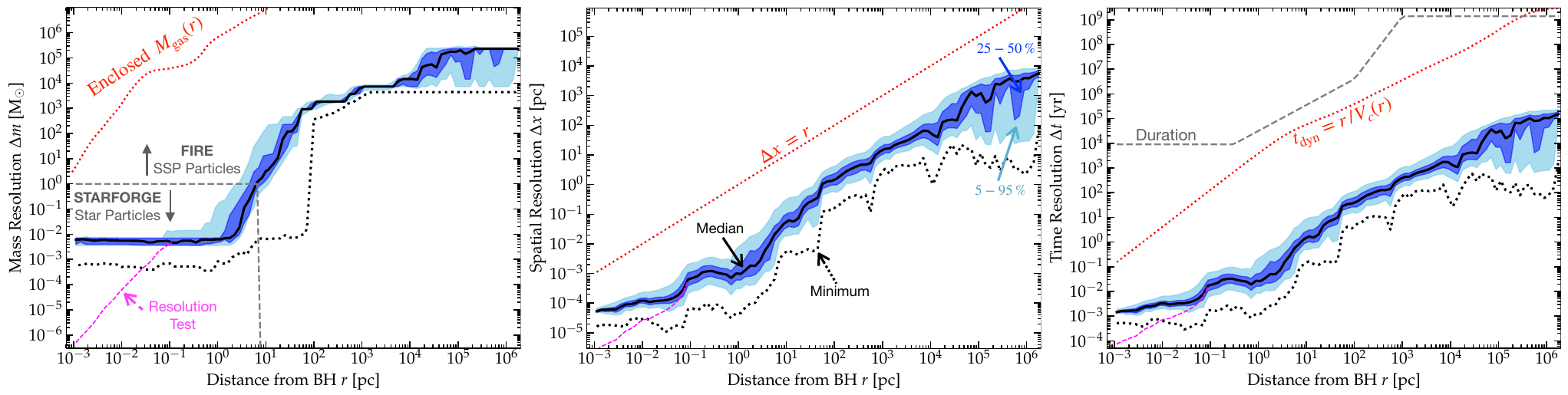}
	\caption{Effective resolution (\S~\ref{sec:methods}) of the simulation as a function of radial distance from the SMBH ($r$). We show the minimum ({\em dotted black}), median ({\em solid}), $50\%$ and $90\%$ inclusion intervals ({\em shaded}) of the mass resolution (gas cell mass $\Delta m$; {\em left}), spatial resolution (equivalent cell size $\Delta x \equiv (\Delta m/\rho)^{1/3}$; {\em center}), and time resolution (numerical timestep $\Delta t$; {\em right}), at each $r$.  
	For reference, we compare the total enclosed gas mass ($M_{\rm gas}(<r)$), radius ($r$), dynamical time ($t_{\rm dyn}\equiv \Omega^{-1} = r/V_{c}(r)$), and how long the simulation was run after reaching maximum refinement at each radius (``duration''). 
	We also show the resolution test with additional refinement in the QAD from \papertwo\ ({\em dashed magenta}), run for much shorter duration.
	We denote where the simulation treats star formation using the un-resolved FIRE ``single stellar population particle'' (SSP) approach (sampling an assumed IMF), versus STARFORGE individual star particles. The scales of interest here (interior to the BHROI at a few pc) are uniformly resolved at the target $\Delta m \sim0.001-0.01\,M_{\odot}$, with spatial (time) resolution reaching $\sim 10$\,au (hours). 
	The duration at highest resolution $\sim 10^{4}$\,yr corresponds to $\sim 10^{5}$ orbital dynamical times of the innermost QAD. 
	Adapted from Fig.~6 in \paperone, but remade for the times analyzed here adding the resolution test and minimum-resolution information.
	\label{fig:res}}
\end{figure*}

We discussed the typical simulation numerical resolution in Fig.~\ref{fig:res} and \S~\ref{sec:methods}. Of course, ideally one would always like to push to even higher-resolution. However given the extremely-small timesteps of $\sim$\,days involved in these simulations (owing to the short dynamical and radiation timescales), it is not possible to do so while running for even a modest fraction of the (already-limited) run-time at highest resolution ($\sim 10^{4}\,{\rm yr}$; see Fig.~\ref{fig:res}). It is also worth stressing that our ``default'' resolution is already significantly finer than any of the historical, idealized circum-SMBH IMF studies referenced in \S~\ref{sec:intro}. 

More extensive STARFORGE resolution tests for LISM conditions (where such tests are feasible) are presented in \citet{guszejnov:2020.starforge.jets,grudic:starforge.methods,guszejnov:environment.feedback.starforge.imf}. These studies argued that the predicted IMFs (the subject of our discussion in \S~\ref{sec:imf}) were reasonably well-converged for ${\rm M_{\odot}} \gtrsim 100\,\Delta m$ ($100\times$ the mass resolution), or $\gtrsim 0.5\,{\rm M_{\odot}}$ here at the median resolution in the QAD/CQM. And we stress that the global, integrated quantities in these simulations which we study (e.g.\ the total mass of sinks, star formation efficiency, locations within GMCs where most stars formed, bounded-ness of the stars, etc., and the subjects of our discussion in \S~\ref{sec:ism.diff}-\ref{sec:feedback}) are converged in these tests even at much poorer resolution than our fiducial simulation, as low as $\sim 10^{-2}-10^{-1}\,M_{\odot}$, suggesting the simulations should be robust in these respects.

For the IMF specifically, our primarily focus on the ``top-heavy'' behavior in \S~\ref{sec:imf} was at much larger masses going up to $\sim 100\,M_{\odot}$, where the key behaviors (and e.g.\ Salpeter slope) are reasonably well-converged in the LISM-like resolution tests in \citet{guszejnov:2020.starforge.jets,grudic:starforge.methods,guszejnov:environment.feedback.starforge.imf}. Moreover, the offsets we study in the IMF (even around the $\sim 0.1-1\,M_{\odot}$ mass range) are generally much larger than the effects from incomplete convergence at this ($\sim$\,few $\times10^{-3}\,{\rm M_{\odot}}$) resolution in \citet{grudic:starforge.methods}. For example, the systematic differences we predict between the IMFs with the ``fast'' versus ``slow'' accretion models (\S~\ref{sec:imf:accmodel}), or at different SMBH-centric radii (\S~\ref{sec:imf:distance}) or in simulations with or without magnetic fields \S~\ref{sec:imf:bfields}, are all much larger than the difference between the IMF at resolutions from $\sim 10^{-5}-10^{-2}\,{\rm M}_{\odot}$ in \citet{grudic:starforge.methods}. Furthermore, there are good physical reasons to think the convergence may be better here than in the tests in small LISM clouds presented in those papers: for example, most of the resolution dependence of the IMF in those tests owed to the ability to resolve more fragmentation of individual cores into multiple (lower-mass) stars at higher resolution. But owing the extreme tidal fields and magnetic fields, we argued that such fragmentation is and must be very strongly suppressed here (see \S~\ref{sec:imf:multiplicity} \&\ \ref{sec:imf:physics}). Thus our key conclusions should be robust.

That said, we have still attempted to check for additional resolution dependence. In \papertwo, we describe a set of resolution tests for both our default simulation and the ``no magnetic fields'' simulations, where we apply an additional refinement layer applied interior to $<0.1\,$pc, with mass resolution reaching as small as $\sim 10^{-7}\,{\rm M}_{\odot}$ at $\lesssim 80\,$au, shown in Fig.~\ref{fig:res}. By necessity owing to the increased computational expense this could only be run for $\sim 10^{3}$ dynamical times at the innermost radii, or $\sim 100\,$yr, and the increased resolution was focused on the smallest radii where the enclosed gas mass was small (reaching such resolution inside the entire BHROI would require $\sim 10^{14}$ gas cells, an infeasible number). The primary goal of that study was (as discussed in \papertwo) not to study the IMF, but to confirm that the global QAD properties in the inner regions were converged. But this still allows us to confirm a few important conclusions of ours: (1) the global properties of the QAD at $\lesssim 0.1\,$pc are robust to resolution; (2) the strong suppression of SF at small radii is also robust -- we see zero sinks form for the duration of this higher-resolution run interior to $\lesssim 0.05\,$pc; (3) the weak effects of stellar ejecta which reach this radii, and the dynamics of the (older) massive stars on eccentric orbits going through the inner disk, are essentially identical. However, because of the refinement scheme adopted, this only applies a modest additional refinement in the outermost QAD where star formation occurs (factor of $\sim 4$ in mass resolution at $\sim 0.05\,$pc), and none in the CQM (see Fig.~\ref{fig:res} for details), and the duration of the additional hyper-refinement ($\sim 100\,$yr) is just $\sim 1$ dynamical time in the outer QAD at $\sim 0.1\,$pc. So this is not so useful for studying the effects of resolution on the IMF and multiplicity statistics. We do see a few sinks ($6$, specifically) form in the outer QAD ($\sim 0.05-0.1\,$pc) after the additional refinement layer is applied, and can confirm (1) they form exclusively at the toroidal magnetic field switch locations as we saw above (\S~\ref{sec:where}); (2) include zero binaries, consistent with the low binary fraction seen above (\S~\ref{sec:imf:multiplicity}); and (3) are statistically consistent with being drawn from the young ($<100\,$yr-old) IMF of stars formed at the same radii in our ``default'' run (\S~\ref{sec:imf:distance}) -- but we stress that the latter two statements in particular are not highly statistically significant, given the tiny sample size.

Unfortunately, refining even one-level higher in the CQM is extremely computationally expensive, as the total gas mass enclosed in the BHROI ($\sim 10^{7}\,{\rm M_{\odot}}$) means this requires $\sim 10^{10}$ gas cells. We attempted such a refinement but given the extreme cost could only run this test for a small fraction of one dynamical time at these larger radii, so the system could not relax and we do not consider this a particularly useful test. So we suspect the sub-Solar IMF at these radii $\gtrsim 1\,$pc, where tidal effects from the BH may not suppress fragmentation as severely and the IMF begins to approach a more LISM-like form, to perhaps be the most uncertain of the predictions we present here. In future work, dedicated hyper-refinement centered on a smaller total gas mass (as opposed to refining on everything within some radius of the SMBH) could be used to study better the behavior of the IMF at these intermediate radii, as well as the IMF in the ISM of the starburst galaxy (at $\sim $\,kpc scales, where our current resolution and FIRE treatment prevent any meaningful IMF predictions).

\end{appendix}

\end{document}